\documentclass[12pt]{article}
\usepackage{amsmath, latexsym, amssymb, hyperref, graphicx, slashed, color, multirow}
\definecolor{nicered}{rgb}{.1,.2,.8}
\definecolor{nicegreen}{rgb}{.1,.5,.7}
\definecolor{darkblue}{rgb}{0,0,.5}
\hypersetup{colorlinks, citecolor=nicegreen,linkcolor=nicered, urlcolor=darkblue}

\usepackage[T1]{fontenc}
\usepackage{slashed}
\usepackage{epstopdf}
\usepackage{booktabs}
\usepackage{mathrsfs}
\usepackage[export]{adjustbox}
\usepackage{epsfig}
\usepackage{multirow}
\usepackage{cite}

\newcommand{\bea}{\begin{eqnarray}}
\newcommand{\eea}{\end{eqnarray}}

\numberwithin{equation}{section}

\begin{document}

%
\begin{titlepage}
\begin{flushright}
OSU-HEP-18-02 
\end{flushright}
\vspace*{10mm}
\begin{center}
\baselineskip 25pt 
{\Large\bf
Displaced vertex signature of type-I seesaw
}
\end{center}
\vspace{1mm}
\begin{center}
{\large
Sudip Jana$^{a,b,}$\footnote{\color{blue} {sudip.jana@okstate.edu}}, 
Nobuchika Okada$^{c,}$\footnote{\color{blue} {okadan@ua.edu}}
and
Digesh Raut$^{c,}$\footnote{\color{blue} {draut@crimson.ua.edu}}
}

\vspace{.5cm}

{\baselineskip 20pt \it
$^{a}$Theory Department, Fermi National Accelerator Laboratory, \\ P.O. Box 500, Batavia, IL 60510, USA \\
$^{b}$Department of Physics, Oklahoma State University, \\ Stillwater, OK 74078-3072, USA \\
$^{c}$Department of Physics and Astronomy, University of Alabama,  \\
Tuscaloosa, Alabama 35487, USA
} 

\end{center}
\vspace{0.5cm}
\begin{abstract}
A certain class of new physics models includes long-lived particles which are singlet 
  under the Standard Model (SM) gauge group. 
A displaced vertex is a spectacular signature to probe such particles productions at the high energy colliders, 
  with a negligible SM background. 
In the context of the minimal gauged $B-L$ extended SM, we consider a pair creation of Majorana right-handed neutrinos (RHNs) 
  at the high energy colliders through the production of the SM and the $B-L$ Higgs bosons and their subsequent decays into RHNs. 
With parameters reproducing the neutrino oscillation data, we show that the RHNs are long-lived and their displaced vertex signature
  can be observed at the next generation displaced vertex search experiments, such as the HL-LHC, the MATHUSLA, the LHeC, and the FCC-eh.
We find that the lifetime of the RHNs is controlled by the lightest light neutrino mass, which leads to a correlation 
  between the displaced vertex search and  the search limit of the future neutrinoless double beta-decay experiments.  
\end{abstract}
\end{titlepage}

\section{Introduction}
\label{sec:1}
The last missing piece of the Standard Model (SM) is finally supplemented with the discovery of the Higgs boson 
  at the Large Hadron Collider (LHC) in 2012 \cite{Aad:2012tfa, Chatrchyan:2012xdj}, 
  which itself is not that surprising given the tremendous success of the SM to explain observed elementary particle phenomena. 
However, the SM is not complete in its current form,  because, for example, the neutrinos are massless 
  in the framework which is not consistent with the experimental evidence of the neutrino oscillation phenomena \cite{PDG}, 
  indicating that the neutrinos have tiny non-zero masses and flavor mixings. 
Hence, we need to extend the current framework of the SM.

Unfortunately, ever since the discovery of the SM Higgs boson, no new signature of new physics beyond the SM has been observed. 
It may indicate that the current energy and luminosity of the LHC are not sufficient to directly probe new particles. 
If so, we can just hope for new particle signals to be observed at the future LHC after the planned upgrade, 
  or at a future collider with energies higher than the LHC. 
However, there is another possibility: if new particles are completely singlet under the SM gauge group, 
  it can naturally explain the null search results at the LHC because SM singlet particles cannot be directly produced 
  at the LHC through the SM interactions. 
Such particles may be produced through new interactions and/or rare decay of the SM particles. 
At a first glance, it seems that such a scenario is even more challenging to test. 
However, if a new particle is long-lived, it can leave a displaced vertex signature at the collider experiments. 
Since the displaced vertex signatures are generally very clean, they allow us to search for such a particle 
  with only a few events at the LHC or future colliders.  
For the current status of displaced vertex searches at the LHC, see, for example, Refs.~\cite{ATLAS:2012av, Chatrchyan:2012sp, Aad:2012kw, Aad:2012zx, Chatrchyan:2012jna, Aad:2014yea, CMS:2014wda, CMS:2014hka, Aaij:2014nma, Aad:2015asa, Aad:2015uaa, Aad:2015rba,Aaboud:2016dgf, Aaboud:2016uth, ATLAS:2016jza, ATLAS:2016olj, Aaboud:2017iio,Aaboud:2017mpt}. 
The search reach will be dramatically improved in the future planned/proposed experiments, 
  such as the High-Luminosity LHC (HL-LHC), the MATHUSLA \cite{Chou:2016lxi}, 
  the Large Hadron electron Collider (LHeC)\cite{AbelleiraFernandez:2012cc} 
  and the the Future Circular electron-hadron Colliders (FCC-eh) \cite{Kuze:2018dqd}.

In this paper, we first review the current theoretical and experimental studies focusing on the displaced vertex searches 
  at the future collider experiments and express their results in a model independent form. 
As a concrete example, we consider a well-motivated simple extension of the SM, 
  namely the minimal $B-L$ (baryon minus lepton number) model  \cite{mBL1,mBL2,mBL3,mBL4,mBL5,mBL6}, 
  where the global ${B-L}$ symmetry in the SM is promoted to the gauge symmetry. 
A minimal $B-L$ model includes an additional electrically neutral gauge boson ($Z^\prime$ boson) 
  as well as a $B-L$ Higgs boson which breaks the ${B-L}$ symmetry. 
In addition, the model also includes three right-handed neutrinos (RHNs) to cancel all the gauge and mixed-gravitational anomalies. 
After the $B-L$ symmetry breaking, the RHNs acquire Majorana masses, and the tiny neutrinos masses 
  are automatically generated through the so-called type-I seesaw mechanism \cite{Weinberg:1979sa,    
  seesaw1,seesaw2,seesaw3,seesaw4,seesaw5} after the electro-weak symmetry breaking. 
In this model context, we investigate the displaced vertex signature of the Majorana RHNs at the future high energy colliders 
   through the production of the SM and the $B-L$ Higgs bosons and their subsequent decays into RHNs.\footnote{
   The collider signatures pertaining to the RHNs pair production through the production of the $Z^\prime$ boson 
   and the Higgs bosons and their subsequent decays into RHNs have been studied before in the literature\cite{Basso1,Basso2,Basso3,Basso4,Basso5,
Basso6,Basso7,Basso8,Accomando:2013sfa,Accomando:2015cfa,Accomando:2015ava,Accomando1,Okada:2016gsh,Caputo:2016ojx,Nemevsek:2016enw,Caputo:2017pit,dig1,dig2}.} 
Since the Majorana RHNs decay into the SM particles through small light-heavy neutrino mixings from type-I seesaw, the RHNs are likely to be long-lived.\footnote{One can also consider a  displaced vertex signature from RHN decay in type-III seesaw scenario \cite{futurework}.}

This paper is organized as follows. 
In Sec.~\ref{sec:2}, we review the prospect of the search reach of displaced vertex signatures at the future high energy colliders.  
Employing the (2$\sigma$) search reach of the displaced vertex signatures obtained in various analysis, 
  we present a model-independent formula for the search reach 
  in terms of the production cross section of a long-lived particle 
  as a function of its lifetime, mass and its mother particle mass whose decay products are the long-lived particle.  
In Sec.~\ref{sec:3}, we give a review on the minimal $B-L$ extended SM. 
In Sec.~\ref{sec:4}, we consider the pair production of RHNs through the production 
  of the Higgs bosons and their subsequent decays into RHNs. 
We apply the best reach values for the production cross section obtained in Sec.~\ref{sec:2} 
   to the RHNs production, assuming a suitable lifetime of the RHNs. 
For benchmark mass values of the $B-L$ Higgs boson and RHNs, 
   we determine the corresponding parameter space for RHNs Majorana Yukawa couplings 
   and a mixing between the SM and the $B-L$ Higgs bosons. 
In Sec.~\ref{sec:5}, we calculate the lifetime of the RHNs for realistic parameters 
   to reproduce the neutrino oscillation data. 
Using this realistic value for the lifetime, we repeat the analysis in Sec.~\ref{sec:4} 
   to determine the  parameter space corresponding to the search reach. 
In Sec.~\ref{sec:6} we discuss the correlation between the displaced vertex search 
   and the search limit of the future neutrinoless double beta-decay experiments.  
Sec.~\ref{sec:7} is devoted to conclusions.

\section{Displaced vertex search at the future colliders}
\label{sec:2}
An electrically neutral particle with a sufficiently long lifetime (for example, its decay length is of $\mathcal{O}$(1 mm) or larger), 
  once produced at the colliders, displays a signature of the displaced vertex, 
  namely the vertex  created by the decay of the particles is located away 
  from the collision point where the particle is produced. 
The final state charged leptons and/or jets from a displaced vertex can be reconstructed 
  by a dedicated displaced vertex analysis. 
Since displaced vertex signatures from SM particles are very well understood, 
  the signature from a new long-lived particle can be easily distinguished, 
  making it a powerful probe to discover such particles.

Let us first review the search reach of displaced vertex signatures at the future colliders which have been investigated in Refs.~\cite{Chou:2016lxi, math2, math3}. 
In Refs.~\cite{Chou:2016lxi, math2}, the authors have proposed the MATHUSLA detector 
  which is specifically designed to explore the long lifetime frontier; 
  the plan is to build a detector on the ground, about 100 m away from the HL-LHC detector. 
The authors have also considered displaced vertex using the inner-detector of the HL-LHC. 
Similarly, in the Ref.~\cite{math3}, the authors have studied the prospect of a dedicated displaced vertex search 
  at the future electron-proton colliders, such as the LHeC and the FCC-eh. 
In their analysis, they consider a pair production of a long-lived particle ``X" created from the rare decay of the SM Higgs boson. 
They have shown the search reach for the branching ratio of the SM Higgs boson 
  to a pair of X particles as a function of X particle's decay length ($c\tau$) ranging from sub-millimeter to $10^7$ m.

\begin{figure}[h!]
\begin{center}
\includegraphics[width=0.6\textwidth, height=7cm]{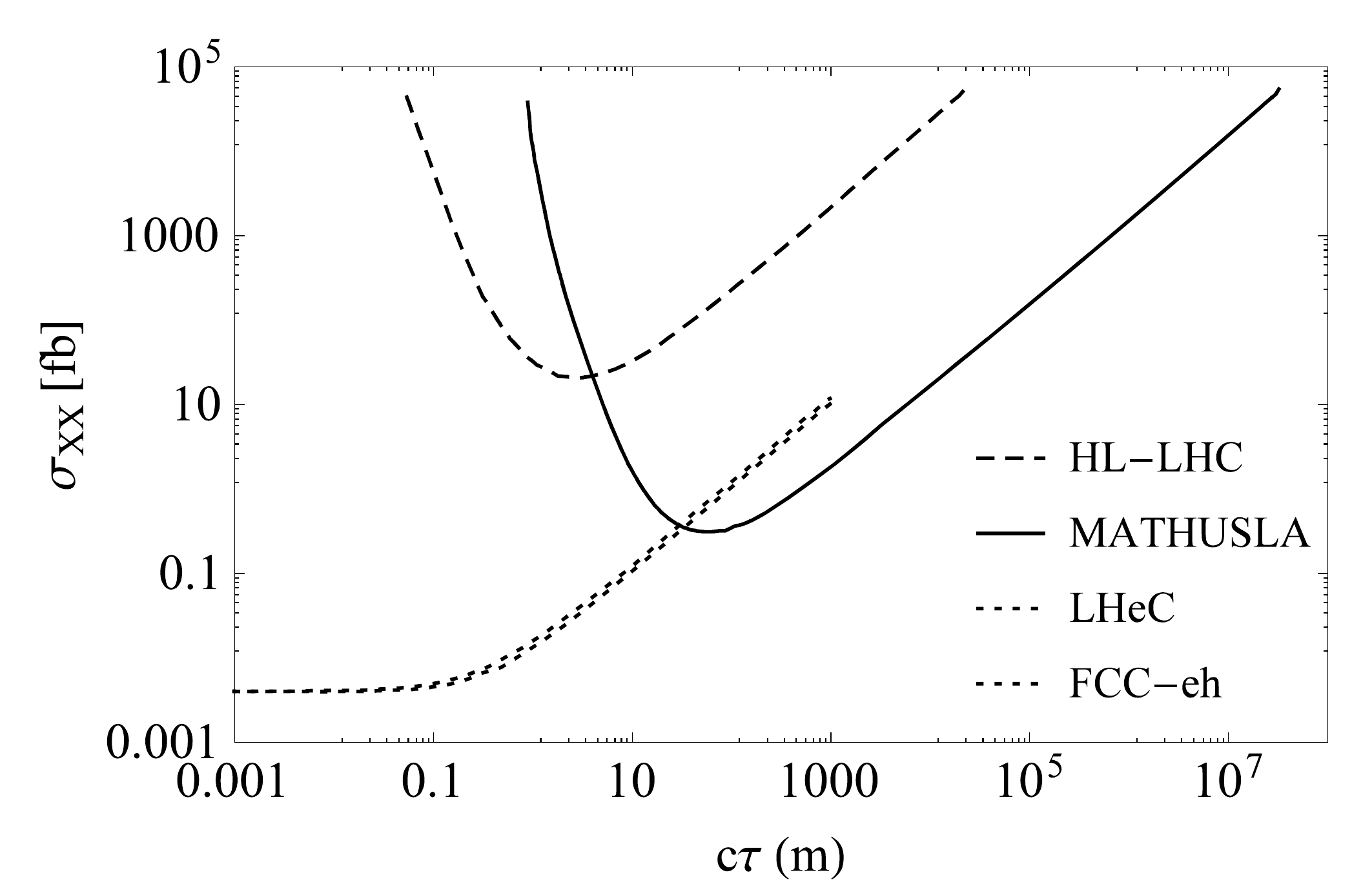} 
\end{center}
\caption{The plot shows the discovery reach for a dedicated displaced vertex searches at the future experiments and newly proposed extensions to the current LHC experiment. 
The lines corresponds to the total production cross section to produce a pair of ``X'' particles in the final states as a function of the X particle decay length, where the mass of the X particle and its mother-particle are fixed to be 20 GeV and 125 GeV, respectively. 
The region above the dashed (solid) line corresponds to the search reach at the HL-LHC and the MATHUSLA experiements. The dotted curved lines correspond to the search reach of the various proposed electron-proton collider upgrade of HL-LHC.}
\label{fig: dvsearch}
\end{figure}

We first summarize the results in Refs.~\cite{Chou:2016lxi, math2, math3} in Fig.~\ref{fig: dvsearch}. 
Here, for a fixed mass of the X particle ($m_X = 20$ GeV), we show the search reach 
   for the X particle pair production cross section ($\sigma_{X X}$) at the future colliders 
   as a function of the lifetime of X particle $c\tau$. 
The dashed and solid lines show the search reach for the displaced vertex signatures  
   at the HL-LHC and the MATHUSLA experiments, respectively, with a 3 ab$^{-1}$ luminosity.  
The dotted lines (almost degenerate) correspond to the discovery reach at the FCC-eh with a 3 ab$^{-1}$ with electron beam energies of 60 GeV (top) and the LHeC with a 1 ab$^{-1}$  (bottom).

\begin{figure}[h!]
\begin{center}
\includegraphics[width=0.6\textwidth, height=7cm]{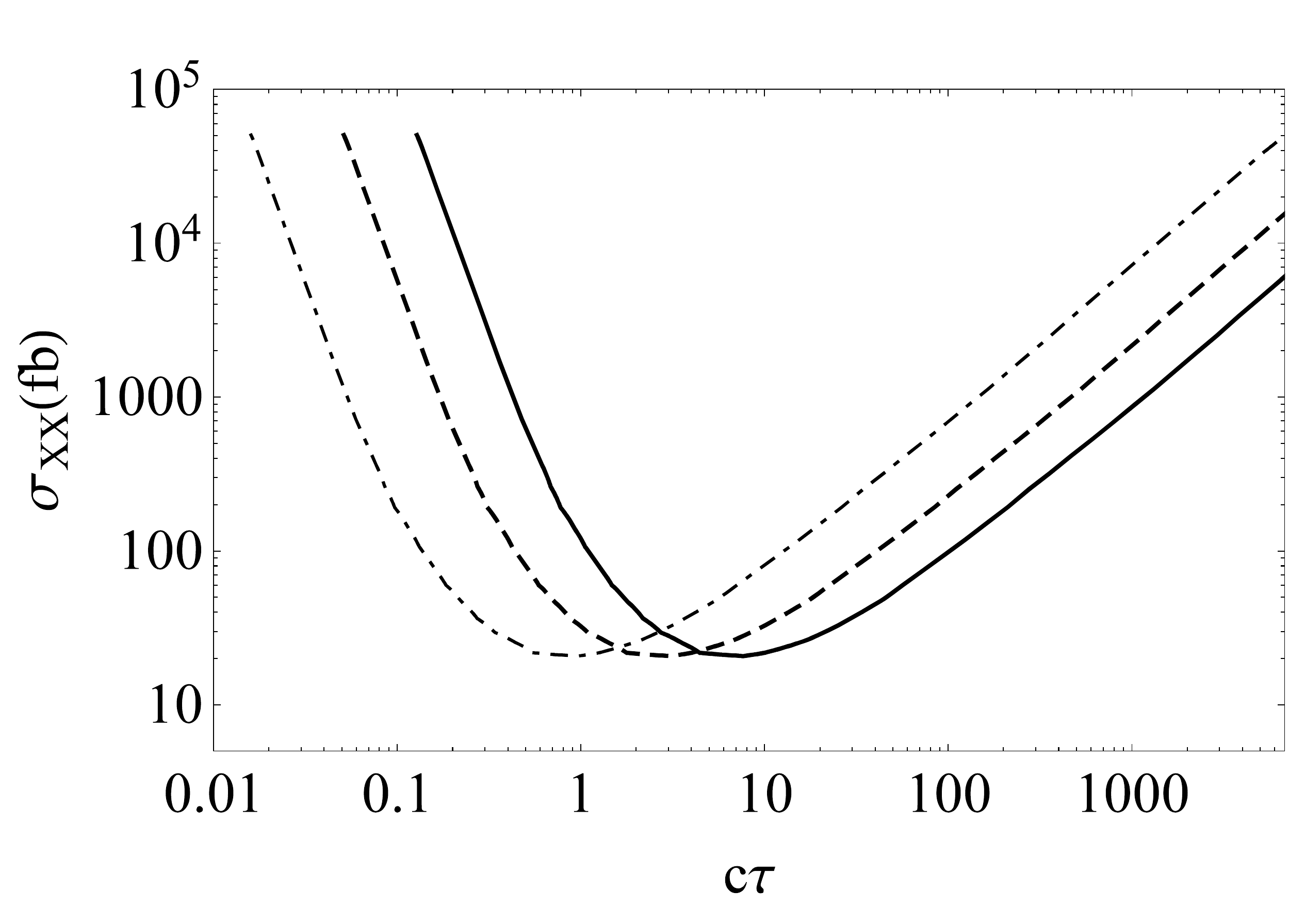} 
\end{center}
\caption{
The search reach of the displaced vertex signatures at the HL-LHC (dashed lines in Fig.~\ref{fig: dvsearch}) 
  for $m_S=50$ (solid), $125$ (dashed) and $400$ (dot-dashed) GeV with $m_X=20$ GeV. 
We have employed Eq.~(\ref{eq: sigS}) to plot the lines for $m_S=50$ and $400$ GeV, 
  based on the dashed line for $m_S=125$ GeV.  
}
\label{fig: dvsearchshift}
\end{figure}

In Fig.~\ref{fig: dvsearch}, the process $pp/ep\to h \to XX$ is considered. 
We generalize the process to $pp/ep\to S \to XX$, where the mother-particle ($S$) 
   is a boson, but not the SM Higgs boson, with a mass $m_{S}$. 
In order to make the results in Fig.~\ref{fig: dvsearch} to applicable to this general case, 
   note that the search reach shown in Fig.~\ref{fig: dvsearch} is model-independent 
   if the curves are plotted as a function of the lifetime of X in the laboratory frame. 
For the process $pp/ep\to h \to XX$, 
   the lifetime of the particle X in the laboratory frame ($\tau ^\prime$) is given by 
   its proper lifetime ($\tau$) as 
\bea
\tau^\prime = \left( \frac{m_h}{2 m_X} \right) \tau, 
\eea
because of the Lorentz boost. 
We then express the search reach of the cross sections in Fig.~\ref{fig: dvsearch} 
   as $\sigma_{XX}(c \tau)= \sigma_{XX}\left(\frac{2 m_X}{m_h} c \tau^\prime\right)$.  
The model-independent search reach can be obtained as a function of $c \tau^\prime$   
   for the fixed values of $m_h=125$ GeV and $m_X=20$ GeV.  
Now it is easy to convert the search reach results in Fig.~\ref{fig: dvsearch} to our general case:  
\bea
\sigma(pp/ep\to S \to XX) = \sigma_{XX}\left(c \tau\times \frac{m_S}{125 \, {\rm GeV}} \times \frac{20 \, {\rm GeV}}{ m_X} \right), 
\label{eq: sigS}
\eea
where $\sigma_{XX}(c\tau)$ represents different curves shown in Fig.~\ref{fig: dvsearch}.  
Hence, depending on the choice of masses for $m_S$ and $m_X$, 
   the curves shown in Fig.~\ref{fig: dvsearch} shifts either to the left or to the right. 
In Ref.~\cite{math2}, the search reach of the cross sections are shown for three benchmark values 
   of $m_X=5$, $20$ and $40$ GeV. 
We have checked that our formula of Eq.~(\ref{eq: sigS}) can reproduce the results 
   for $m_X=5$ and $40$ GeV in  Ref.~\cite{math2} from the result for $m_X=20$ GeV. 
In Fig.~\ref{fig: dvsearchshift}, we show the search reach at the HL-LHC for $m_S=50$ GeV (solid line) 
   and $m_S=400$ GeV (dot-dashed line) 
   by employing the result for $m_S=125$ GeV (dashed line) and Eq.~(\ref{eq: sigS}). 
Here, we have fixed $m_X=20$ GeV. 
As we raise/lower $m_S$ for $m_X=20$ GeV,  the line shifts to the left/right, 
   since the created X particle is more/less boosted. 
In the following, we employ the generalized formula to investigate the search reach 
   of long-lived RHNs at the future high energy colliders.

\section{The minimal $B-L$ extended Standard Model}
\label{sec:3}
\begin{table}[h]
\begin{center}
\begin{tabular}{c|ccc|c}
            & SU(3)$_c$ & SU(2)$_L$ & U(1)$_Y$ & U(1)$_{B-L}$  \\
\hline
$ q_L^i $    & {\bf 3}   & {\bf 2}& $+1/6$ & $+1/3$  \\ 
$ u_R^i $    & {\bf 3} & {\bf 1}& $+2/3$ & $+1/3$  \\ 
$ d_R^i $    & {\bf 3} & {\bf 1}& $-1/3$ & $+1/3$  \\ 
\hline
$ \ell^i_L$    & {\bf 1} & {\bf 2}& $-1/2$ & $-1$  \\ 
$ N_R^i$   & {\bf 1} & {\bf 1}& $ 0$   & $-1$  \\ 
$ e_R^i  $   & {\bf 1} & {\bf 1}& $-1$   & $-1$  \\ 
\hline 
$ H$         & {\bf 1} & {\bf 2}& $-1/2$  &  $ 0$  \\ 
$ \varphi$      & {\bf 1} & {\bf 1}& $  0$  &  $+2$  \\ 
\end{tabular}
\end{center}
\caption{
Particle content of the minimal $B-L$ model. 
In addition to the SM particle content, three RHNs ($N_R^i$, $i=1,2,3$ denotes the generation index)  
  and a complex scalar ($\varphi$) are introduced. 
}
\end{table}
Here we review the minimal $B-L$ extended SM (the minimal $B-L$ model).  
The particle content of the model is listed in Table~1. 
In this model, the global $B-L$ symmetry in the SM is gauged, and in addition to the SM particle content, 
  three RHNs and a complex scalar ($B-L$ Higgs field) are introduced. 
While the $B-L$ Higgs field spontaneously breaks the $B-L$ symmetry by its vacuum expectation value (VEV),  
  the three RHNs are necessary to cancel all the gauge and mixed-gravitational anomalies.

The Yukawa sector of the SM is extended to include
\bea
\mathcal{L} _{Y}\supset -\sum_{i, j=1}^{3} Y_{D}^{ij} \overline{\ell_{L}^{i}} H N_{R}^{j}-\frac{1}{2} \sum_{k=1}^{3} Y_{N}^{k} \Phi \overline{N_{R}^{k \ c}} N_{R}^{k}+ \rm{h. c.}, 
\label{U1XYukawa}
\eea 
where the first and second terms are the Dirac and Majorana Yukawa couplings. 
Here, we work on a diagonal basis for the Majorana Yukawa couplings ($Y_{N}$) without loss of generality. 
Associated with the $B-L$ gauge symmetry breaking, 
  the $B-L$ gauge boson ($Z^\prime$ boson) and the RHNs acquire their masses as follows:
\bea 
m_{Z^\prime}= 2 \, g \, v_{BL},  \; \; 
m_{N^i}=\frac{Y_N^i}{\sqrt{2}} v_\Phi, 
\label{masses}
\eea
where $v_{BL}= \sqrt{2} \langle \varphi\rangle$ is the VEV of the $B-L$ Higgs field.

\begin{figure}[t]
\begin{center}
\includegraphics[width=0.6\textwidth, height=7cm]{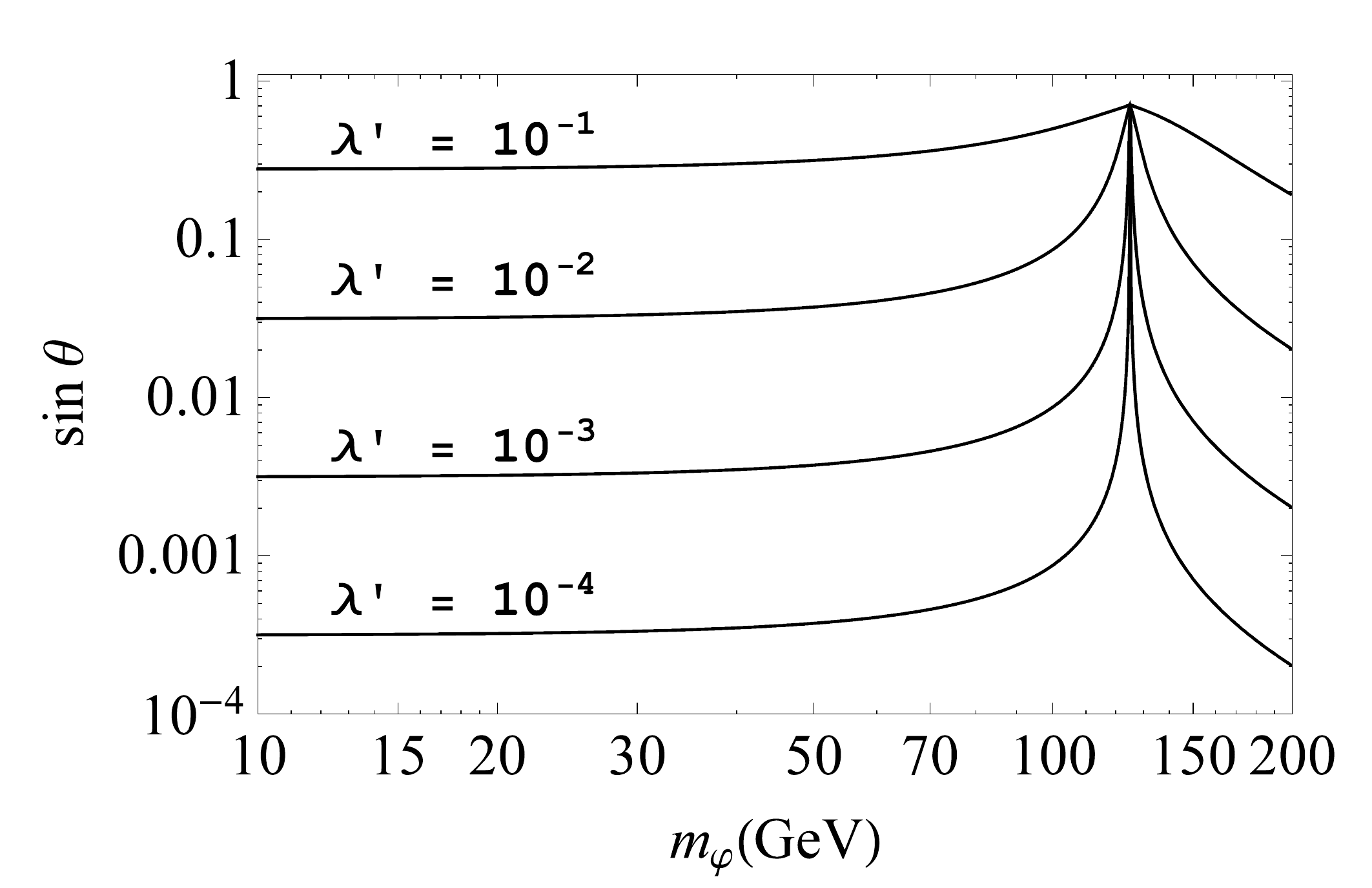}
 \caption {
 The mixing angle as a function of $m_{\varphi}$ for $v_{BL} = 200$ GeV and 
   various values of $\lambda^\prime$. 
The solid lines from top to bottom corresponds to $\lambda^\prime =10^{-1}, 10^{-1}, 10^{-2}, 10^{-3}$, and $10^{-4}$.
} 
 \end{center}
\label{fig:mix}
 \end{figure}

A renormalizable scalar potential for the $B-L$ Higgs field ($\varphi$) and the SM Higgs doublet ($H$) is given by  
\bea
V(|H|,|\varphi|) =  \lambda \left(|\varphi|^2 -\frac{v^2_{BL}}{2}\right)^2 + \lambda_H \left(|H|^2 -\frac{v^2_{SM}}{2}\right)^2 +  \lambda^{\prime} \left(|H|^2 -\frac{v^2_{SM}}{2}\right)\left(|\varphi|^2 -\frac{v^2_{BL}}{2}\right), 
\label{InfPot}
\eea 
where $v_{SM}=246$ GeV is the VEV of the SM Higgs doublet, and we take $\lambda^{\prime}>0$ 
  which introduces a mixing between the two scalar fields. 
In the unitary gauge, we expand the SM and $B-L$ Higgs fields around their VEVs, 
   $\langle H\rangle = (\frac{v_{SM}}{\sqrt{2}} \;\;0)^T$ and $\langle\varphi\rangle = v_{BL}/\sqrt{2}$, 
   to identify $\phi_{SM}$ and $\phi_{BL}$ being the SM and the $B-L$ Higgs bosons in the original basis. 
The mass matrix for the Higgs bosons is given by 
\begin{eqnarray}
{\cal L}  \supset -
\frac{1}{2}
\begin{bmatrix}
\phi_{SM}  & \phi_{BL}
\end{bmatrix}
\begin{bmatrix} 
m_{H}^2 &  \lambda^{\prime} v_{BL} v_{SM} \\ 
 \lambda^{\prime} v_{BL} v_{SM} & m_{\varphi}^2
\end{bmatrix} 
\begin{bmatrix} 
\phi_{SM} \\ \phi_{BL} 
\end{bmatrix}, 
\label{eq: massmatrix}
\end{eqnarray} 
where $m_H = \sqrt{2 \lambda_{H}} v_{{SM}}$, and $m_{\varphi}^2 = 2 \lambda v_{BL}^2$. 
We diagonalize the mass matrix by 
\begin{eqnarray}
\begin{bmatrix}
\phi_{SM} \\ \phi_{BL}  \end{bmatrix}
=
\begin{bmatrix} 
\cos\theta &   \sin\theta \\ 
-\sin\theta & \cos\theta  
\end{bmatrix} 
\begin{bmatrix} 
h \\ \phi 
\end{bmatrix}  ,
\label{eq: eigenstate}
\end{eqnarray} 
where $h$ and $\phi$ are the mass eigenstates. 
The relations among the mass parameters and the mixing angle ($\theta$) are the following: 
\bea
&2 v_{BL} v_{{SM}}  \lambda^\prime= ( m_H^2 -m_\varphi^2) \tan2\theta,   \nonumber  \\
 &m_{h}^2 = m_H^2     - \left(m_\varphi^2  - m_H^2 \right) \frac{\sin^2\theta}{1-2 \sin^2\theta} , \nonumber \\
 &m_{\phi}^2 = m_\varphi^2 + \left(m_\varphi^2 - m_H^2 \right) \frac{\sin^2\theta}{1-2 \sin^2\theta} \ .
\label{eq: mixings} 
\eea 
The properties of the Higgs boson measured at the LHC are consistent with the SM predictions, 
  so that the mixing angle $\theta$ must be small. 
In this case, we identify the mass eigenstates $h \simeq \phi_{SM}$ and $\phi \simeq \phi_{BL}$ 
  as mostly the SM and $B-L$ Higgs bosons with masses $m_h \simeq m_H$ and $m_\varphi \simeq m_\phi$, respectively.
We set $m_H = 125$ GeV in the following. 
For completeness, we show in Fig.~\ref{fig:mix} the relation between $m_\varphi$ and $\sin \theta$ 
  for a fixed $v_\varphi = 200$ GeV and various values of $\lambda^\prime$. 
As can be understood from the first line in Eq.~(\ref{eq: mixings}), 
  $m_H = m_\phi$ is a singular point where the mixing becomes large. 

\section{Displaced vertex signature of heavy neutrinos}
\label{sec:4}
Let us now consider the displaced vertex signatures of the heavy neutrinos $N^i$ 
  in the minimal $B-L$ model.\footnote{
  See \cite{Accomando1} for the previous work on the displaced vertex signature 
  of the heavy neutrinos at the LHC in the context of the minimal $B-L$ model}. 
For the main process for a pair production of $N_i$, we can consider two cases: 
one is through the $Z^\prime$ boson production \cite{(see, for example, Ref.~\cite{})} and its decay to $N_i$s, 
and the other is through the production of the Higgs bosons ($h$ and $\phi$) 
  and their decays.\footnote{
Without the interactions through the $Z^\prime$ boson and the Higgs boson, 
   the heavy neutrinos can be produced through the heavy-light neutrino mixings. 
The study of the displaced vertex signature for this case, 
   see, for example, Refs.~\cite{Gago:2015vma, Antusch:2016vyf, Antusch:2017hhu}}  
The LHC results on the search for the $Z^\prime$ boson resonance of the minimal $B-L$ model severely constrain the $B-L$ gauge coupling to be very small (see, for example, Ref.~\cite{Okada:2018ktp}), so that the heavy neutrino production cross section from the $Z^\prime$ boson decay is expected to be small. 
Hence, we focus on the heavy neutrino production through the Higgs bosons in this paper\footnote{In our work, we focus on production of heavy RHNs with mass greater than 20 GeV, see also Ref.~\cite{}. The case for the production of RHNs of masses less than 10 GeV has been recently considered in Ref.~\cite{faser}, where the authors have considered MATHUSLA, FASER and CODEX-b collider experiments}.

Because of its representation under the gauge groups, $\phi_{BL}$ has no tree level coupling with the SM particles 
  at the renormalizable level, and hence it cannot be directly produced at the high colliders.  
However, as described in  Eqs.~(\ref{eq: eigenstate}) and (\ref{eq: mixings}), 
  the mass eigenstates are mixture of  $\phi_{BL}$ and $\phi_{SM}$ and through the mixing,   
  the Higgs boson $\phi$ can be produced through the same process as the SM Higgs boson. 
At the LHC, among a variety of the production processes of the SM Higgs boson, 
  such as  gluon-gluon fusion (ggF), vector boson fusion (VBF), and 
  the productions associated with $W/Z$ bosons ($Vh$) and with $t\bar{t}$ ($t\bar{t}h$), 
  the ggF channel dominates the production cross section.\footnote{
  At the 13 TeV LHC, the SM Higgs boson with mass of around 125 GeV, 
  the Higgs boson production cross sections through these channels are evaluated as \cite{hwg,j1}:  
  $\sigma_h^{ggF}=43.92$ pb, $\sigma_h^{VBF}=3.748$ pb, $\sigma_h^{Wh}=1.38$ pb, 
  $\sigma_h^{Zh}=0.869$ pb, and $\sigma_h^{t\bar{t}h}=508.5$ fb.} 
For a small mixing, the production cross section of the SM-like Higgs boson ($h$) 
   is given by 
\bea
\sigma(pp \to h) = \cos^2\theta \times \sigma_h (m_h), 
\label{eq: hcs}
\eea 
where $\sigma_h (m_h)$ is the SM Higgs boson production cross section with the Higgs boson mass of $m_h = 125$ GeV. 
In the limit of $\theta \to 0$, $\sigma(pp \to h)$ reduces to the SM case. 
Similarly, the production cross section for the $B-L$-like Higgs boson ($\phi$) is expressed as 
\bea
\sigma(pp \to \phi ) = \sin^2\theta \times \sigma_h (m_\phi), 
\label{eq: phics}
\eea 
where $\sigma_h (m_\phi)$ is the SM Higgs boson production cross section if the Higgs boson mass were $m_\phi$
As discussed before, we are interested in a small mixing, and hence for the remainder of this paper, 
  we shall simply refer to the mass eigenstates $h$ and $\phi$ as the SM(-like) Higgs boson 
  and the $B-L$ Higgs, respectively.

\begin{figure}[t]
\begin{center}
\includegraphics[width=0.8\textwidth, height=7cm]{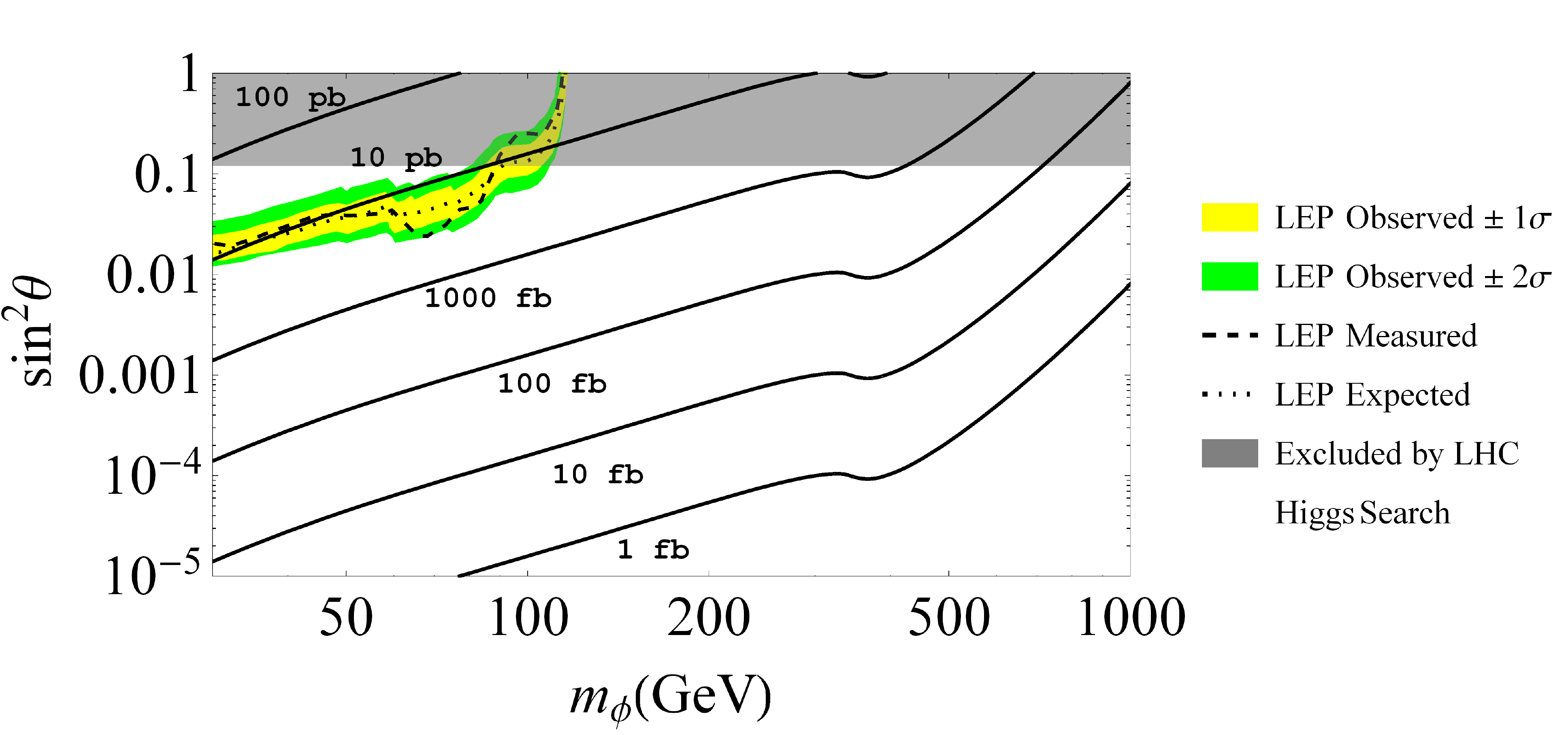}
 \caption {
The $B-L$ Higgs production cross section at the 13 TeV LHC in $(m_\phi, \sin^2\theta)$-plane, 
  along with the upper bounds on $\sin^2 \theta$ from the LEP and the LHC experiments. 
The region above the dashed curve is excluded by the LEP experiments, 
  while the gray shaded region is excluded by the LHC experiments. 
} 
\label{fig:prod}
 \end{center}
 \end{figure}

Using Eq.~(\ref{eq: phics}), we show in Fig.~{\ref{fig:prod}} the contour plot 
   for the production cross section of $\phi$ at the 13 TeV LHC ($\sigma(pp \to \phi)$)   
   in the ($m_\phi, \sin^2\theta$)-plane. 
Here, we also show the constraint obtained by the LEP experiments on the search 
   for the SM Higgs boson through its production associated with the $Z$ boson \cite{lep}. 
For a given Higgs boson mass, no evidence of the SM Higgs boson production sets an upper bound 
  on the $Z$-boson associate production cross section, which is interpreted to be 
  an upper bound on the anomalous SM Higgs coupling with $Z$ boson 
  as a function of the Higgs boson mass. 
In our model, this upper bound is interpreted as an upper bound on $\sin \theta$ as a function of $m_\phi$, 
  which is shown as the dashed curve in Fig.~{\ref{fig:prod}}. 
%
The gray shaded region in the figure is excluded by the current measurement of the SM Higgs boson properties 
  by the ATLAS and CMS collaborations. 
The Higgs boson properties are characterized by the signal strengths defined as   
\begin{equation}
\mu^i_f =  \frac{\sigma^i \cdot BR_f}{(\sigma^i)_{SM} \cdot (BR_{f})_{SM}} = \mu^i\cdot\mu_f, 
\label{eq:muif}
\end{equation}
where $\sigma^i$ are the Higgs boson production cross sections  
  through $i$-channel with $i= ggF, VBF,$ $Wh,$ $Zh,$ $t\bar{t}h$, 
  and $BR_f$ are the Higgs boson branching ratios into final states $f = ZZ^{\star},$ $WW^{\star},$  
$\gamma \gamma,$ $\tau^+ \tau^-,$ $b\bar{b},$ $\mu^+ \mu^-$. 
The cross sections and branching ratios with the subscript ``$SM$'' denote the SM predictions.  
The latest updates of the signal strengths from the ATLAS and the CMS experiments at the LHC Run II 
  with a $37$ fb$^{-1}$ luminosity are listed in Refs.~\cite{econf, j1}, along with the references. 
The averaged signal strength is obtained as $\mu \simeq  0.925 \pm 0.134$, which is consistent 
  with the SM prediction $\mu=1$. 
In our model,  the couplings of the SM-like Higgs boson is modified by a factor $\cos \theta$ 
  from the SM case. 
Hence, the production cross section of $h$ is scaled by $\cos^2 \theta$ 
  while its branching ratios remain the same as the SM one. 
Therefore, the signal strengths are given by $\mu^i_f = \cos^2{\theta} \times \mu_f$. 
In anticipation of the analysis presented below, 
   we also consider the constraint from the invisible SM Higgs decay modes, 
   and the current upper bound on the branching ratio of 
   the Higgs invisible decay mode is given by ${BR}_{\rm inv}^{\rm higgs} < 0.23$ \cite{invisible}. 
Together with the averaged signal strength, we obtain an upper bound 
  on the mixing angle as $\sin^2{\theta} \leq 0.12$ at 95$\%$ C.L 
  and the excluded region is depicted by the gray shaded region in Fig.\ref{fig:prod}. 
Hence, for the entire range of $m_\phi$ shown in the figure, the mixing angle is small.

\begin{figure}[t]
\begin{center}
\includegraphics[width=0.6\textwidth, height=7cm]{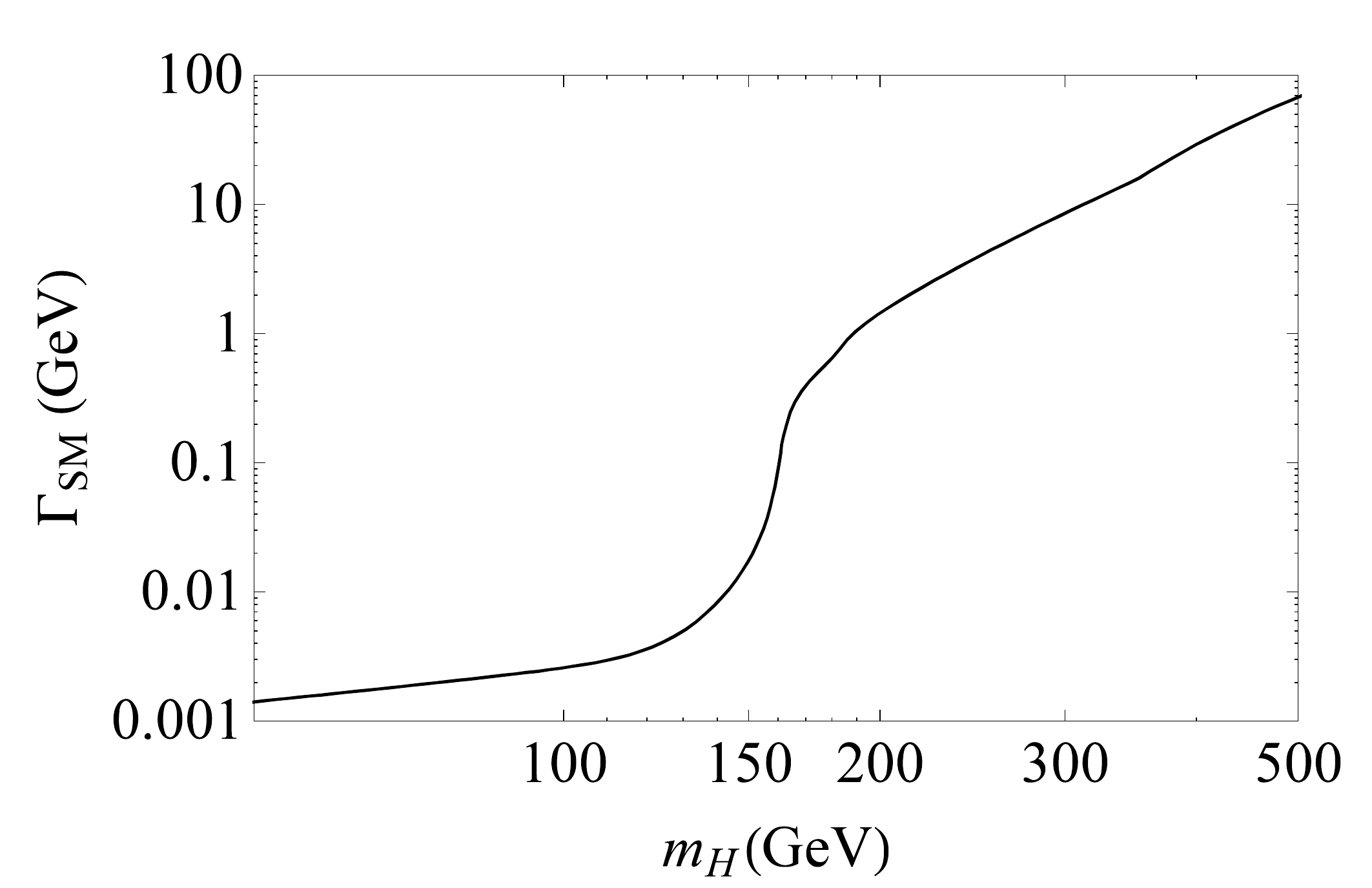} \;  
\end{center}
\caption{Total decay width of SM Higgs as a function of the Higgs mass.}
\label{fig: higgsdecay}
\end{figure}

We now consider the decay of the Higgs bosons. 
In the presence of the mixing, the total decay width of the SM-like Higgs and the $B-L$ Higgs 
  into the SM final states are given by 
\bea
\Gamma (h\to SM) &=& \cos^2  \theta \times \Gamma_ {SM} (m_h), \nonumber \\
\Gamma(\phi \to SM) 
 &=& \sin^2\theta \times \Gamma_ {{SM}} (m_\phi) ,
\label{eq: phitoNN}
\eea
respectively, where $\Gamma_ {SM}(M)$ is the total decay width of the SM Higgs boson 
   if the SM Higgs mass were $M$. 
The list of the partial decay widths of the SM-like Higgs boson is given in the Appendix. 
In Fig.~\ref{fig: higgsdecay} we show the total decay width $\Gamma_ {SM}(m_H)$ as a function $m_H$. 
The partial decay widths of $\phi$ and $h$ into a pair of heavy neutrinos are given by 
\bea
\Gamma(\phi \to NN)
 &=& \cos^2\theta  \times \Gamma_{NN} (m_\phi), \nonumber \\
\Gamma (h \to NN)
 &=& \sin^2\theta  \times \Gamma_{NN} (m_h) ,
\label{eq: phitoNN}
\eea
respectively, where $\Gamma_{NN}(M)$ is the total decay width of $\phi_{BL}$ with a mss $M$ 
  into heavy neutrinos in the limit of $\theta \to 0$, which is given by  
\bea
\Gamma_{NN} (M) =   \frac{3  Y^2}{16\pi}  M  \left( 1- \frac{4m_N^2}{M^2} \right)^{3/2}. 
\label{eq: dwphi}
\eea
Here, we have considered only one RHN with a Majorana Yukawa coupling $Y$ and its mass $m_N$, for simplicity.  
If $m_\phi >2 m_h$, $\phi$ can also decay into a pair of SM-like Higgs bosons, 
  and the partial decay width of this process is expressed as 
\bea
\Gamma(\phi \to hh)  =\frac{C_{\phi h h}^2}{32\pi} \frac{1}{m_\phi} \left(1- \frac{4 m_h^2}{m_\phi^2}\right)^{1/2},
\eea 
where 
\bea
 C_{\phi h h} = \frac{\lambda^\prime}{4}\left( v_{BL} (\cos\theta +  3 \cos3\theta) + v_{SM} (\sin\theta - 3 \sin3\theta) \right). 
\eea
For a small mixing angle $\theta \lesssim 0.1$ and $v_{SM} < v_{BL}$, 
  $C_{\phi h h}$ can be approximated as
\bea
 C_{\phi h h} \simeq  \frac{\lambda^\prime}{4}\left(4 v_{BL} -  2 v_{SM} \theta  \right) \simeq   \lambda^\prime v_{BL} = \frac{ m_h^2 -m_\phi^2}{2 v_{SM}} \tan2\theta, 
\eea
where we have used Eq.~(\ref{eq: mixings}). 
Hence, $\Gamma(\phi \to hh)$ is determined by $m_\phi$ and  $\theta$. 
Using the decay widths of $\phi$ and $h$, the branching ratio of $\phi$ and $h$ into a pair of RHNs are given by
\bea
BR(\phi \to NN) &=& \frac{\cos^2\theta  \times \Gamma_{NN} (m_\phi) }{\cos^2\theta  \times \Gamma_{NN} (m_\phi) 
\;+\; \sin^2\theta \times \Gamma_ {SM} (m_\phi) 
\;+\; \Gamma(\phi \to hh) },  \nonumber \\
BR(h \to NN) &=& \frac{\sin^2\theta  \times \Gamma_{NN} (m_h) }{\sin^2\theta  \times \Gamma_{NN} (m_h)
\;+\; \cos^2\theta \times \Gamma_ {SM}(m_h)},
\label{eq: branch}
\eea 
respectively.

\begin{figure}[t]
\begin{center}
\includegraphics[scale =0.5]{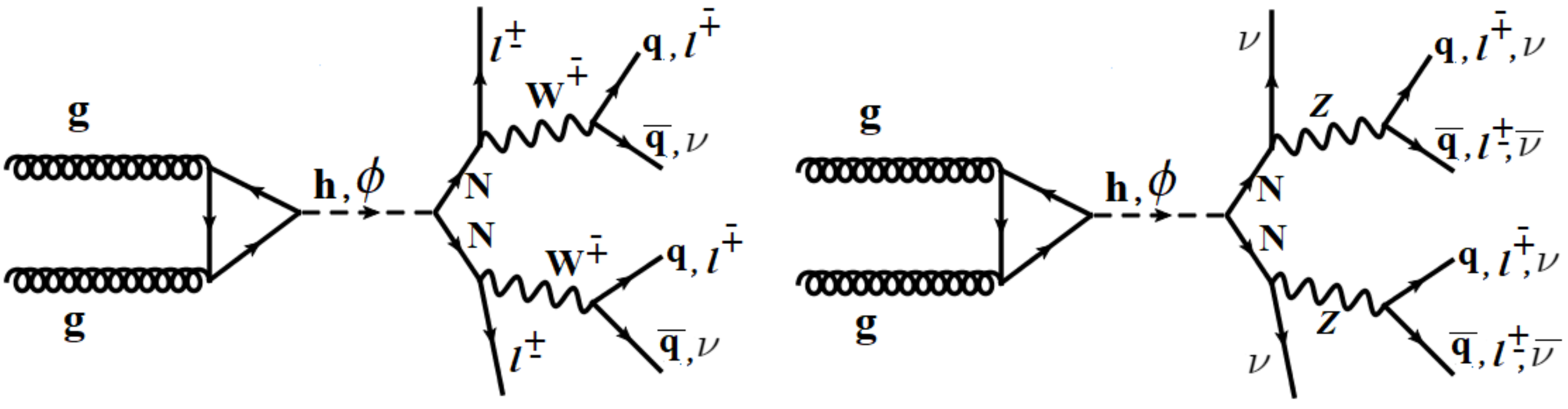} \;  
\end{center}
\caption{
Representative Feynman diagrams for the Higgs portal Majorana neutrino pair production and subsequent decay modes.}
\label{fig: feyn}
\end{figure}

Let us now consider the production cross section for the RHNs at the LHC 
  from the $\phi$ and $h$ productions and their decays. 
Using Eqs.~(\ref{eq: hcs}), (\ref{eq: phics}) and (\ref{eq: branch}), 
  the cross section formulas are given by  
\bea
\sigma(pp\to \phi \to NN) &=& \sin^2\theta  \times  \sigma_h (m_\phi) \times  BR(\phi\to NN),  \nonumber \\
\sigma(pp\to h \to NN) &=&  \cos^2\theta  \times  \sigma_h (m_h) \times  BR(h\to NN), 
\label{eq: sighphi} 
\eea 
respectively, and they are controlled by four parameters, $Y$, $\theta$, $m_\phi$ and $m_N$. 
Throughout this section, we fix $m_N = 20$ GeV, for simplicity. 
The representative diagrams of the RHN productions including their decays 
   are shown in Fig.~\ref{fig: feyn}.  
We will discuss the decay of RHNs into the SM final states in details in Sec.~\ref{sec:5}. 
In the remainder of the analysis in this section, we fix the lifetime of RHNs to yield 
    the best reach of $\sigma_{XX}$ in Fig.~\ref{fig: dvsearch} 
    for both the future HL-LHC and MATHUSLA displaced vertex searches, 
    namely, $\sigma_{\rm min} ({\rm HL-LHC}) = 20.7$ and $\sigma_{\rm min} ({\rm MATH}) =0.3$ fb, 
    which corresponds to $c\tau = 3.1$ and $58.4$ m, respectively.  
Here, we identify $X$ with the RHN while $S$ is either $h$ or $\phi$.

\begin{figure}[h!]
\begin{center}
\includegraphics[width=0.6\textwidth, height=7cm]{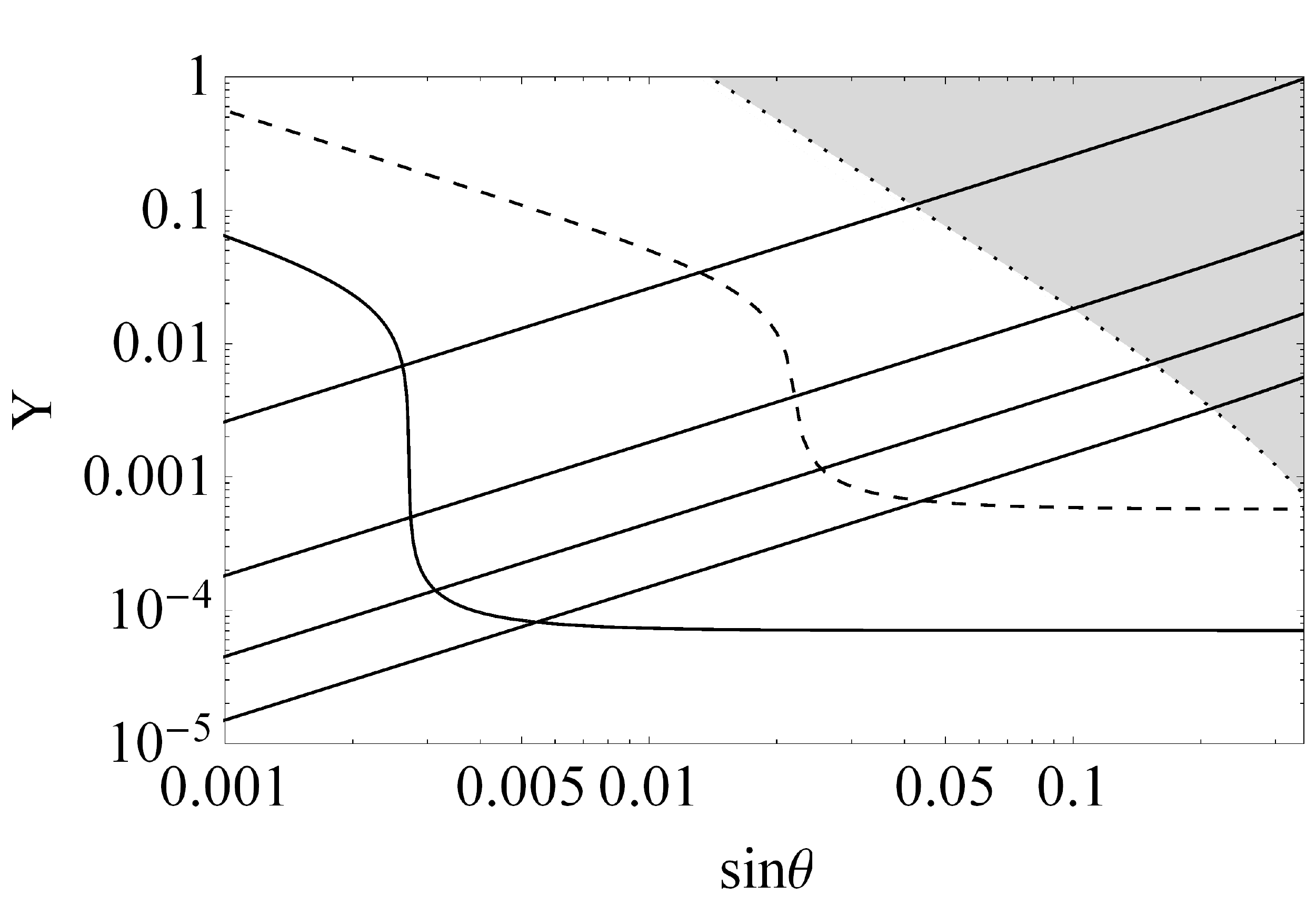}
\end{center}
\caption{
The plots show
(i) the best reaches of displaced vertex searches at the HL-LHC (dashed curve) and 
    the the MATHUSLA (dotted curve); 
(ii) branching ratios of $\phi \to NN$ denoted as the diagonal solid lines 
    ($BR(\phi \to NN) = 99.99\%$, $98\%$, $75\%$, and $25\%$, respectively, from top to bottom); 
(iii) the excluded region (gray shaded) from the LHC constraint 
     on the Higgs branching ratio into the invisible decay mode  \cite{invisible}. 
}
\label{fig: yVsin125}
\end{figure}

We first consider the case where $h$ and $\phi$ masses are almost degenerate, $m_h \simeq m_\phi = 126$ GeV. 
In this case, the total cross section $\sigma_{XX}$ is given by the sum of the productions from $\phi$ and $h$.\footnote{
   Although $\phi$ and $h$ are almost degenerate, we do not consider the interference between the processes, 
   $pp \to \phi \to NN$ and $pp \to h \to NN$, since their decay width is much smaller (a few MeV) 
   than their mass differences. 
   Hence, in evaluation the total cross section, we simply add the individual production cross section in Eq.~(\ref{eq: sighphi}). }
The best search reach of the displaced vertex signatures at the HL-LHC or the MATHUSLA 
   are expressed as 
\bea 
  \sigma_{min} &=& \sigma(pp\to \phi \to NN) + \sigma(pp\to h \to NN)  \nonumber \\
  &\simeq&  \left[ \sin^2\theta \times BR(\phi\to NN) + \cos^2\theta  \times  BR(h\to NN) \right] {\sigma_h (m_h)},   
\eea   
where we have used the approximation $\sigma_h (m_\phi) \simeq \sigma_h (m_h)$.
Hence, the best search reach is expressed as a function of $Y$ and $\theta$ 
  for the fixed values of $m_N=20$ GeV, $m_h=125$ GeV and $m_\phi=126$ GeV. 
In Fig.~\ref{fig: yVsin125}, our results are shown in $(Y, \sin \theta)$-plane. 
This plots show 
(i) the best reaches of displaced vertex searches at the HL-LHC (dashed curve) and 
    the MATHUSLA (dotted curve); 
(ii) branching ratios of $\phi \to NN$ denoted as the diagonal solid lines 
    ($BR(\phi \to NN) = 99.99\%$, $98\%$, $75\%$, and $25\%$, respectively, from top to bottom); 
(iii) the excluded region (gray shaded) from the LHC constraint 
     on the Higgs branching ratio into the invisible decay mode,  
     namely, 
\bea
   {BR}_{\rm inv}^{\rm higgs} = \sin^2\theta \times BR(\phi\to NN) + \cos^2\theta  \times  BR(h\to NN) < 0.23. 
\eea
Note that along the dashed curve, the RHN pair production is dominated by the decay of $h$ ($\phi$) 
   for $\sin \theta <0.02$ ($\sin \theta > 0.02$). 
Similarly,  the RHN pair production is dominated by the decay of $h$ ($\phi$) 
   for $\sin \theta <0.002$ ($\sin \theta > 0.002$) along the solid curve.

\begin{figure}[t]
\begin{center}
\includegraphics[width=0.49\textwidth, height=5.5cm]{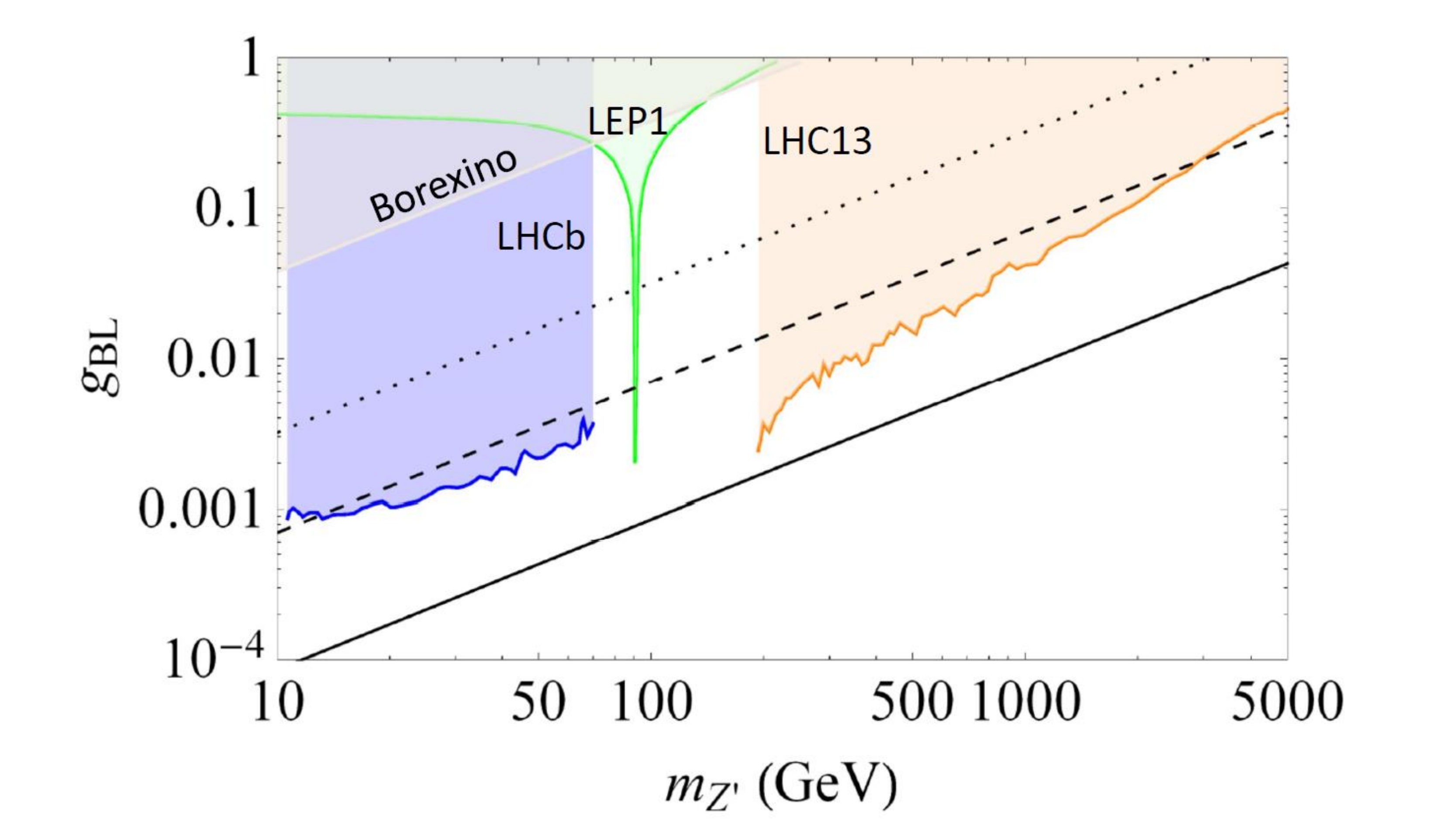}
\includegraphics[width=0.49\textwidth, height=5.5cm]{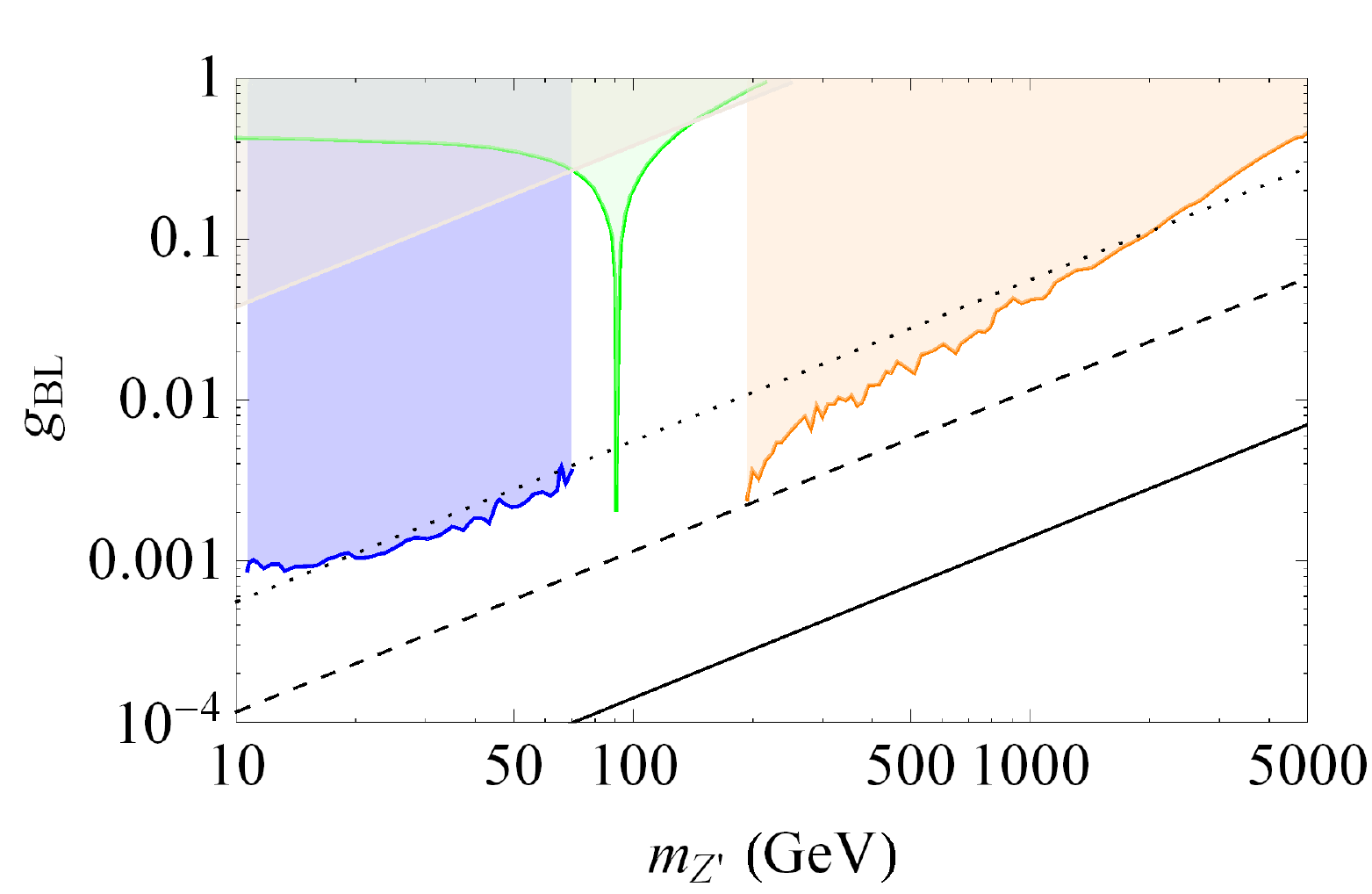}
\end{center}
\caption{
For fixed $Y$ values, the diagonal lines show the  the $Z^\prime$ boson gauge coupling values as a function of $m_{Z^\prime}$, 
  along with the excluded shaded region from various $Z^\prime$ boson searches. 
In the left panel, the three diagonal lines show the results 
  for $Y=0.0181$ (dotted), $3.97 \times 10^{-3}$ (dashed) and $4.85 \times 10^{-4}$ (solid), respectively, 
  which are chosen from the intersections of  the diagonal line for $BR(\phi \to NN) = 98\%$ 
  with the solid curve, the dashed curve and dotted line in Fig.~\ref{fig: yVsin125}. 
The results for the case with $BR(\phi \to NN) = 25\%$ are shown in the right panel, 
  where the three diagonal lines correspond to 
  $Y\simeq 3.15\times 10^{-3}$ (dotted), $6.50\times 10^{-4}$ (dashed) and $7.95\times 10^{5}$ (solid), respectively. 
}
\label{fig: gVmZ125}
\end{figure}

In our model, once $Y$ and $m_N$ are fixed, the relation between the $B-L$ gauge coupling 
  and the $Z^\prime$ boson mass is determined by (see Eq.~(\ref{masses})) 
\bea
g_{BL} = \frac{1}{2 \sqrt{2}} \frac{Y}{m_N} m_{Z^\prime}. 
\label{eq: gauge}
\eea
The $Z^\prime$ boson has been searched by various experiments, 
   and the upper bound on the $B-L$ gauge coupling as a function of its mass 
   is obtained for a wide mass range of ${\cal O}(1) \lesssim m_{Z^\prime}[{\rm GeV}] \leq 5000$. 
For a $Y$ value chosen in Fig.~\ref{fig: yVsin125}, we examine the consistency 
   with the current constraints from the $Z^\prime$ boson search. 
For several benchmark $Y$ values, we show in  Fig.~\ref{fig: gVmZ125} 
  the relation of Eq.~(\ref{eq: gauge}) 
  along with the current experimental constraints from the $Z^\prime$ boson searches 
  (the shaded regions are excluded from the result in Ref.~\cite{Batell:2016zod}, the LHCb results \cite{Aaij:2017rft,
Ilten:2018crw}, 
   and the resent ATLAS results \cite{ATLAS_Z_Search}). 
In the left panel, we show the relation for $Y=0.0181$ (dotted), $3.97 \times 10^{-3}$ (dashed) and $4.85 \times 10^{-4}$ (solid), 
  which are chosen from the intersections of  the diagonal line for $BR(\phi \to NN) = 98\%$ 
  with the solid curve, the dashed curve and dotted line in Fig.~\ref{fig: yVsin125}. 
We find the current constraints from the $Z^\prime$ boson search are very severe 
  and complementary to the future search reach of the displaced vertex signatures. 
In the right panel, we show the relation for 
  $Y\simeq 3.15\times 10^{-3}$ (dotted), $6.50\times 10^{-4}$ (dashed) and $7.95\times 10^{5}$ (solid), 
  which are chosen from the intersections of  the diagonal line for $BR(\phi \to NN) = 25\%$ 
  with the solid curve, the dashed curve and dotted line in Fig.~\ref{fig: yVsin125}. 
From the results, we see that if the RHN pair production is dominated by the SM-like Higgs boson decay,
   the allowed parameter space is very severely constrained except for a window around $m_{Z^\prime} = 100$ GeV.  
However,  Fig.~\ref{fig: gVmZ125} also shows that we can avoid the constraints by lowering $g_{BL}$.

\begin{figure}[t]
\begin{center}
\includegraphics[width=0.49\textwidth, height=5.5cm]{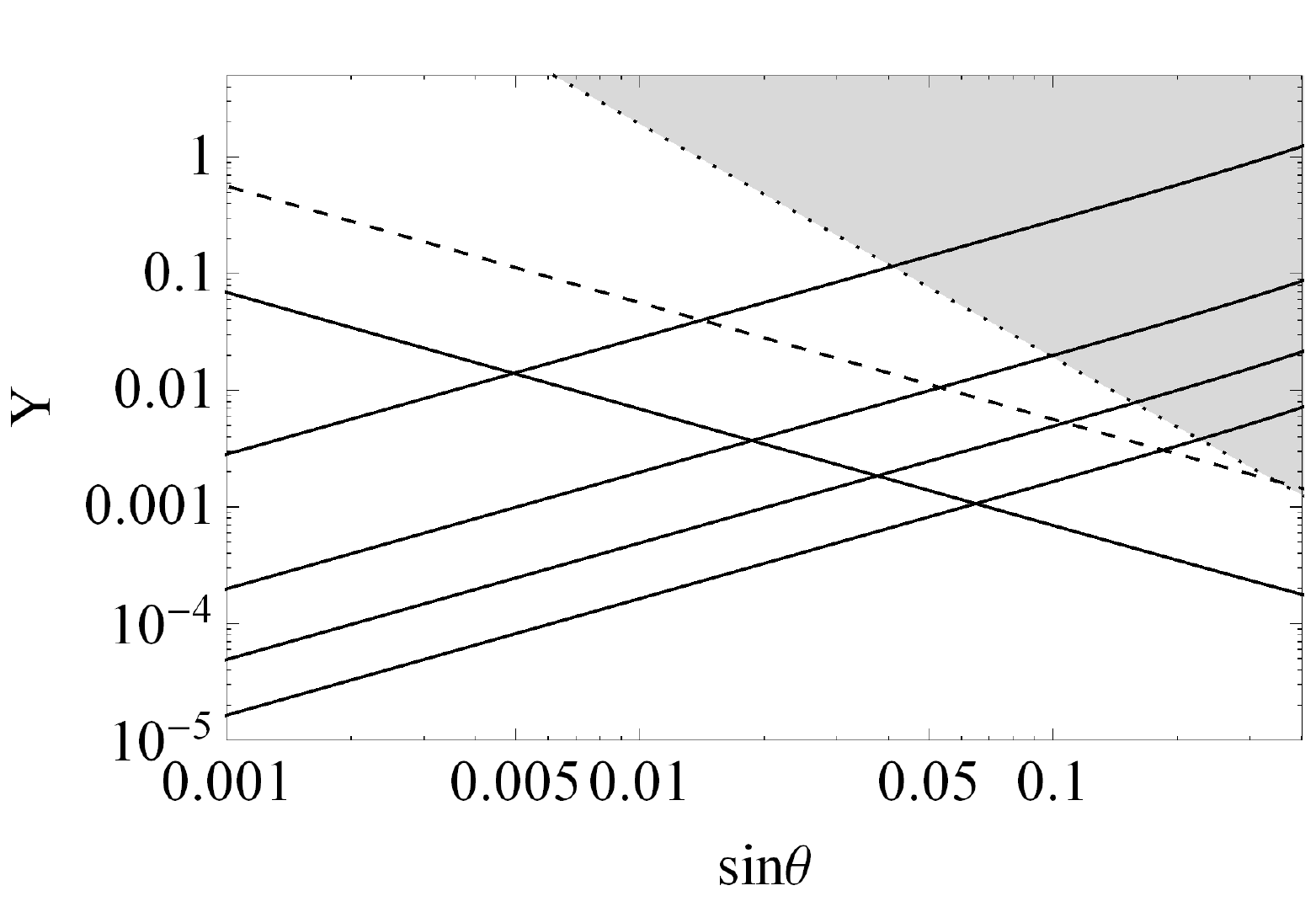}
\includegraphics[width=0.49\textwidth, height=5.5cm]{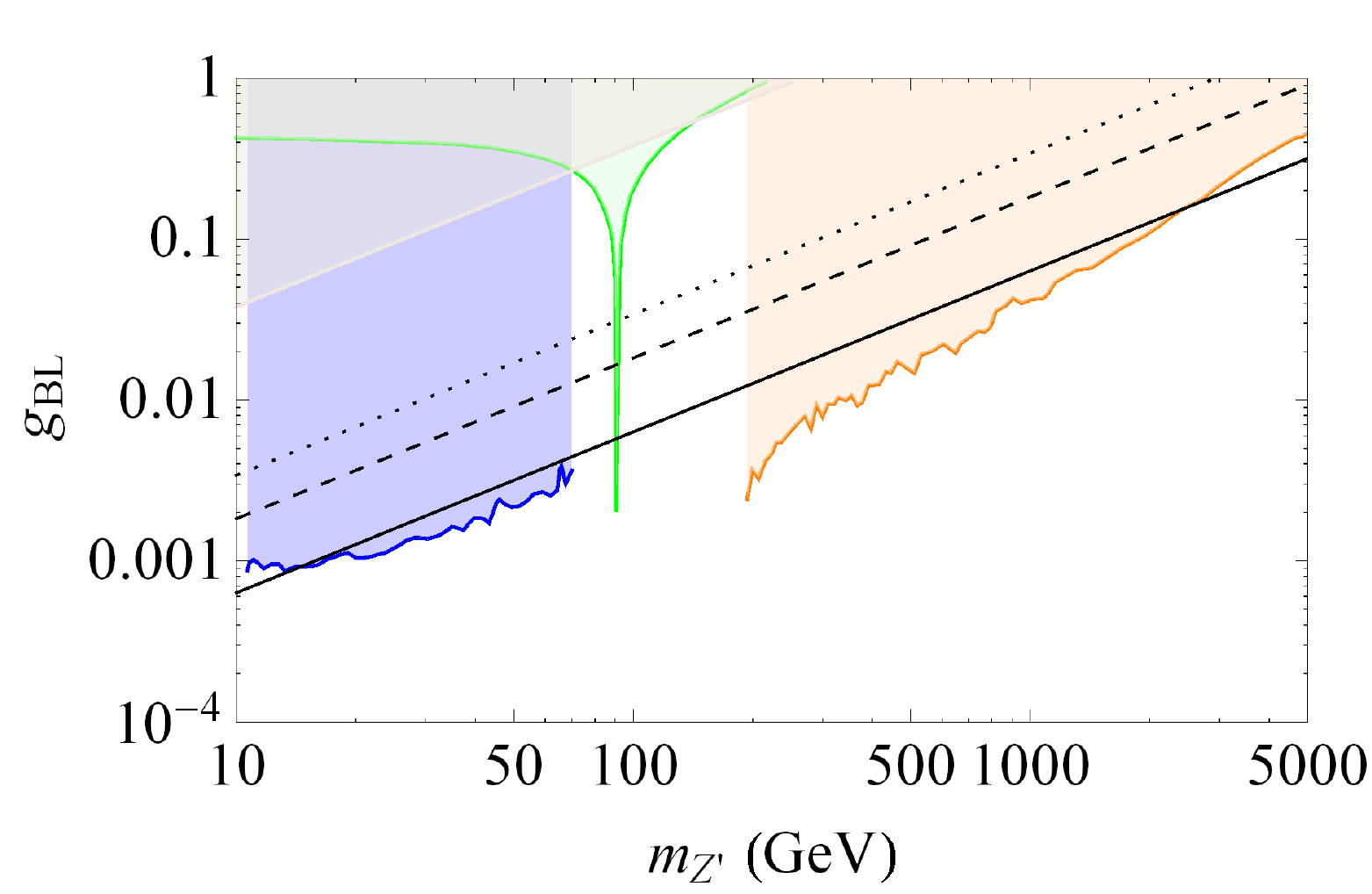}
\label{fig1: mphi70}
\end{center}
\caption{The plots shows the parameter space when the RHNs production at the LHC is dominated by SM Higgs decay for fixed $m_\phi = 70$ GeV and $m_N = 20$ GeV.
In the left panel, along dashed (solid) diagonal lines  with a negative slope, the RHN production cross section at the LHC from the SM Higgs decay is fixed to be the best search reach value for the HL-LHC  (MATHUSLA) displaced vertex searches $\sigma(pp \to XX) = 20.7 (0.3)$ fb in Fig.~\ref{fig: dvsearch}. 
The gray shaded region are excluded region by the SM Higgs boson invisible decays searches \cite{invisible}. 
The plot also shows $BR(\phi \to NN)$ lines, diagonal solid lines with positive slope. 
From top to bottom, along the line,  $BR(\phi \to NN) = 99.99\%$, $98\%$, $75\%$, and $25\%$, respectively.  
In the right panel, for a fixed Y value, the diagonal lines show the  the $Z^\prime$ boson gauge coupling values as a function of $m_Z^\prime$, along with the excluded shaded regions from various $Z^\prime$ boson searches. 
The Yukawa values are chosen to satisfy $BR(\phi \to NN) = 98\%$ in the left panel. 
The dotted diagonal line correspond to $Y\simeq 1.90\times 10^{-2}$ is fixed using the intersection of the $BR(\phi \to NN) = 98\%$ with the dotted line. 
Similarly, the dashed (solid) line correspond to $Y\simeq 1.00\times 10^{-2}$ ($3.59\times 10^{-3}$) fixed using the intersection of  $BR(\phi \to NN) = 98\%$ with the dashed (HL-LHC) and solid (MATHUSLA) lines.}
\label{fig1: mphi70}
\end{figure}

\begin{figure}[h!]
\begin{center}
\includegraphics[width=0.49\textwidth, height=5.5cm]{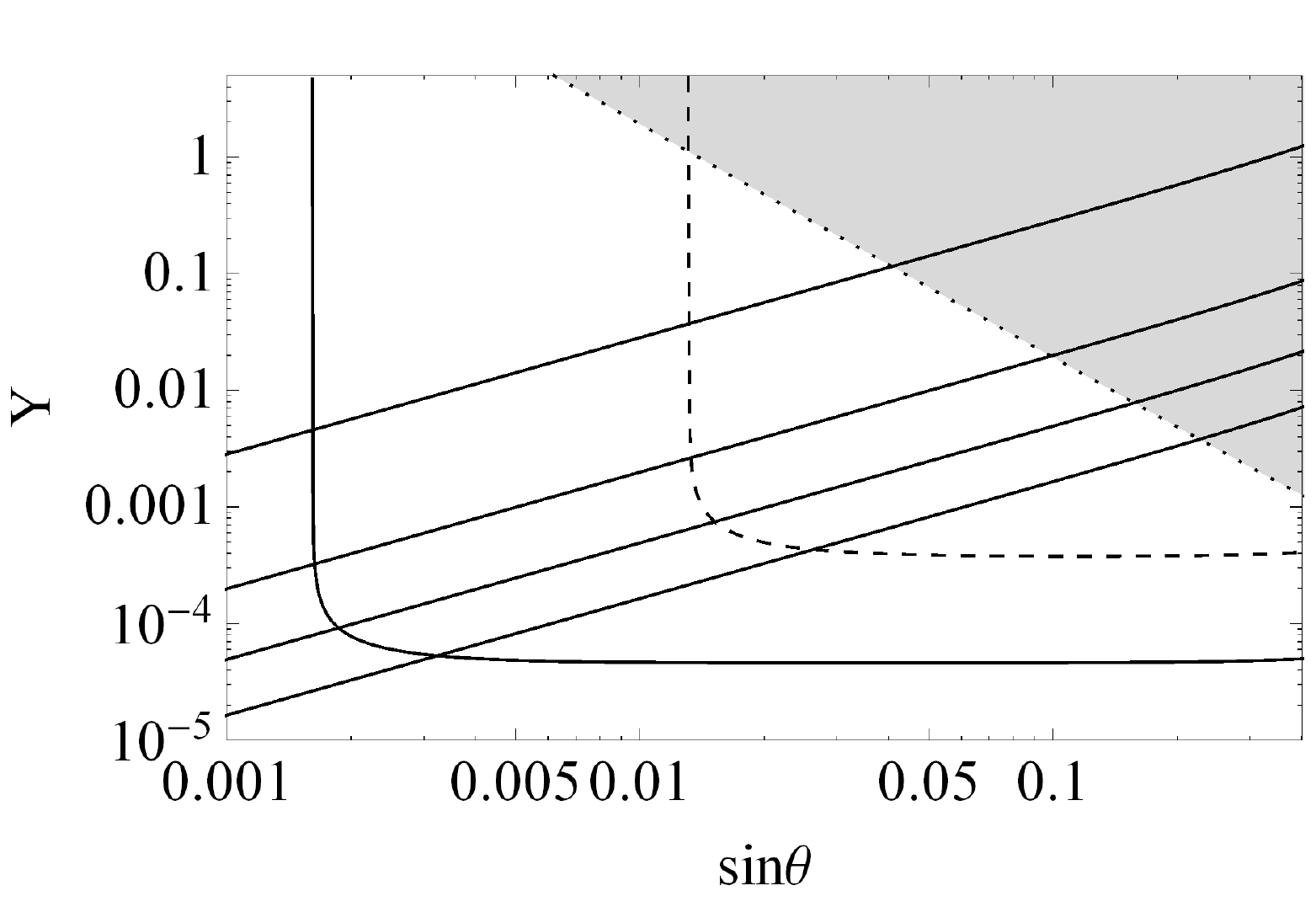}
\includegraphics[width=0.49\textwidth, height=5.5cm]{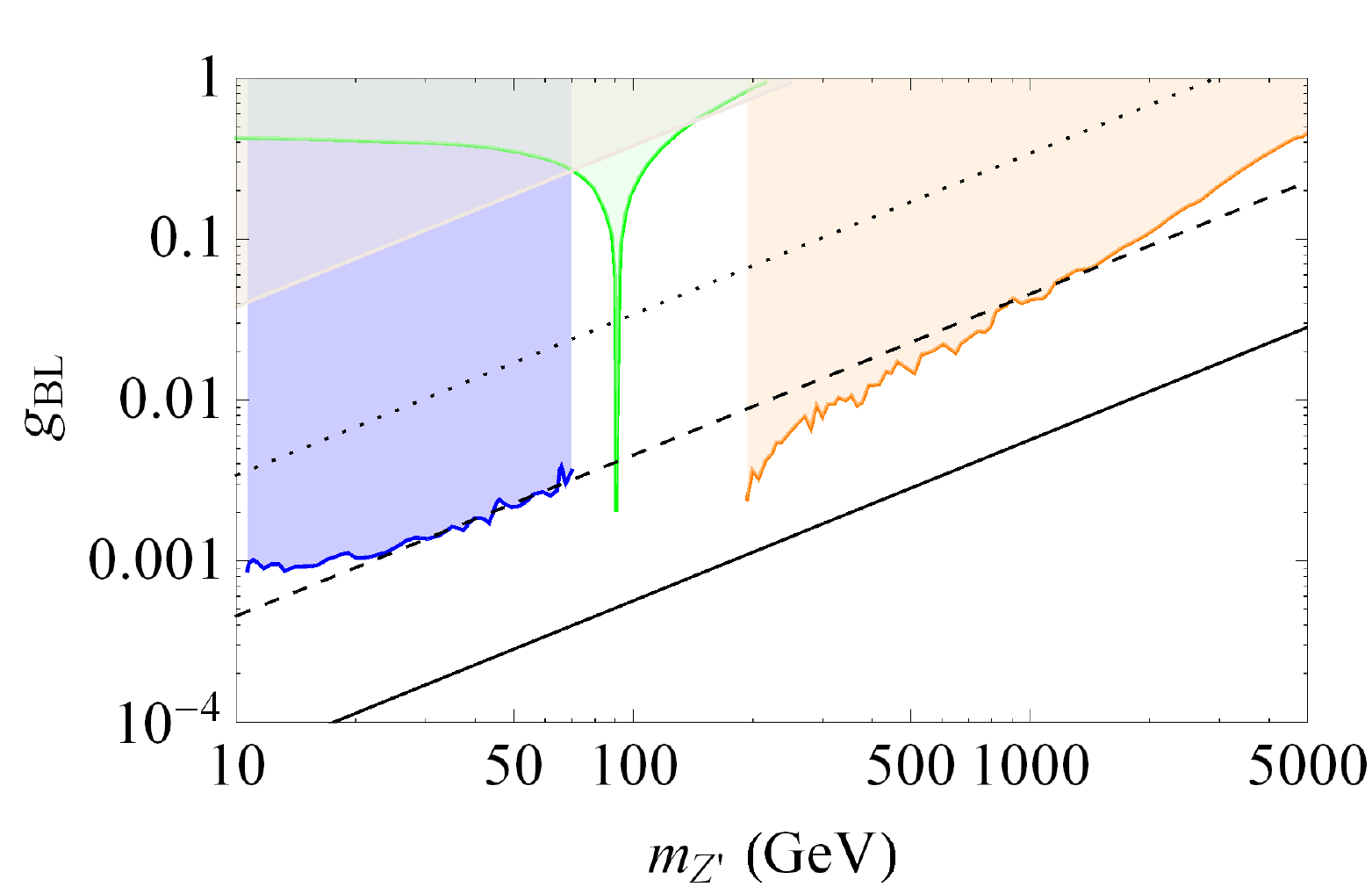}
\label{fig: yVmz}
\end{center}
\caption{The plots shows the parameter space when the RHNs production at the LHC is dominated by $B-L$ Higgs decay for fixed $m_\phi = 70$ GeV and $m_N = 20$ GeV. 
In the left panel, along dashed (solid) curved lines, the RHN production cross section at the LHC from the SM Higgs decay is fixed to be the best search reach value for the HL-LHC  (MATHUSLA) displaced vertex searches $\sigma(pp \to XX) = 20.7 (0.3)$ fb in Fig.~\ref{fig: dvsearch}. 
Along the diagonal solid lines with positive slope. $BR(\phi \to NN)$ is fixed.  
From top to bottom, $BR(\phi \to NN) = 99.99\%$, $98\%$, $75\%$, and $25\%$, respectively.  
The gray shaded region are excluded region by the SM Higgs boson invisible decays searches \cite{invisible}. 
In the right panel, for a fixed Y value, the diagonal lines show the  the $Z^\prime$ boson gauge coupling values as a function of $m_Z^\prime$, along with the excluded shaded regions from various $Z^\prime$ boson searches. 
The Yukawa values are chosen to satisfy $BR(\phi \to NN) = 98\%$ in the left panel. 
The dotted diagonal line correspond to $Y\simeq 1.92\times 10^{-2}$ is fixed using the intersection of the $BR(\phi \to NN) = 98\%$ with the dotted line in the left panel. 
Similarly, the dashed (solid) line correspond to $Y\simeq 2.58\times10^{-3}$ and $3.20\times10^{4}$ fixed using the intersection of  $BR(\phi \to NN) = 98\%$ with the dashed (HL-LHC) and dashed (MATHUSLA) lines in the left panel}
\label{fig2: mphi70}
\end{figure}

Let us next consider the case that $m_\phi$ and $m_h$ are well separated. 
In this case, we calculate the RHN pair productions from the decays of $h$ and $\phi$ separately. 
For each case, we assume the lifetime of the RHN to yield the best search reach 
  at the HL-LHC or the MATHUSLA shown in Fig.~\ref{fig: dvsearch}. 
The best search reach of the displaced vertex signatures at the HL-LHC or the MATHUSLA 
   are expressed as $\sigma_{min} = \sigma(pp\to S \to NN)$, where $S$ is either $h$ or $\phi$.  
Once $m_\phi$ is fixed, this equation leads to a relation between $Y$ and $\theta$ 
   as shown in Fig.~\ref{fig: yVsin125}  for $m_\phi=126$ GeV. 
Using Eqs.~(\ref{eq: sigS}), (\ref{eq: branch}) and (\ref{eq: sighphi}), we express 
  $\sigma_{min} = \sigma(pp\to S \to NN)$ 
  for $S=\phi$ and $h$, separately, as  
\bea
 \Gamma_{NN} (m_\phi) &=& \frac{R(m_\phi) \tan^2\theta}{\sin^2 \theta - R(m_\phi)} \left(\Gamma_{SM}(m_\phi) + \frac{\Gamma(\phi \to hh)}{\sin^2\theta}\right),
\nonumber \\
 \Gamma_{NN} (m_h) &=& \frac{R(m_h) \cot^2\theta}{\cos^2 \theta - R(m_h)} \Gamma_ {SM} (m_h), 
\label{eq: dw}
\eea 
where $R$ is defined as 
\bea
R(m_S) 
 =  \frac{\sigma_{min}}{\sigma_h (m_S)}. 
\eea
Using Eqs.~(\ref{eq: dwphi}) and $(\ref{eq: dw})$, we then obtain the relation between $Y$ and $\theta$ 
  for each case as follows: 
\bea
Y^2 &=& 
\frac{16\pi}{3} 
\frac{ \tan^2\theta}{\sin^2 \theta - R(m_\phi)}\left(\frac{R(m_\phi)}{ m_\phi}\right) 
\left(\Gamma_ {{SM}} (m_\phi) + \frac{\Gamma(\phi \to hh)}{\sin^2\theta}\right)  \left( 1- \frac{4m_N^2}{m_\phi^2} \right)^{-3/2}, 
\nonumber \\
Y^2 &=&\frac{16\pi}{3} 
\frac{\cot^2\theta}{\cos^2 \theta - R(m_h)}
 \left(\frac{R(m_h)}{ m_h}\right)  \; \Gamma_ {{SM}} (m_h)  \left( 1- \frac{4m_N^2}{m_h^2} \right)^{-3/2}. 
\label{eq: yukawa}
\eea 
The first equation in Eq.~(\ref{eq: yukawa}) indicates that for a fixed $R(m_\phi) <1$, 
  $Y^2$ becomes singular for $\sin^2\theta = R(m_\phi)$. 
Thus, there is a lower bound on $\sin \theta$ to achieve the best reach cross section $\sigma_{\rm min}$. 
For $\sin \theta \sim 1$, $Y^2$ becomes singular in both the equations. 
However, such a large mixing angle is excluded by the measurement of the SM Higgs boson properties at the LHC.

In Fig.~\ref{fig1: mphi70}, we show our results for the case that the RHNs are produced from the SM-like Higgs boson decay. 
Here, we have fixed $m_\phi = 70$ GeV and $m_N = 20$ GeV.  
In the left panel, the best reach cross section at the HL-LHC  (MATHUSLA) is achieved 
  along the dashed (solid) diagonal line with a negative slope. 
The four solid diagonal lines with positive slopes denote the relations between $Y$ and $\sin \theta$ 
  to yield  $BR(\phi \to NN) = 99.99\%$, $98\%$, $75\%$, and $25\%$, respectively, from top to bottom.   
The gray shaded region is excluded by the LHC constraint on the Higgs branching ratio into the invisible decay mode \cite{invisible}, 
  which is simply given by 
\bea  
 {BR}_{\rm inv}^{\rm higgs}  = \cos^2\theta  \times  BR(h \to NN) <  0.23, 
\eea
for the present case. 
The right panel corresponds to Fig.~\ref{fig: gVmZ125}. 
The three diagonal lines corresponds $Y\simeq 1.90\times 10^{-2}$ (dotted), 
  $Y\simeq 1.00\times 10^{-2}$ (dashed) and $3.59\times 10^{-3}$ (solid), respectively, 
  which are chosen from the intersections of the diagonal line for $BR(\phi \to NN) = 98\%$ 
  with the solid, dashed and dotted lines with negative slopes in the left panel.

Fig.~\ref{fig2: mphi70} shows the results corresponding to Fig.~\ref{fig: gVmZ125}, 
  but for the case that the RHNs are produced from the $B-L$ Higgs boson decay, 
  with $m_\phi = 70$ GeV and $m_N = 20$ GeV.  
Along the dashed and solid curves, the best reach cross section is achieved 
  at the HL-LHC and the MATHUSLA, respectively. 
The four solid diagonal lines with positive slopes denote the relations between $Y$ and $\sin \theta$ 
  to yield  $BR(\phi \to NN) = 99.99\%$, $98\%$, $75\%$, and $25\%$, respectively, from top to bottom.   
The gray shaded region is excluded region by the SM Higgs boson invisible decays search. 
As we have discussed, the curves show the singularities for small $\sin \theta$ values. 
In the right panel, the three diagonal lines corresponds 
  $Y\simeq 1.92\times 10^{-2}$ (dotted), 
  $Y\simeq  2.58\times10^{-3}$ (dashed) and $3.20\times10^{-4}$ (solid), respectively,    
  which are chosen from the intersections of the diagonal line for $BR(\phi \to NN) = 98\%$ 
  with the solid curve, the dashed curve and the dotted line in the left panel.

\begin{figure}[t]
\begin{center}
\includegraphics[width=0.49\textwidth, height=5.5cm]{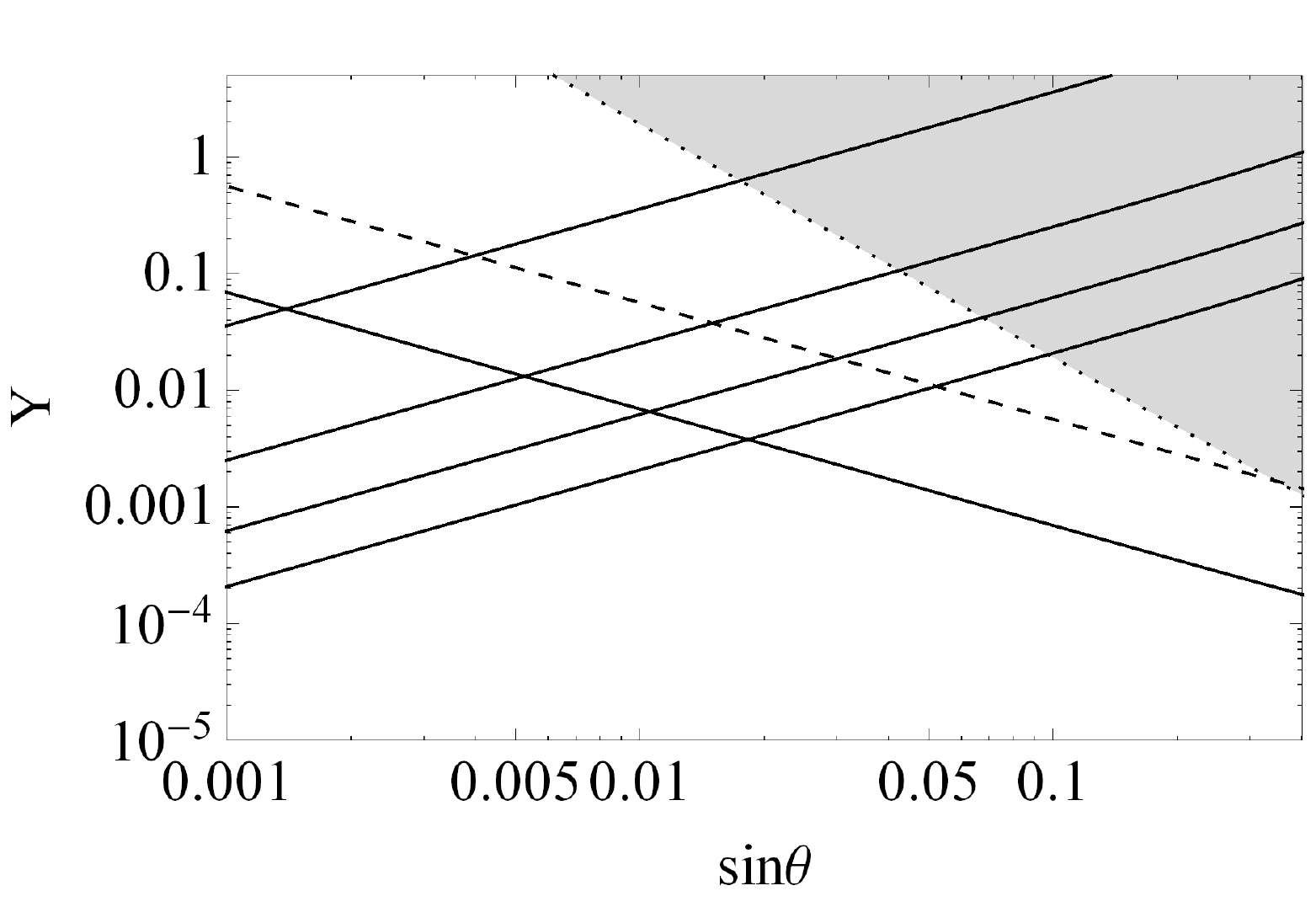}
\includegraphics[width=0.49\textwidth, height=5.5cm]{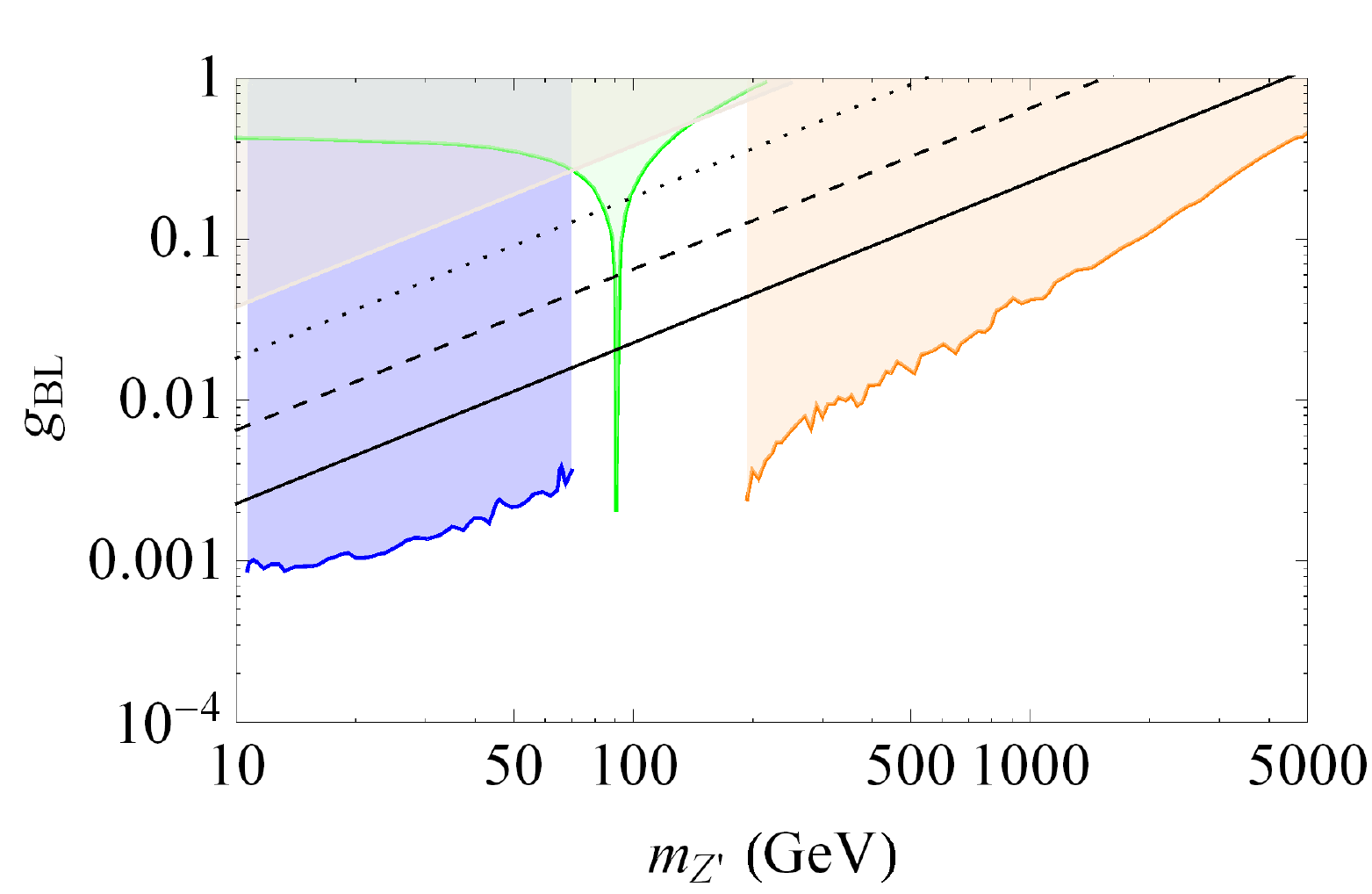}\\
\includegraphics[width=0.49\textwidth, height=5.5cm]{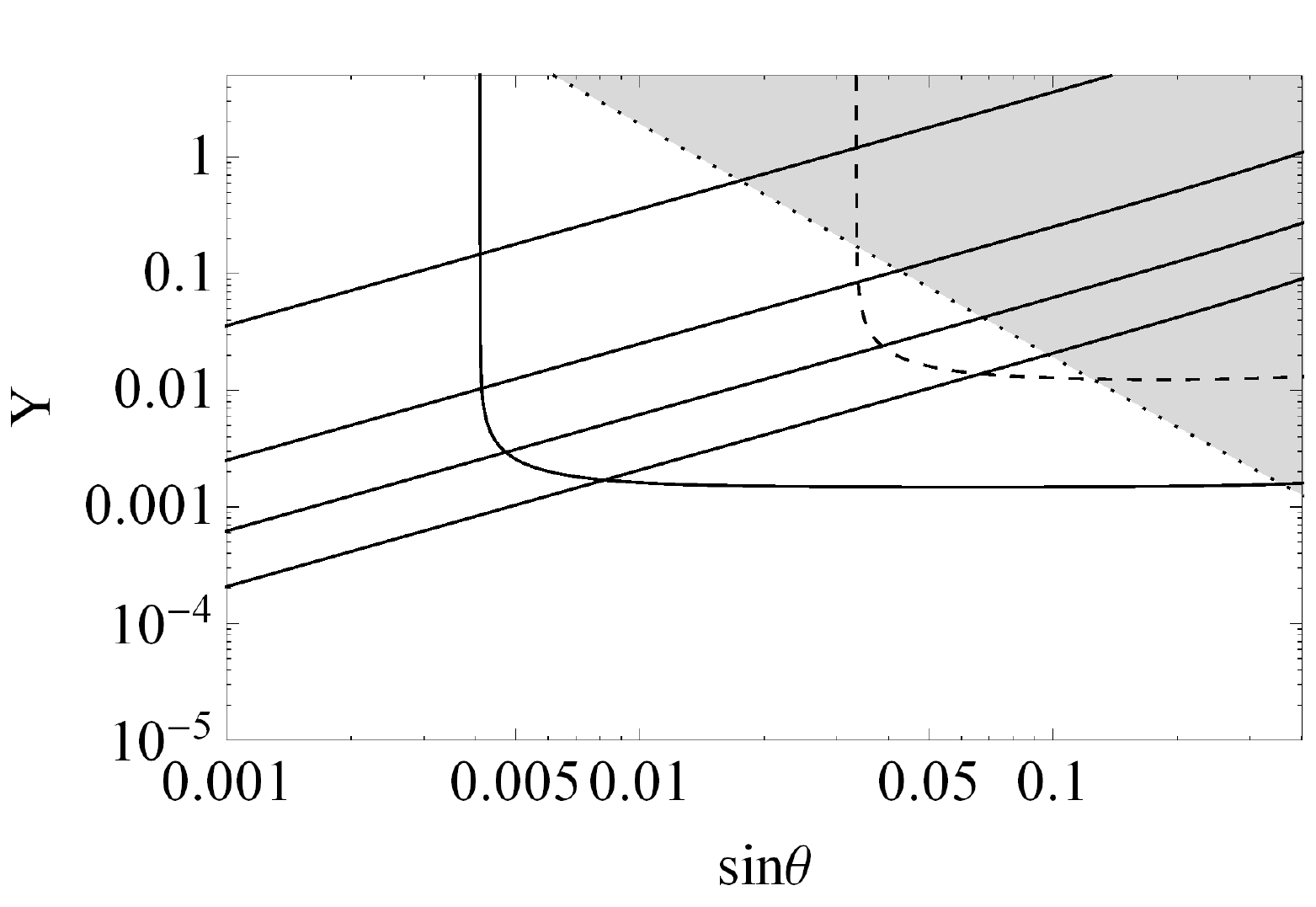}
\includegraphics[width=0.49\textwidth, height=5.5cm]{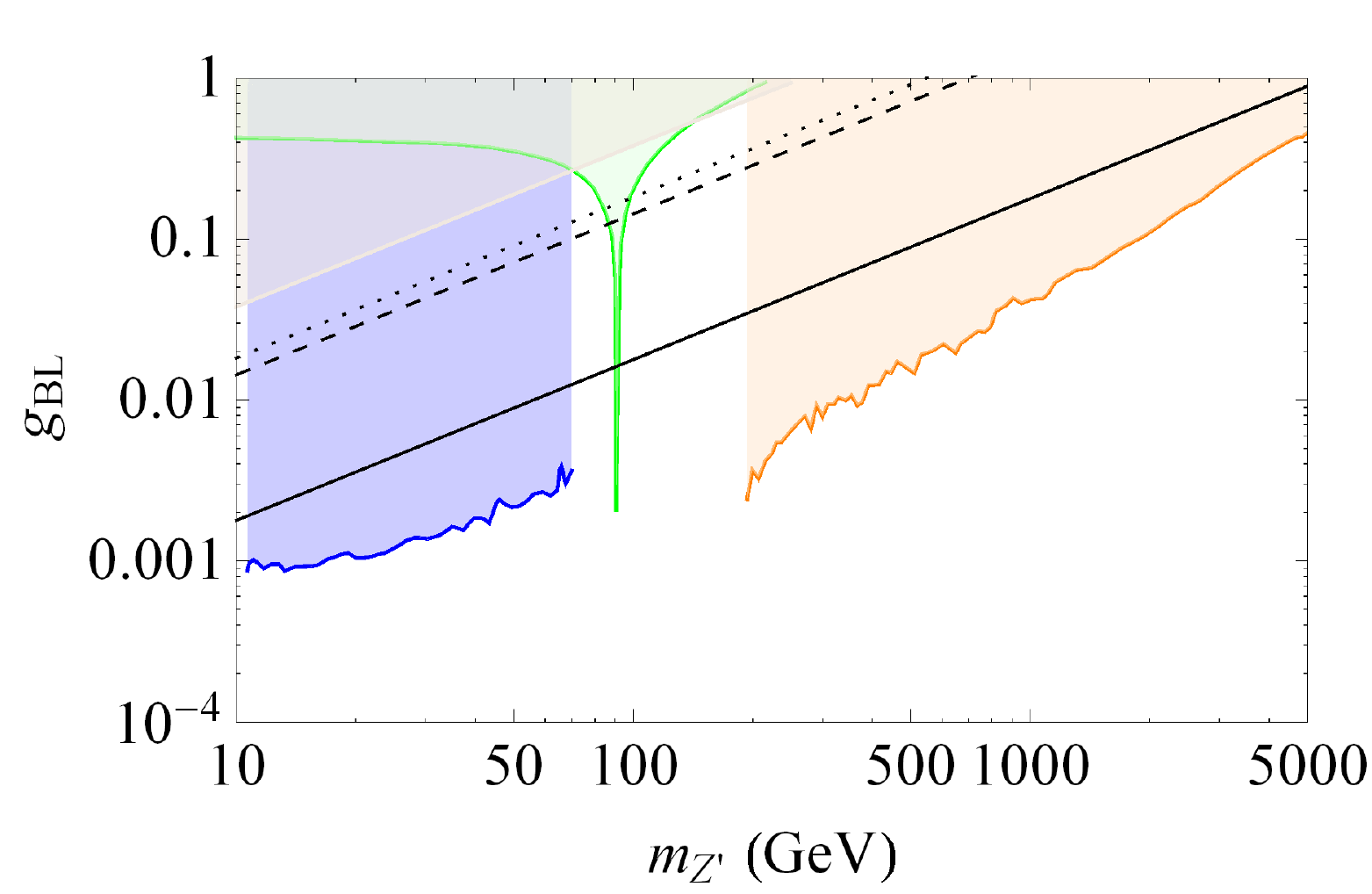}
\label{fig: yVmz}
\end{center}
\caption{For fixed $m_\phi = 200$ GeV and $m_N = 20$ GeV, the plots shows the parameter space when the RHNs production at the LHC is from the: SM Higgs decay (top panel) and $B-L$ decay (bottom panel). 
The line coding for plots in the top and bottom panels is the same as Figures~\ref{fig1: mphi70} and  \ref{fig2: mphi70}, respectively. 
For the plots in the top (bottom) right column, the branching ratio in fixed to be, $BR(\phi \to NN) = 98\%$ with a corresponding $Y\simeq 1.03\times10^{-1}, 3.68\times 10^{-2}$ and $1.28079\times10^{-2}$ ($Y\simeq 1.03\times10^{-1}, 8.12\times10^{2}$ and $1.01\times10^{-2}$), for dotted, dashed, and solid lines, respectively.}
\label{fig: mphi200}
\end{figure}

\begin{figure}[t]
\begin{center}
\includegraphics[width=0.49\textwidth, height=5.5cm]{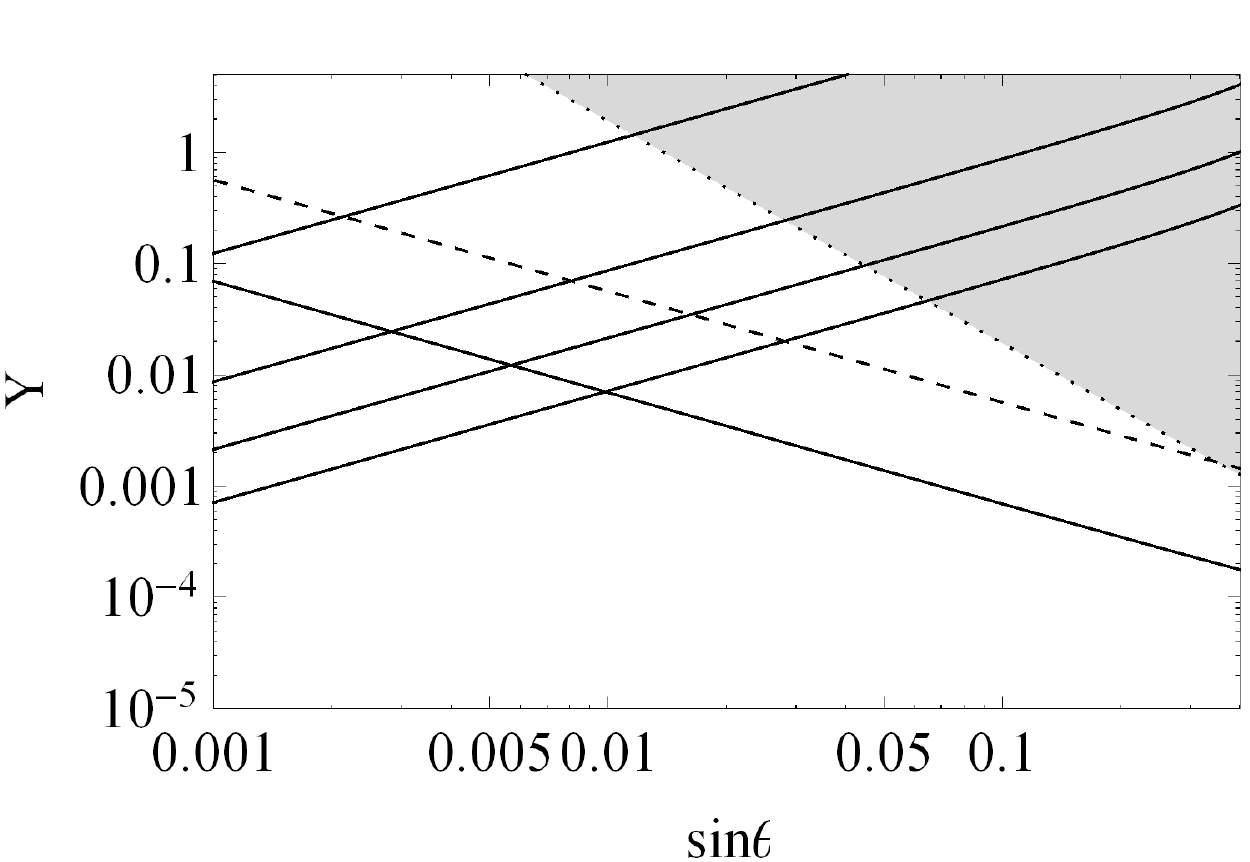}
\includegraphics[width=0.49\textwidth, height=5.5cm]{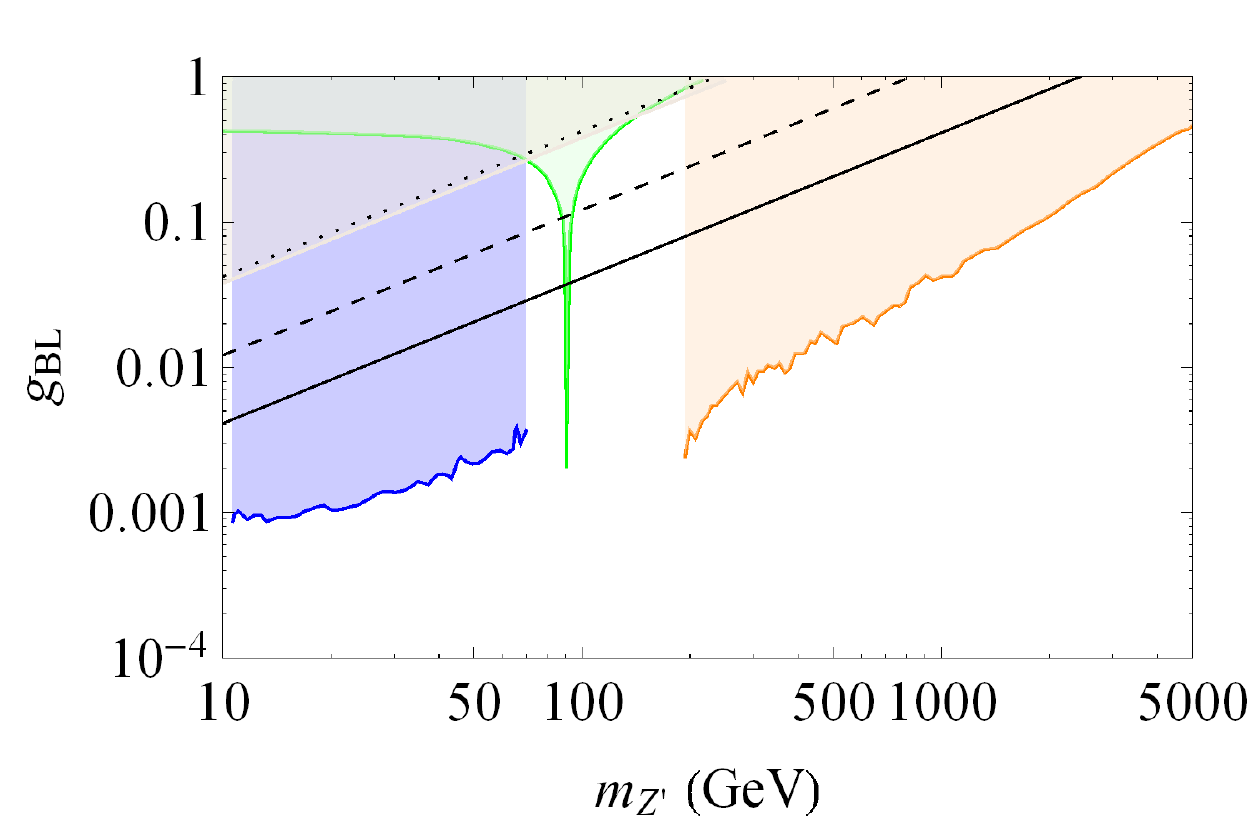}\\
\includegraphics[width=0.49\textwidth, height=5.5cm]{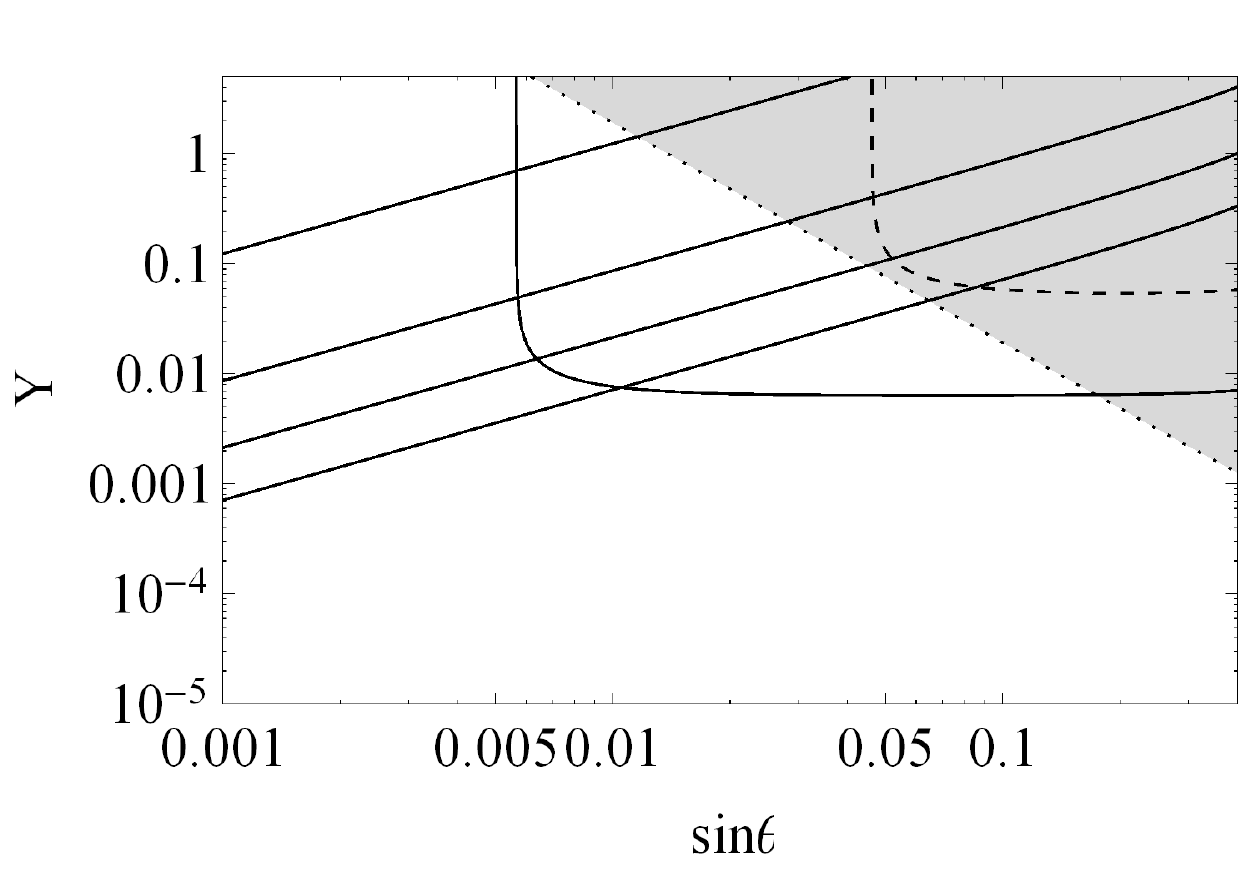}
\includegraphics[width=0.49\textwidth, height=5.5cm]{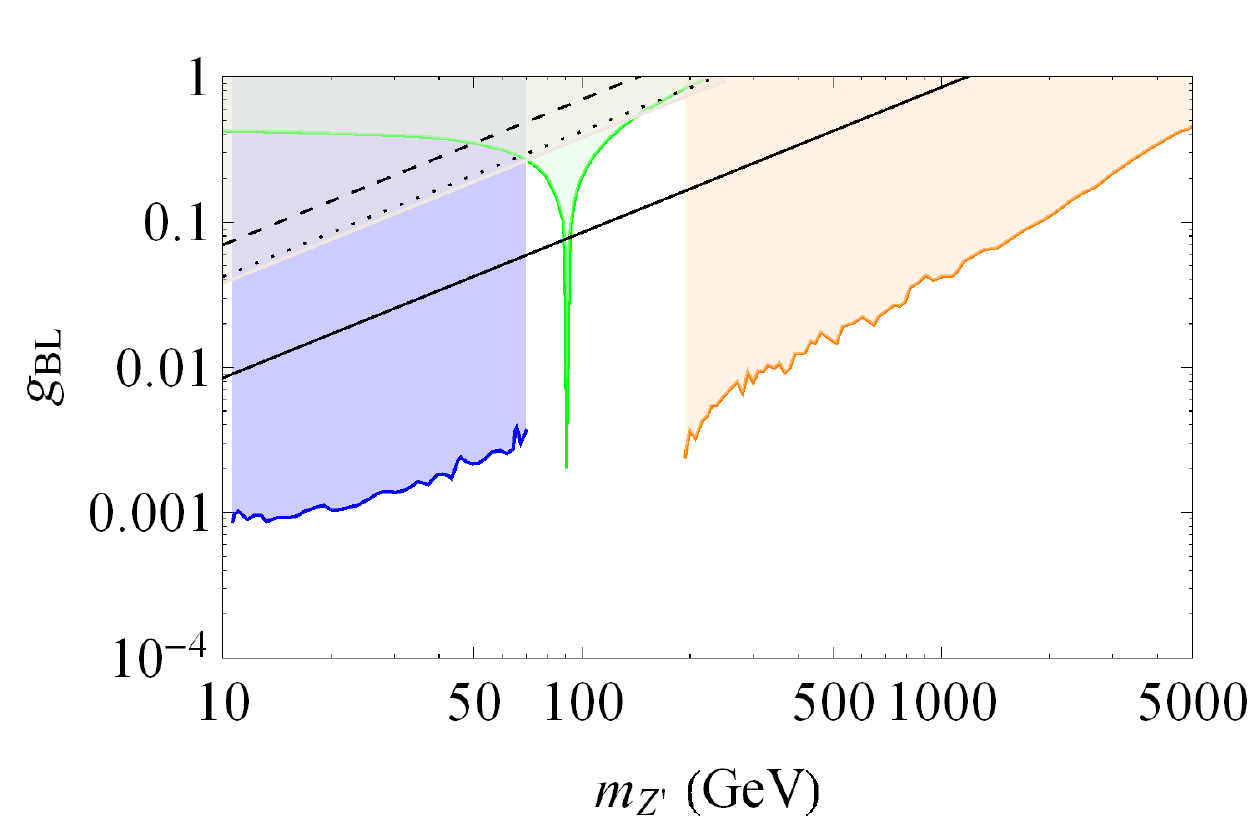}
\end{center}
\caption{For fixed $m_\phi = 400$ GeV and $m_N = 20$ GeV, the plots shows the parameter space when the RHNs production at the LHC is from the: SM Higgs decay (top panel) and $B-L$ decay (bottom panel). 
The line coding for the top and bottom panels is the same as Figures \ref{fig1: mphi70} and  \ref{fig2: mphi70}, respectively. For the plots in the top (bottom) right column, the branching ratio in fixed to be, $BR(\phi \to NN) = 98\%$ with a corresponding $Y\simeq 2.38\times10^{-1}, 6.86\times10^{-2}$ and $2.32\times10^{-2}$ ($Y\simeq 2.38\times10^{-1}, 3.94\times10^{-1}$ and $4.79\times10^{-2}$), for dotted, dashed, and solid lines, respectively.}
\label{fig: mphi400}
\end{figure}

Our results for $m_\phi=200$ GeV corresponding to Figs.~\ref{fig1: mphi70} and \ref{fig2: mphi70} 
  are shown, respectively, in the top panels and the bottom panels of Fig.~\ref{fig: mphi200}. 
All line and color codings are the same as Figs.~\ref{fig1: mphi70} and \ref{fig2: mphi70}. 
In the top-right (bottom-right) panel, three diagonal lines correspond to 
    $Y\simeq 1.03\times10^{-1}$, $Y=3.68 \times10^{-2}$ and $Y=1.28\times10^{-2}$ 
   ($Y\simeq 1.03\times10^{-1}$, $Y=8.12\times10^{-1}$ and $1.01\times10^{-2}$), respectively,  
   from top to bottom. 
Fig.~\ref{fig: mphi400} is the same as Fig.~\ref{fig: mphi200} but for $m_\phi=400$ GeV. 
Note that in this case, the total decay width also includes the decay mode of $\phi \to hh$.  
In the top-right (bottom-right) panel, three diagonal lines correspond to 
  $Y\simeq 2.38\times10^{-1}$, $Y=6.86\times10^{-2}$ and $Y=2.32\times10^{-2}$ 
   ($Y\simeq 2.38\times10^{-1}$, $Y=3.94\times10^{-1}$ and $4.79\times10^{-2}$), respectively,  
   from top to bottom. 
Qualitative behaviors of various curves and lines shown in this figure 
   are similar to those in Fig.~\ref{fig: mphi200}.

Let us here summarize our results as $m_\phi$ is increased. 
From the left panel in Fig.~\ref{fig2: mphi70}, the bottom-left panels in Fig.~\ref{fig: mphi200} and 
 Fig.~\ref{fig: mphi400}, 
  we can see that the resultant curves and the diagonal lines are shifting upward to the right 
  as $m_\phi$ is increased. 
This is because 
 (i) as $m_\phi$ is increased, the partial decay widths of $\phi$ into the SM particles 
  become larger and as a result, $Y$ is increased to yield a fixed branching ratio to $\phi \to NN$, see Eq.~\ref{eq: yukawa}; 
 (ii) since $\sigma_h (m_\phi)$ is decreasing as $m_\phi$ is increased, 
  the lower bound on $\sin \theta$ (at which $Y$ becomes singular) is increasing (see the discussion below Eq.~(\ref{eq: yukawa}). 
Hence, the LHC constraint from the invisible decay of the SM Higgs boson relatively becomes more severe. 
In fact, the bottom-left panel of Fig.~\ref{fig: mphi400} shows that  
  the dashed curve appears inside the gray shaded region 
  and thus the entire parameter region which can be explored at the future HL-LHC is already excluded. 
According to Eq.~\ref{eq: gauge}, the $B-L$ gauge coupling becomes larger for a fixed $m_{Z^\prime}$ 
   as $Y$ becomes larger.   
Hence, the current constraints from the $Z^\prime$ boson search become more severe 
   as $m_\phi$ is increased as can be seen from the right panel in Fig.~\ref{fig2: mphi70}, 
   the bottom-right panel in Fig.~\ref{fig: mphi200} and the bottom-right panel in Fig.~\ref{fig: mphi400}. 
Note that if we take $g_{BL}$ small enough, for example, $g_{BL} < 10^{-4}$, 
   all the existing collider constraints from the $Z^\prime$ boson search can be avoided.

\begin{figure}[t]
\begin{center}
\includegraphics[width=0.7\textwidth, height=8cm]{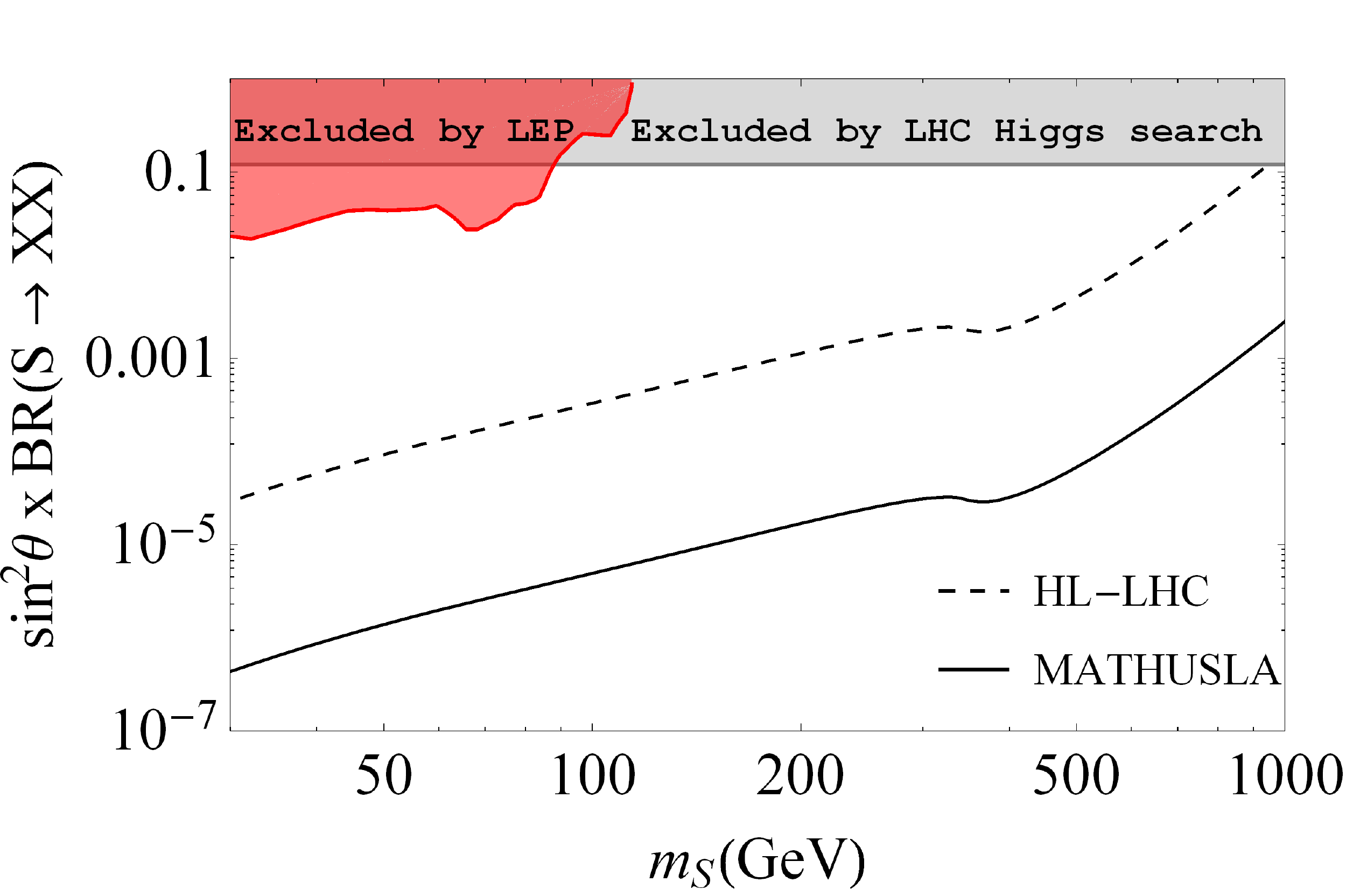}
 \caption {
The search reach of the displaced vertex searches at the HL-LHC and MATHUSLA. 
For $BR(S \to XX) \simeq 100\%$, the red (gray) shaded region is excluded 
  by the LEP (LHC) experiments. 
} 
\label{fig:dis reach}
 \end{center}
 \end{figure}

We conclude this section by generalizing our analysis for the long-lived heavy neutrino 
  to the case for an SM-singlet particle $X$ produced through $pp\to S \to XX$ 
  with an SM-singlet scalar $S$ at the LHC. 
Since $S$ is SM-singlet, it is produced at the LHC through a mixing with the SM Higgs boson, 
  just like $\phi$. 
Hence, the total production cross section of the process, $pp\to S \to XX$,  is given by
\bea
\sigma(pp\to S \to XX) &=& \sigma(pp\to S) \times  BR(S\to XX) \nonumber \\ 
  &=& \sin^2\theta  \times  \sigma_h (m_S) \times  BR(S\to XX),
\eea
  where $\theta$ is the mixing angle, and $\sigma_h (m_S)$ is the production cross section of the SM Higgs boson 
  if the SM Higgs boson mass were the mass of $S$ ($m_S$).  
Let us assume the lifetime of $X$ to yield the best reach values, 
  $\sigma_{\rm min} ({\rm HL-LHC}) = 20.7$ and $\sigma_{\rm min} ({\rm MATHUSLA}) =0.310$ fb, 
  for the displaced vertex searches at the HL-LHC and the MATHUSLA, respectively. 
We then calculate $\sigma_h (m_S)$ at the 13 TeV LHC to obtain a relation 
  between $\sin^2\theta \times BR(S\to XX)$ and $m_S$ 
  to achieve the best reach value $\sigma(pp\to S \to XX) =\sigma_{\rm min} ({\rm HL-LHC})$ 
  or  $\sigma_{\rm min} ({\rm MATHUSLA})$. 
Our results are  shown in Fig. \ref{fig:dis reach}. 
The region above the dashed black line (solid black line) 
   can be explored at the HL-LHC (the MATHUSLA) with a $3 \rm~ab^{-1}$ luminosity. 
For the case of $BR(S \to XX) \simeq 100\%$, 
   the red shaded region is excluded by the LEP results on the search for the SM Higgs boson, 
   while the gray shaded region is excluded by the LHC measurement of the SM Higgs boson properties. 
Assuming  $BR(S \to XX) \simeq 100\%$, 
   we can read off the search reach of $m_S$ from Fig. \ref{fig:dis reach} 
   for a fixed value of $\sin \theta$. 
Our results for three benchmark points are listed in Table~\ref{tab2}.    


\begin{table}[t]
\centering
\label{my-label}
\begin{tabular}{||l|l|l|l||}
\hline
\multirow{2}{*}{\textbf{Benchmark Points}} & \multirow{2}{*}{\textbf{Mixing angle} ($\theta$)} & \multicolumn{2}{l||}{\textbf{Search Reach of $m_S$[GeV]} } \\ \cline{3-4} 
                  &                   & \textbf{ MATHUSLA }         & \textbf{HL-LHC}            \\ \hline
      \quad  \textbf{BP1}          &    8 $\times 10^{-3}$              &    476       & 39          \\ \hline
            \quad    \textbf{BP2}           &   5 $\times 10^{-2}$               &    972       &      293     \\ \hline
                  \quad \textbf{BP3}        &      $1 \times 10^{-2}$              &           556  &      65     \\ \hline
\end{tabular}
\caption{Summary of $\phi$ mass reach at MATHUSLA and HL-LHC experiment.}
\label{tab2}
\end{table}

\section{Lifetime of heavy neutrinos}
\label{sec:5}
We assumed a suitable lifetime of the heavy neutrino in the previous section.  
In this section, we calculate the lifetime of the heavy neutrinos 
  for realistic parameters to reproduce the neutrino oscillation data 
  and investigate the prospect of searching for the displaced vertex signatures 
  of the heavy neutrino productions. 

After the $B-L$ and electroweak symmetry breakings, 
  the neutrino mass matrix is generated to be 
\bea
{\cal M}_{\nu}=\begin{pmatrix}
0&&m_{D}\\
(m_{D})^{T}&&M_{N}
\end{pmatrix},
\label{typeInu}
\eea
where $M_N$ and $m_D$  are the Majorana and the Dirac mass matrices, respectively. 
From Eqs.~(\ref{U1XYukawa}) and (\ref{masses}), 
  we have $M_N={\rm diag}( m_{N^1}, m_{N^2}, m_{N^3})$ and $m_D^{ij} = Y_D^{ij} v_{SM}/\sqrt{2}$. 
Assuming the mass hierarchy $|m_D^{ij} / m_{N^k}| \ll 1$, 
  the seesaw formula for the light Majorana neutrinos is given by
\bea
m_{\nu} \simeq - m_{D}(M_{N})^{-1}m_{D}^{T}. 
\label{seesawI}
\eea 
The light neutrino flavor eigenstate $(\nu)$ can be expressed in terms the light $(\nu_m)$ and heavy $(N_m)$ Majorana neutrino mass eigenstates, 
$\nu \simeq \mathcal{N} \nu_m+\mathcal{R} N_m$, 
  where $\mathcal{R} =m_D (M_N)^{-1}$, $\mathcal{N}=\Big(1-\frac{1}{2}\mathcal{R}^\ast\mathcal{R}^T\Big)\simeq U_{\rm{MNS}}$, and $U_{\rm{MNS}}$ is the neutrino mixing matrix which diagonalizes $m_\nu$ by
\bea
  U_{\rm MNS}^T m_\nu U_{\rm MNS}  = D_\nu = {\rm diag}(m_1, m_2, m_3), 
\label{seesawII}
\eea 
where the neutrino mixing matrix is parameterized as 
\bea
U_{\rm{MNS}} = \begin{pmatrix} c_{12} c_{13}&c_{12}c_{13}&s_{13}e^{-i\delta}\\-s_{12}c_{23}-c_{12}s_{23}s_{13}e^{i\delta}&c_{12}c_{23}-s_{12}s_{23}s_{13}e^{i\delta}&s_{23} c_{13}\\ s_{12}c_{23}-c_{12}c_{23}s_{13}e^{i\delta}&-c_{12}s_{23}-s_{12}c_{23}s_{13}e^{i\delta}&c_{23}c_{13}
\end{pmatrix} 
\begin{pmatrix}
1&0&0\\
0&e^{-i \rho_1}&0\\
0&0&e^{-i \rho_2}
 \end{pmatrix},
\label{pmatrix}
\eea
where $c_{ij}=\cos\theta_{ij}$,  $s_{ij}=\sin\theta_{ij}$, $\delta$ is the Dirac $CP$ phase, 
 and $\rho_1$ and $\rho_2$ are the Majorana $CP$ phases. 
Using the Eqs.~(\ref{seesawI}) and (\ref{seesawII}),  the Dirac mass matrix is parameterized as \cite{Casas:2001sr}
\bea 
  m_D =  U_{\rm{MNS}}^* \sqrt{D_{\nu}} \; O \sqrt{M_N} ,  
\label{mD}
\eea
where $O$ is a general orthogonal matrix, 
  $\sqrt{M_N} \equiv {\rm diag}( \sqrt{m_{N^1}}, \sqrt{m_{N^2}}, \sqrt{m_{N^3}})$ 
  and  $\sqrt{D_\nu} \equiv {\rm diag}( \sqrt{m_1}, \sqrt{m_2}, \sqrt{m_3})$.

The charged current interaction of the neutrino mass eigenstates is expressed as 
\bea 
\mathcal{L}_{CC}= 
 -\frac{g}{\sqrt{2}} W_{\mu}
  \overline{\ell_\alpha} \gamma^{\mu} P_L 
   \left( {\cal N}_{\alpha i} \nu_{m}^i+ {\cal R}_{\alpha i} N_{m}^i \right) + \rm{h.c.}, 
\label{CC}
\eea
where $\ell_\alpha$ are the 3 generations of the charged SM leptons, and $P_L =  (1- \gamma_5)/2$ is the left handed projection operator. 
Similarly, for the the neutral current interaction, we have 
\bea 
\mathcal{L}_{NC}&=& 
 -\frac{g}{2 \cos \theta_{\rm W}}  Z_{\mu} 
\Big[ 
  \overline{\nu_{m}^i} \gamma^{\mu} P_L ({\cal N}^\dagger {\cal N})_{ij} \nu_{m}^j
 +  \overline{N_{m}^i} \gamma^{\mu} P_L ({\cal R}^\dagger {\cal R})_{ij} N_{m}^j \nonumber \\
&+& \Big\{ 
  \overline{\nu_{m}^i} \gamma^{\mu} P_L ({\cal N}^\dagger  {\cal R})_{ij} N_{m}^j 
  + \rm{H.c.} \Big\} 
\Big] , 
\label{NC}
\eea
where $\theta_{\rm W}$ is the weak mixing angle.

With the general parameterization of Eq.~(\ref{mD}), 
  the matrix ${\mathcal R}$ is given by
\bea
R_{\alpha i} =  m_D (M_N)^{-1}= U_{\rm{MNS}}^* \sqrt{D_{\nu}} \; O (\sqrt{M_N})^{-1}.
\label{eq: R}  
\eea
In order to fix $R_{\alpha i}$, we employ the neutrino oscillation data:  
  $\sin^{2}2{\theta_{13}}=0.092$ \cite{Neut4} 
  along with
 $\sin^2 2\theta_{12}=0.87$, $\sin^2 2\theta_{23}=1.0$, as well as the mass squared differences, 
  $\Delta m_{12}^2 = m_2^2-m_1^2 = 7.6 \times 10^{-5}$ eV$^2$ and 
  $\Delta m_{23}^2= |m_3^2-m_2^2|=2.4 \times 10^{-3}$ eV$^2$ \cite{PDG}. 
Motivated by the recent measurements, we also fix the Dirac $CP$ phase as $\delta=\frac{3\pi}{2}$ \cite{CP-phase}, 
  while we simply take $\rho_1=\rho_2=0$ for the Majorana $CP$ phases. 
To simplify our analysis, we set the orthogonal matrix $O$ to be the identity matrix and 
   assume the mass degeneracy for three heavy neutrinos, $m_{N^{1,2,3}} = m_N$. 
For the light neutrino mass spectrum, we consider two cases: 
   the normal hierarchy (NH), $m_1< m_2< m_3$, and the inverted hierarchy (IH), $m_3< m_1< m_2$.

Let us now consider the decay of the heavy neutrinos into the SM particles. 
In our analysis, we set $m_N = 20$ GeV, hence the heavy neutrino decays into the SM quarks and leptons
   via intermediate off-shell $W$ and $Z$ bosons. 
The expression for the decay width of heavy neutrinos into various final states are as follows: 

\noindent
1. Leptons in the final states: 
\bea
\Gamma^{(W)}({N^i \to \ell_L^\alpha \ell_L^\beta \nu^\kappa})
 &=&   \left(\sum_{\alpha=1}^3|R_{\alpha i}|^2\right) \left(\sum_{\beta ,\kappa}|U^{\beta \kappa }_{MNS}|^2\right)
 \Gamma_{\rm N^i},
\nonumber \\
\Gamma^{(Z)}({N^i \to \nu^\alpha \ell_L^{\beta} \ell_L^{\kappa}})
 &=&  \left(\sum_{\alpha=1}^3|R_{\alpha i}|^2\right) \left(\sum_{\beta ,\kappa=1} \delta_{\beta \kappa} \right) \cos^22\theta_W 
 \, \frac{1}{4} \, \Gamma_{\rm N^i},
\nonumber \\
\Gamma^{(Z)}({N^i \to \nu^\alpha \ell_R^{\beta} \ell_R^{\kappa}})
 &=& \left(\sum_{\alpha=1}^3|R_{\alpha i}|^2\right) \left(\sum_{\beta ,\kappa} \delta_{\beta \kappa}\right) \,
  \sin^4\theta_W  \, \Gamma_{\rm N^i}, 
\nonumber \\
\Gamma^{(Z)}({N^i \to \nu^\alpha \nu^\beta \nu^\kappa})
 &=& \left(\sum_{\alpha=1}^3|R_{\alpha i}|^2\right) \left(\sum_{\beta ,\kappa}\delta_{\beta \kappa}\right)  
 \, \frac{1}{4} \, \Gamma_{\rm N^i},
\label{eq: dwlepton1}
\eea
where 
\bea
\Gamma_{\rm N^i} =\frac{G_F^2}{192 \pi^3} m_{N^i}^5
\eea
with the Fermi constant $G_F$, 
  $U_{MNS}^{\alpha \beta}$ is a $(\alpha, \beta)$-element of the neutrino mixing matrix, 
  and $\sum_{\beta ,\kappa}|U^{\beta \kappa }_{MNS}|^2 = 3 = \sum_{\beta ,\kappa} \delta_{\beta \kappa}$. 
In deriving the above formulas, we have neglected all lepton masses. 
For a degenerate heavy neutrino mass spectrum, we obtain $  \sum_{\alpha=1}^3|R_{\alpha i}|^2= \frac{ m_i}{m_N}$. 
For the lepton final states, we have an interference between the $Z$ and $W$ boson mediated decay processes: 
\bea
\nonumber \\
\Gamma^{(Z/W)}({N^i \to \nu \ell \ell})
 &=&
 \left(\sum_{\alpha=1}^3|R_{\alpha i}|^2\right) \times 2{\rm Re} (U_{ii}) \times \Gamma_{\rm N^i}.
\label{eq: dwlepton2}
\eea

\noindent
2. Quarks in the final states: 
\bea
\Gamma^{(W)}(N^i \to \ell^\alpha  q_L^\beta{\bar q}_L^\kappa)
 &=&N_c  \left(\sum_{\alpha=1}^3|R_{\alpha i}|^2\right) 
\left(\sum_{\beta, \kappa}
|V^{q^\beta {\bar q}^\kappa}_{CKM}|^2\right)
\Gamma_{\rm N^i},
\nonumber \\
\Gamma^{(Z)}(N^i \to \nu^\alpha  q_L^\beta{\bar q}_L^\kappa)
 &=&  N_c  \left(\sum_{\alpha=1}^3|R_{\alpha i}|^2\right)
\left(\sum_{\beta, \kappa}
|V^{q^\beta {\bar q}^\kappa}_{CKM}|^2\right)
 \cos^22\theta_W \,  \frac{1}{4} \, \Gamma_{\rm N^i},
\nonumber \\
\Gamma^{(Z)}(N^i \to \nu^\alpha  q_R^\beta{\bar q}_R^\kappa)
 &=&N_c \left(\sum_{\alpha=1}^3|R_{\alpha i}|^2\right)
\left(\sum_{\beta, \kappa}
|V^{q^\beta {\bar q}^\kappa}_{CKM}|^2\right) \,
 \sin^4\theta_W \, \Gamma_{\rm N^i},
\label{eq: dwquark}
\eea
where $N_c = 3$ is the color factor. 
Since we have set $m_N =20$ and $40$ GeV in the following analysis, 
   we only consider the first two generation of quarks in the final states 
   and $\sum_{\beta, \kappa} |V^{q^\beta {\bar q}^\kappa}_{CKM}|^2 = 2$. 
For the final state quarks, there is no interference between $W$ and $Z$ boson mediated processes.

\begin{figure}[t]
\begin{center}
\includegraphics[width=0.49\textwidth, height=5.7cm]{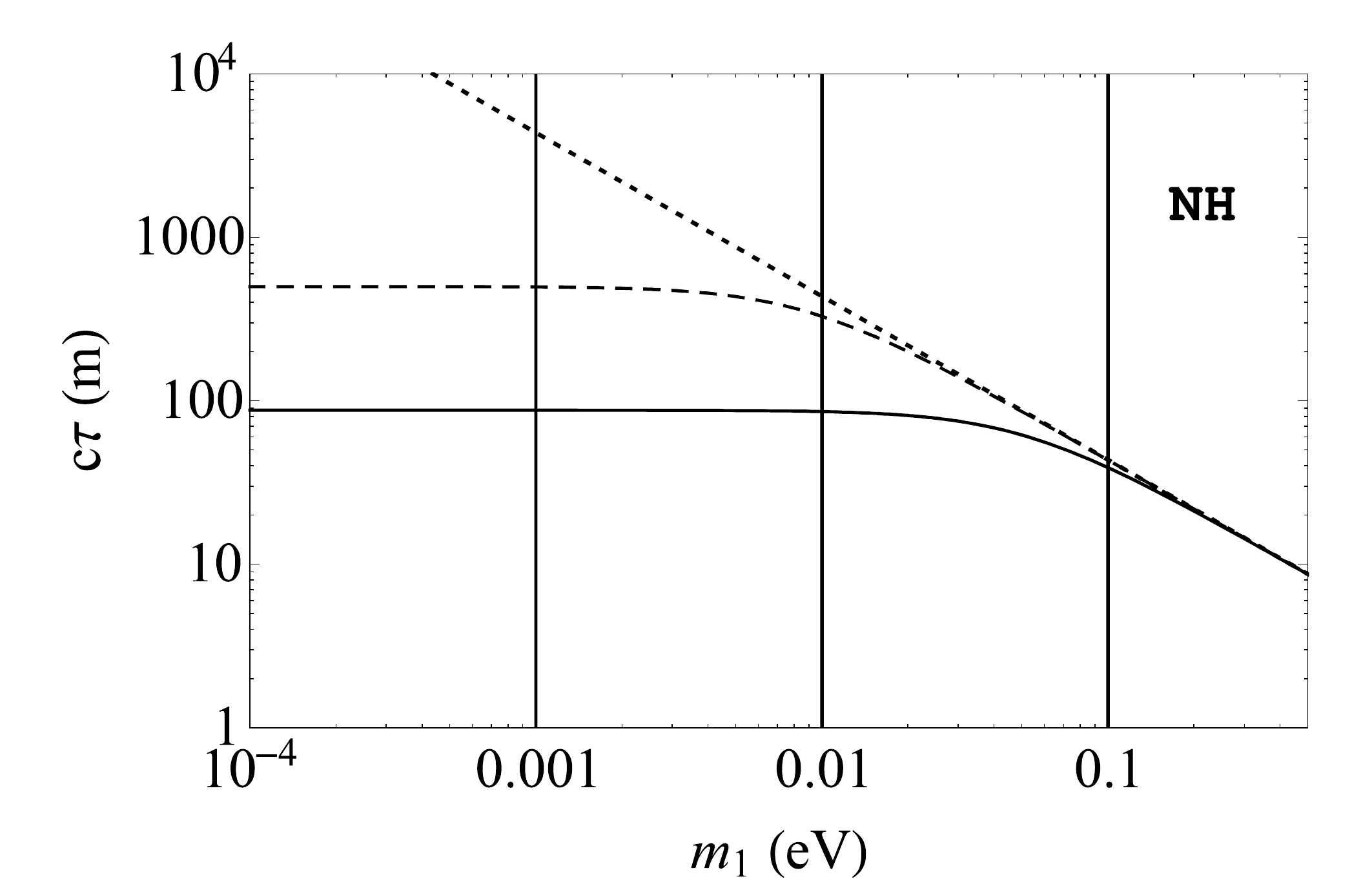}
\includegraphics[width=0.49\textwidth, height=5.7cm]{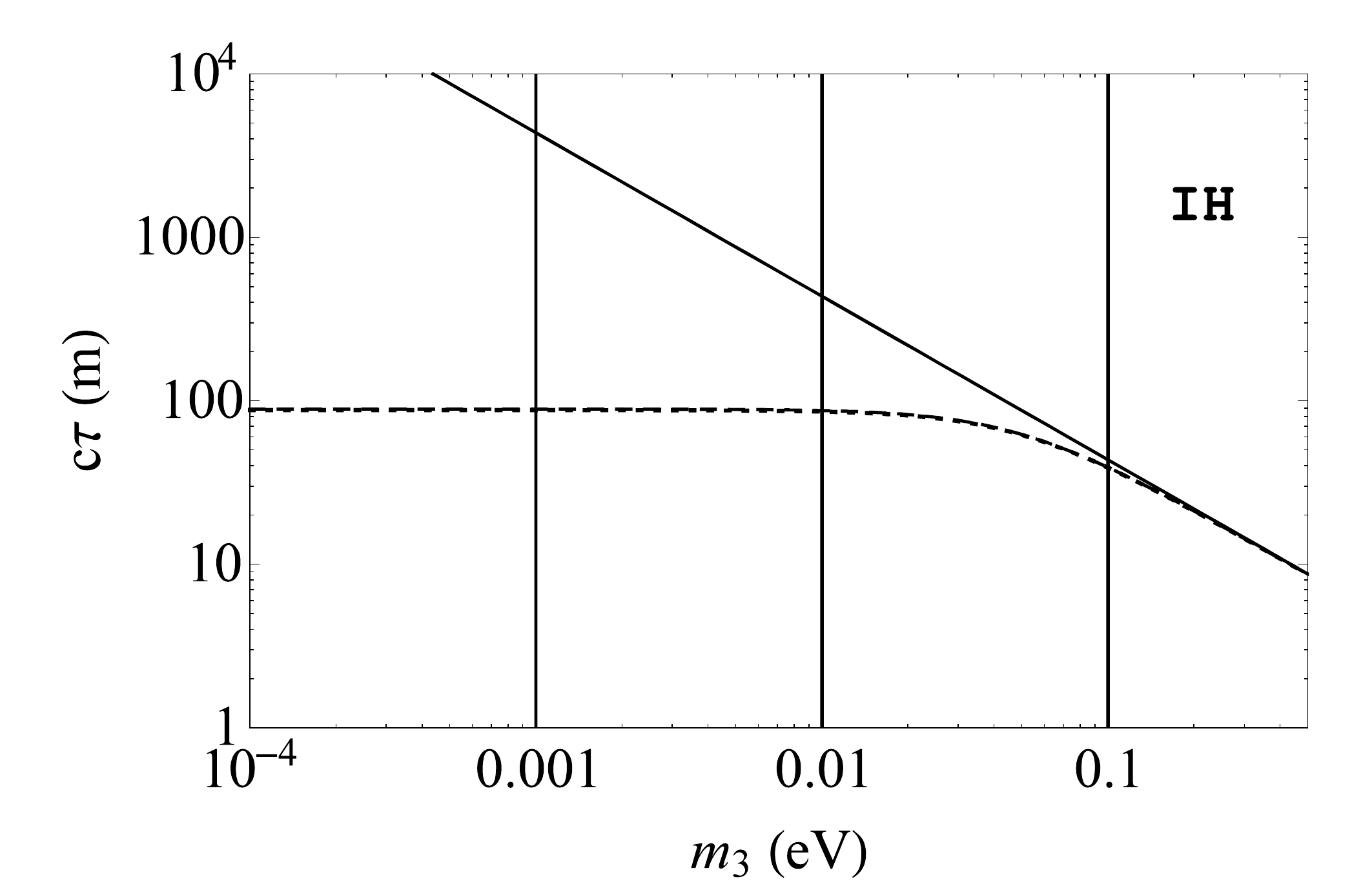}
 \end{center}
\caption{
The left (right) panel shows the decay lengths of heavy neutrinos  
  as a function of the lightest light neutrino mass $m_1$ ($m_3$), 
  for the NH (IH) case. 
In both panels, the dotted, dashed, and the solid lines correspond to the decay lengths 
   of $N^1, N^2$, and $N^3$, respectively, with $m_N= 20$  GeV. 
In the right panel, the dotted and dashed lines are indistinguishable. 
}
\label{fig: decaylength}
\end{figure}

Using Eqs.~(\ref{eq: dwlepton1}), (\ref{eq: dwlepton2}), and (\ref{eq: dwquark}), we evaluate the total decay width 
  of each heavy neutrino ($\Gamma_{N^{i}}$). 
The decay length (in meters) are found to be 
\bea
c\tau_1&=& \frac{1}{\Gamma_{N^1}} \simeq  6.98\times10^{5} \frac{1}{\;m_{N}^4 \;m_1} , \nonumber \\ 
c\tau_2&=& \frac{1}{\Gamma_{N^2}} \simeq  7.25\times10^{5} \frac{1}{\;m_{N}^4 \;m_2} , \nonumber \\ 
c\tau_3&=& \frac{1}{\Gamma_{N^3}} \simeq  7.11\times10^{5} \frac{1}{\;m_{N}^4 \;m_3} ,
\label{eq: dl123} 
\eea
where $m_N$ is in units of GeV while $m_i$ is in units of eV. 
The decay length $c \tau_i$ is inversely proportional to $m_i$ because of 
  $\Gamma_{N^i} \propto  \sum_{\alpha=1}^3|R_{\alpha i}|^2= \frac{ m_i}{m_N}$. 
For the NH and IH cases, 
  the decay lengths for the heavy neutrinos $N^{1,2,3}$ are plotted in Fig.~\ref{fig: decaylength} 
  as a function of the lightest light neutrino mass.

\begin{figure}[h!]
\begin{center}
\includegraphics[width=0.49\textwidth, height=5.7cm]{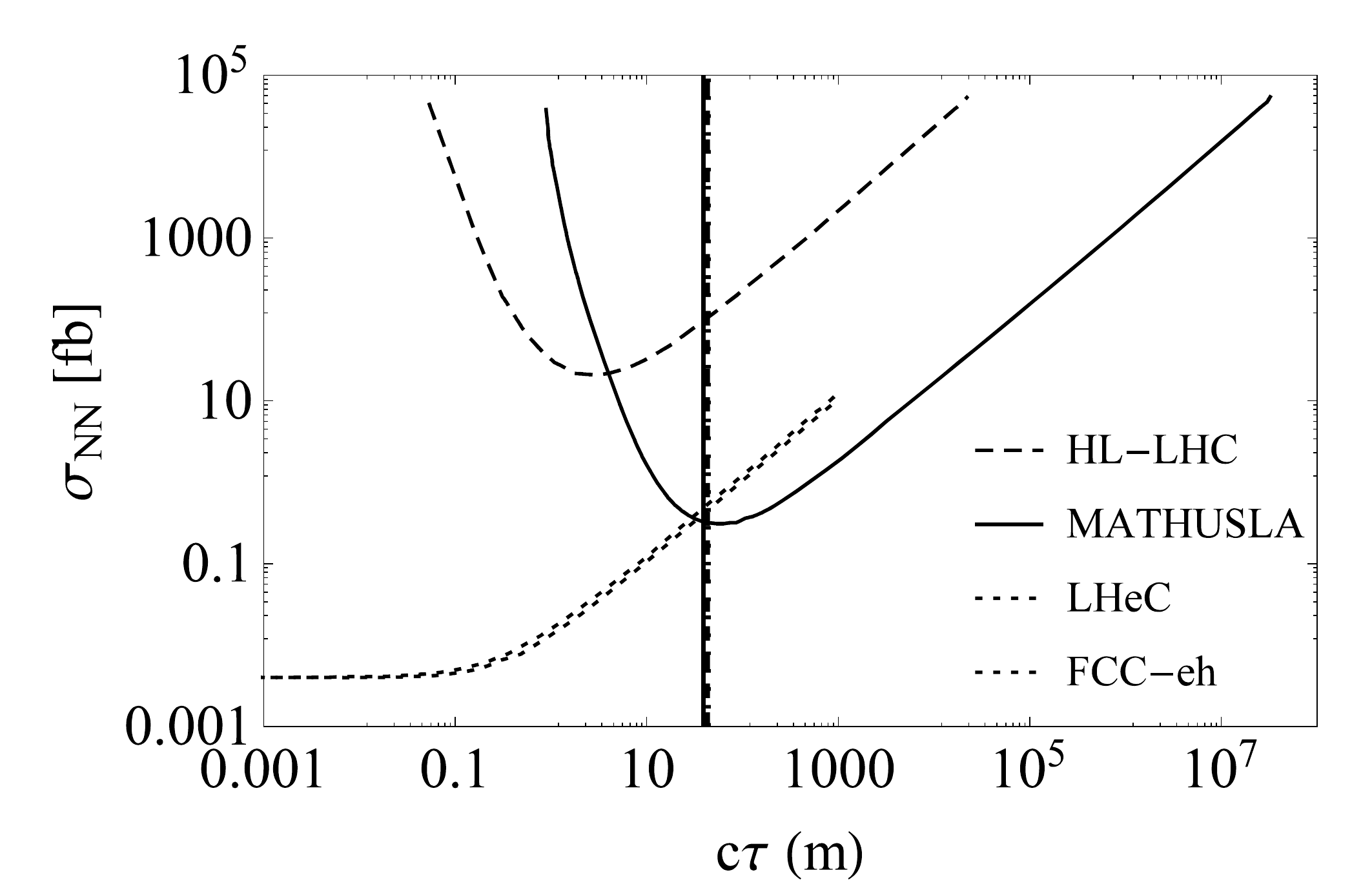}\;
\includegraphics[width=0.49\textwidth, height=5.7cm]{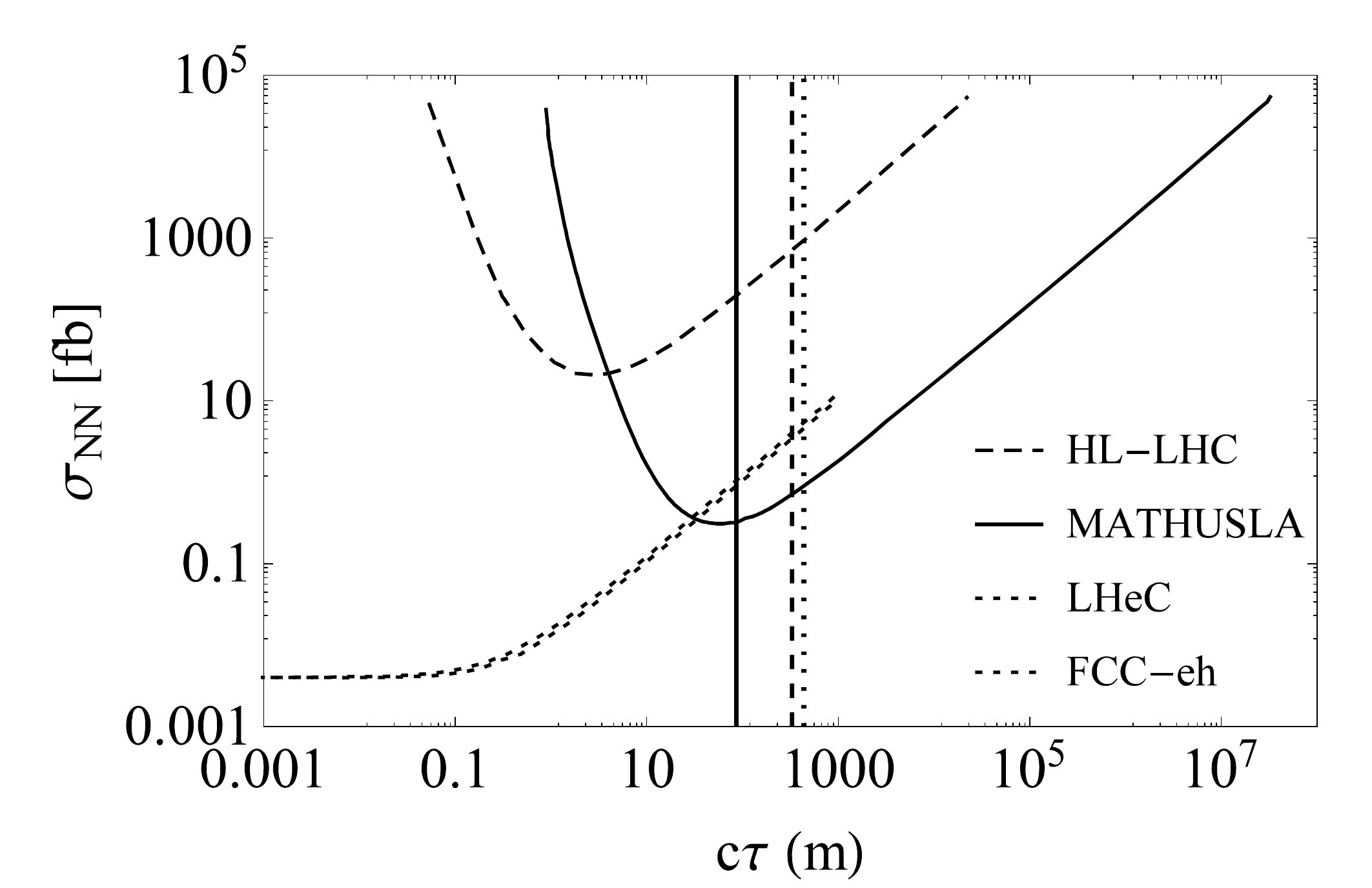}\\
\includegraphics[width=0.49\textwidth, height=5.7cm]{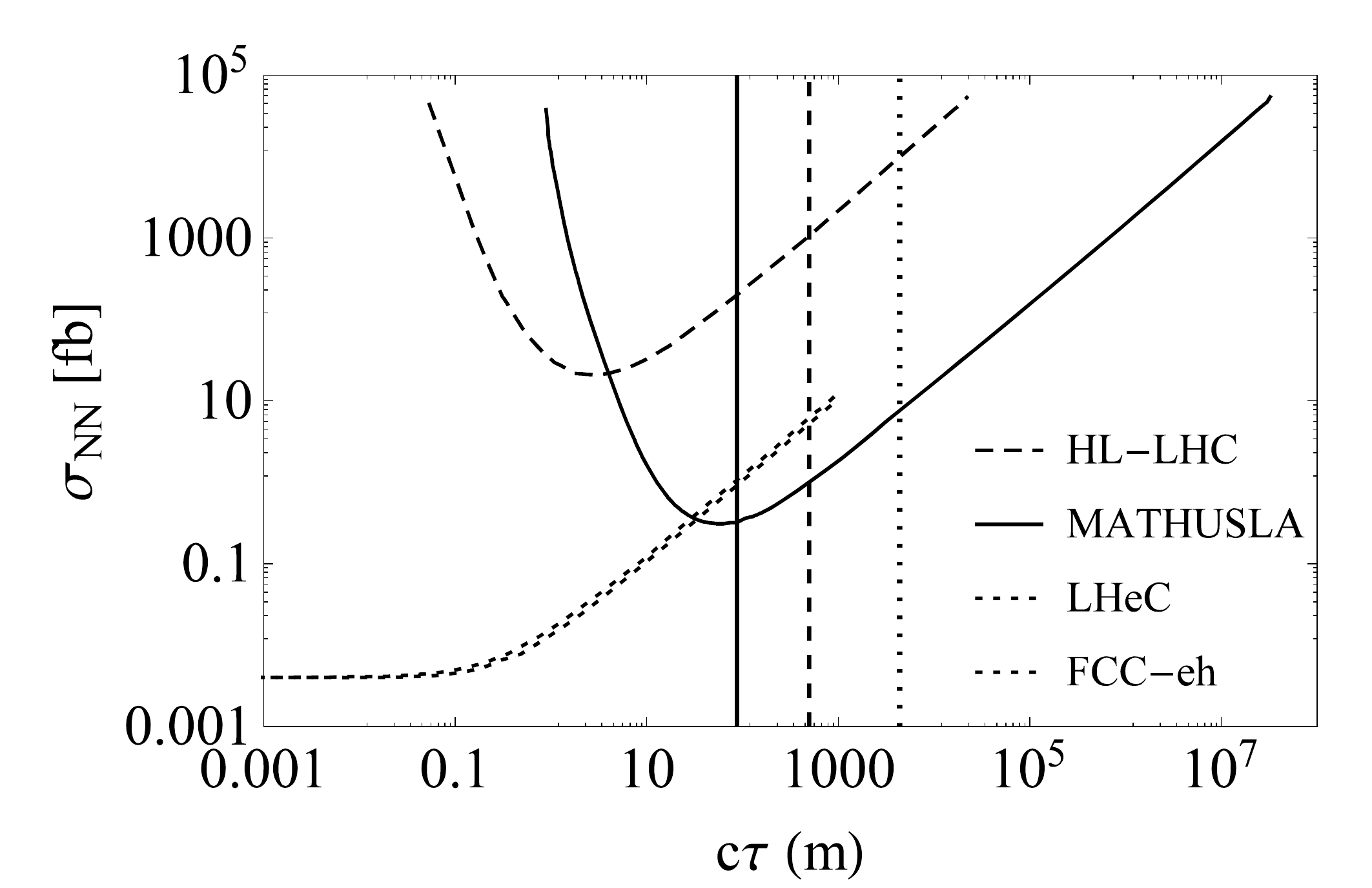}
 \end{center}
\caption{
For the NH case, the search reach cross sections with $m_\phi=126$ GeV and $m_N=20$ GeV 
  along with the heavy neutrino decay lengths for the three benchmark $m_{\rm lightest}$ values (vertical lines).   
The top-left, top-right and bottom panels are for $m_{\rm lightest} = m_1=0.1$, $0.01$ and $0.001$ eV, respectively 
}
\label{fig: dvNH}
\end{figure}

\begin{figure}[h!]
\begin{center}
\includegraphics[width=0.49\textwidth, height=5.7cm]{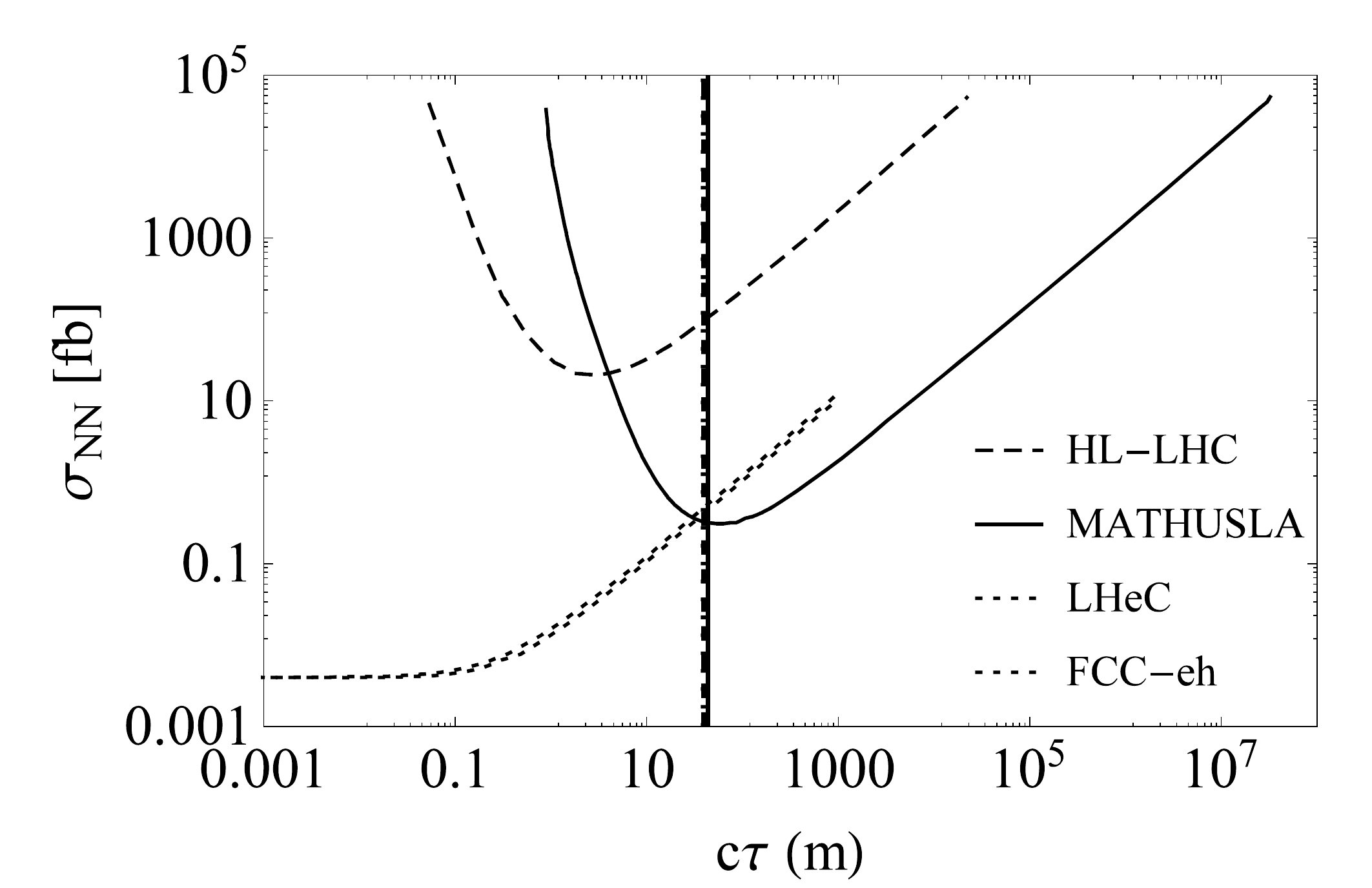}\;
\includegraphics[width=0.49\textwidth, height=5.7cm]{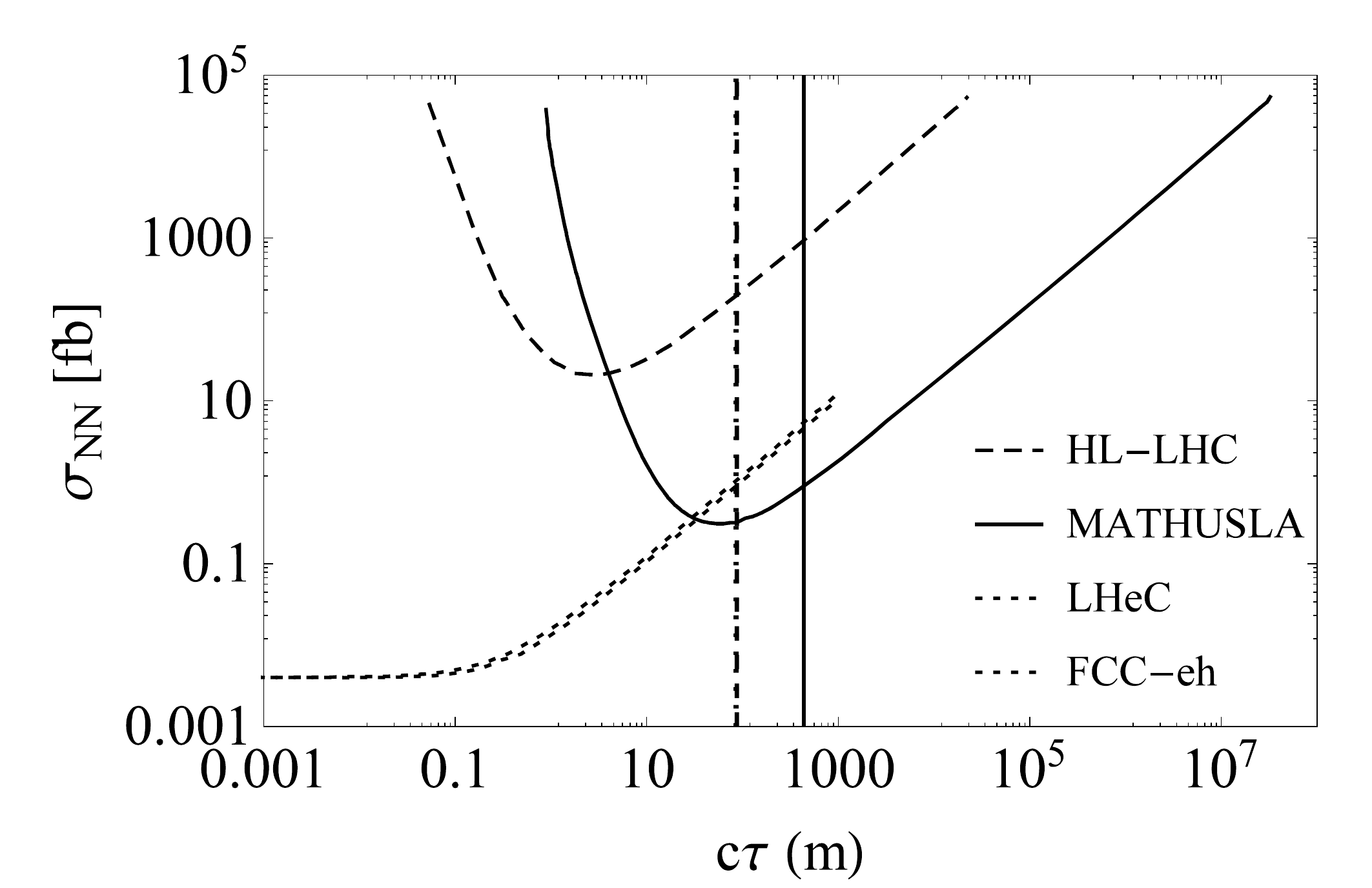}\\ 
\includegraphics[width=0.49\textwidth, height=5.7cm]{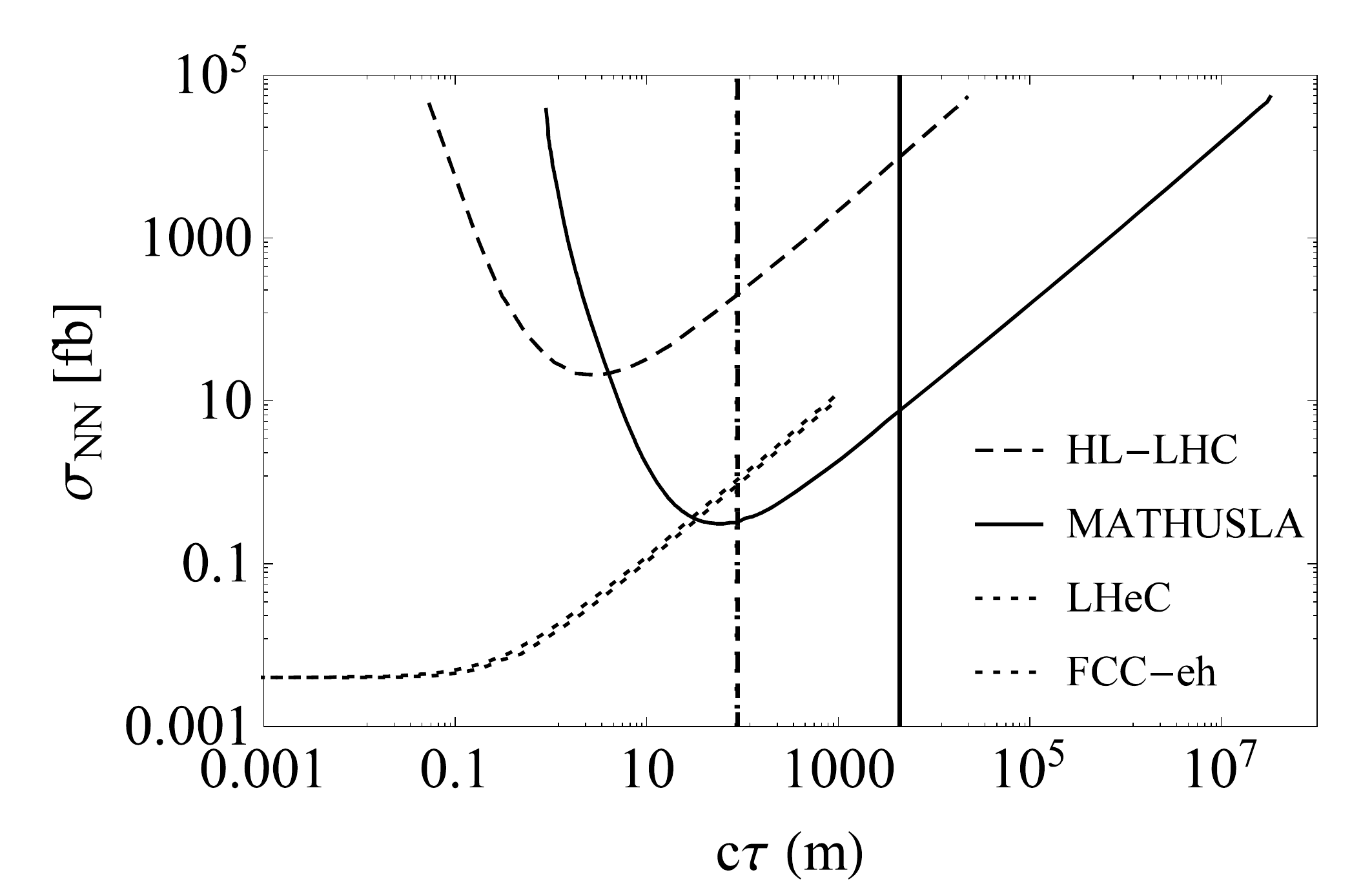}
 \end{center}
\caption{
Same as Fig.~\ref{fig: dvNH} but for the IH case. 
The top-left, top-right and bottom panels are for $m_{\rm lightest} = m_3=0.1$, $0.01$ and $0.001$ eV, respectively 
}
\label{fig: dvIH}
\end{figure}

In Fig.~\ref{fig: decaylength}, let us consider three benchmark values 
  for the lightest light neutrino mass as $m_{\rm lightest} = 0.1$, $0.01,$ and $0.001$ eV. 
For each benchmark $m_{\rm lightest}$ value, the decay lengths for the three heavy neutrinos are fixed. 
It is then interesting to combine the benchmark decay lengths with the search reach of the displaced vertex signatures 
  at the future colliders. 
For the NH case, we show in Fig.~\ref{fig: dvNH} the search reach cross sections with $m_\phi=126$ GeV and $m_N=20$ GeV 
  along with the heavy neutrino decay lengths for the three benchmark $m_{\rm lightest}$ values (vertical lines).   
The top-left, top-right and bottom panels are for $m_{\rm lightest} = m_1=0.1$, $0.01$ and $0.001$ eV, respectively 
Interestingly, for all benchmarks, the value of $c \tau_3$ is very close to the lifetime
   yielding the best reach cross section at the MATHUSLA. 
Same as Fig.~\ref{fig: dvNH} but for the IH case is shown in Fig.~\ref{fig: dvIH}. 
The top-left, top-right and bottom panels are for $m_{\rm lightest} = m_3=0.1$, $0.01$ and $0.001$ eV, respectively 
For all benchmarks, the value of $c \tau_1$ is very close to the lifetime
   yielding the best reach cross section at the MATHUSLA.

The heavy neutrino lifetimes become shorter as $m_{\rm lightest}$ is raised (see Eq.~(\ref{eq: dl123})). 
However, if we consider the cosmological bound on the sum of the light neutrino masses, 
  $\sum m_{\rm lightest} = 0.23$ MeV \cite{Ade:2015xua}, 
  we obtain the lower bound on the decay length to be $c\tau \gtrsim 20$ m. 
In fact, this lower bound can be significantly reduced if we consider the complex orthogonal matrix $O$
  in the general parametrization. 
See, for example, Ref.~\cite{DO}.

\begin{figure}[h!]
\begin{center}
\includegraphics[width=0.49\textwidth, height=5.7cm]{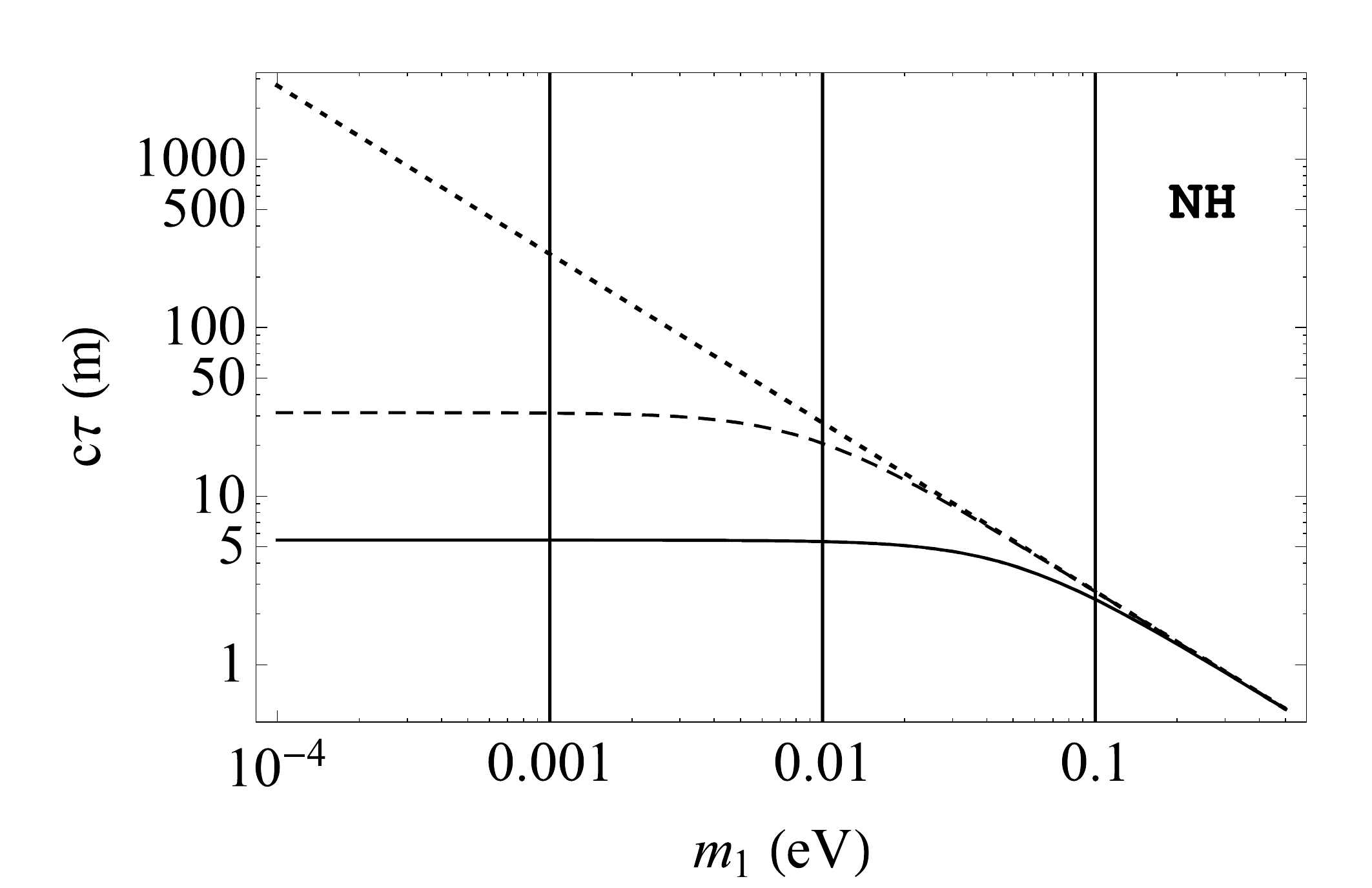}
\includegraphics[width=0.49\textwidth, height=5.7cm]{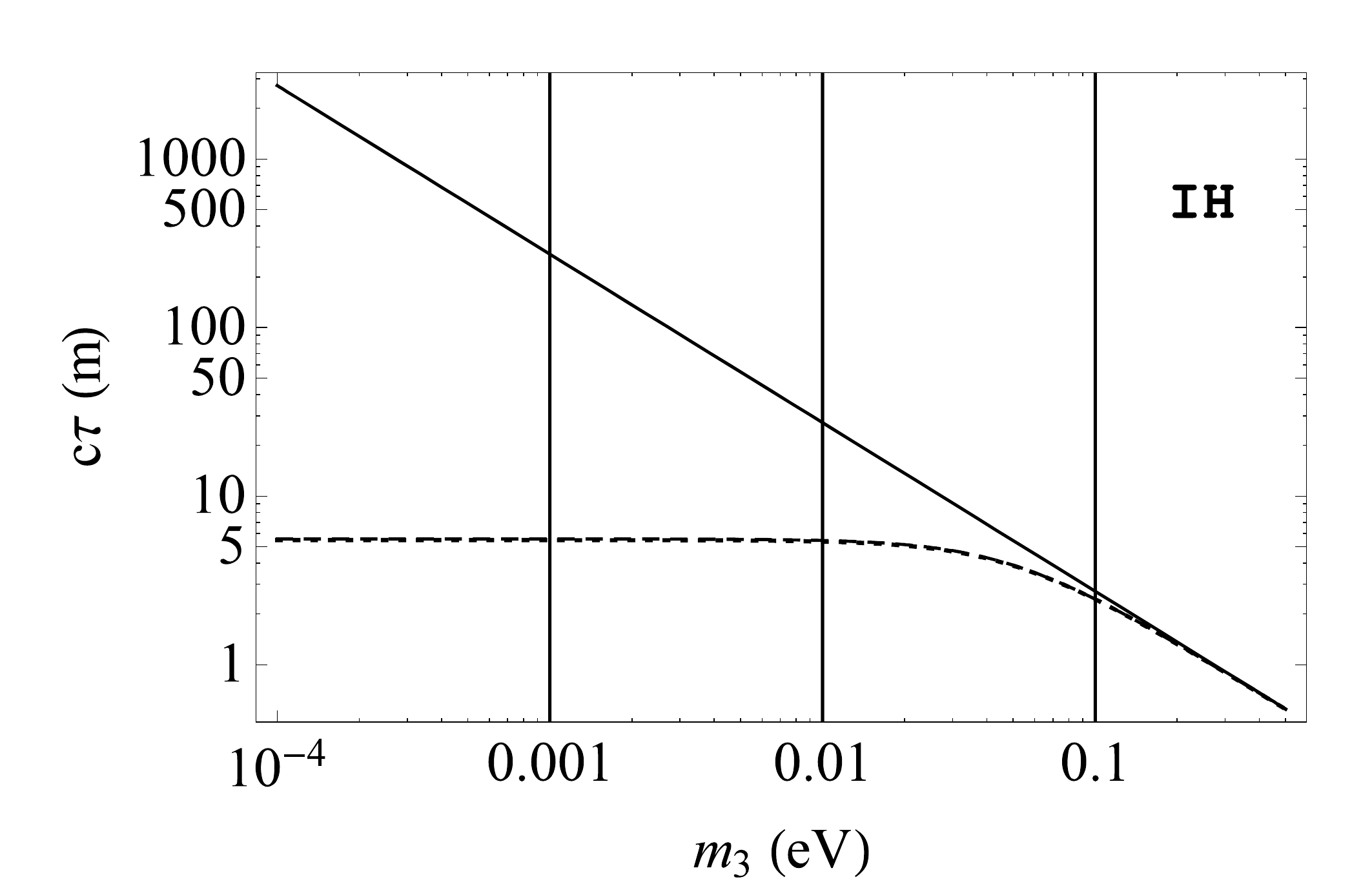}
 \end{center}
\caption{
Same as Fig.~\ref{fig: decaylength} but for the case with $m_N=40$ GeV. 
}
\label{fig: decaylength40}
\end{figure}

In the previous section, we have shown the relation between $Y$ and $\sin \theta$ 
   to achieve the best reach cross section at the HL-LHC/MATHUSLA, 
   assuming the heavy neutrino lifetime to be the best point for each experiment.  
To conclude this section, we repeat the same analysis in the previous section 
   but for various values of $c\tau$ determined by $m_{\rm lightest}$ values. 
For this analysis, we set $m_N=40$ GeV and $m_\phi=150$ GeV. 
The decay lengths of heavy neutrinos for this parameter choice are depicted in Fig.~\ref{fig: decaylength40}. 
In this case, the cosmological lower bound on $c \tau$ is found to be $c\tau \simeq 1.3$ m.

\begin{figure}[h!]
\begin{center}
\includegraphics[width=0.49\textwidth, height=5.7cm]{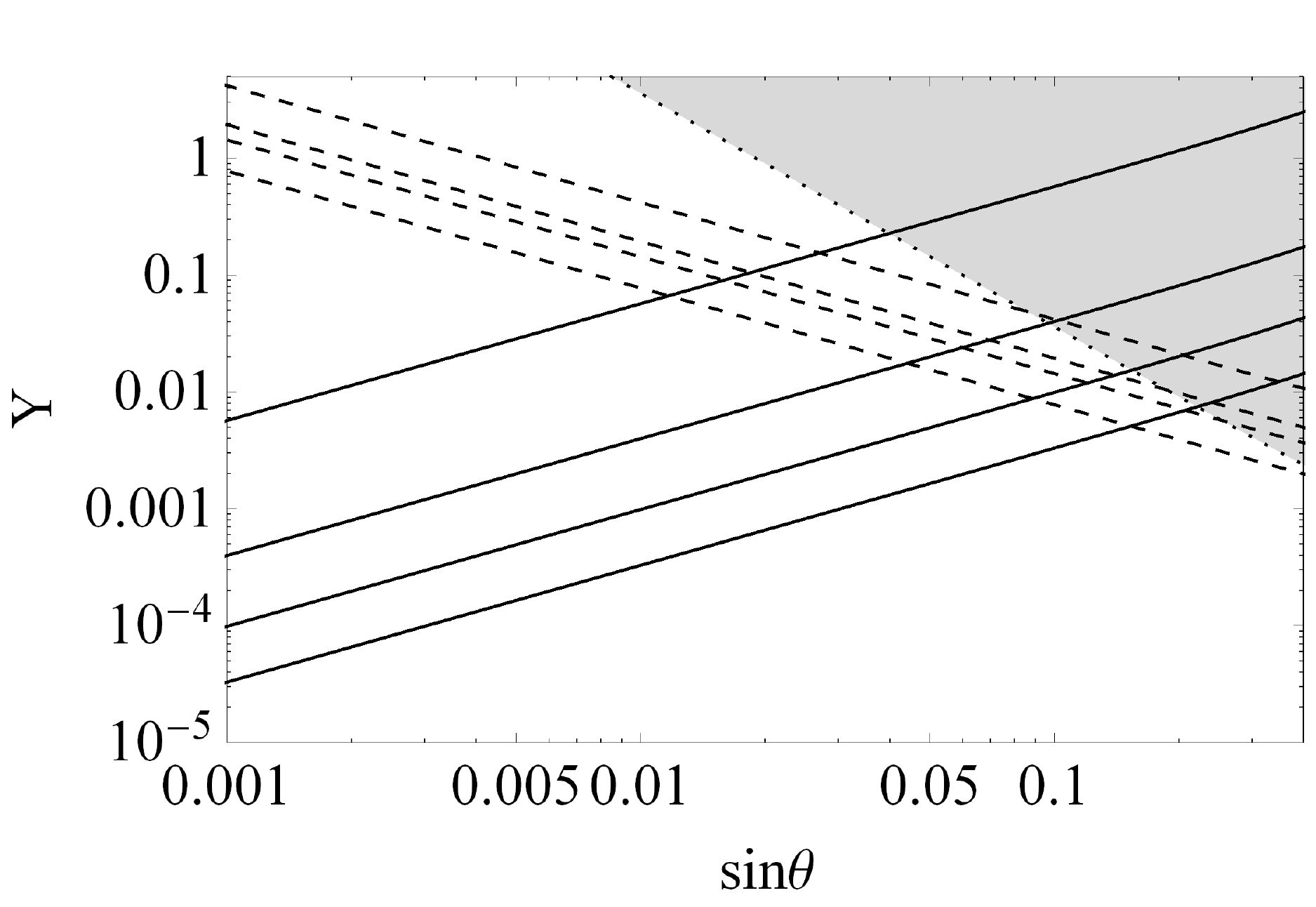}\;
\includegraphics[width=0.49\textwidth, height=5.7cm]{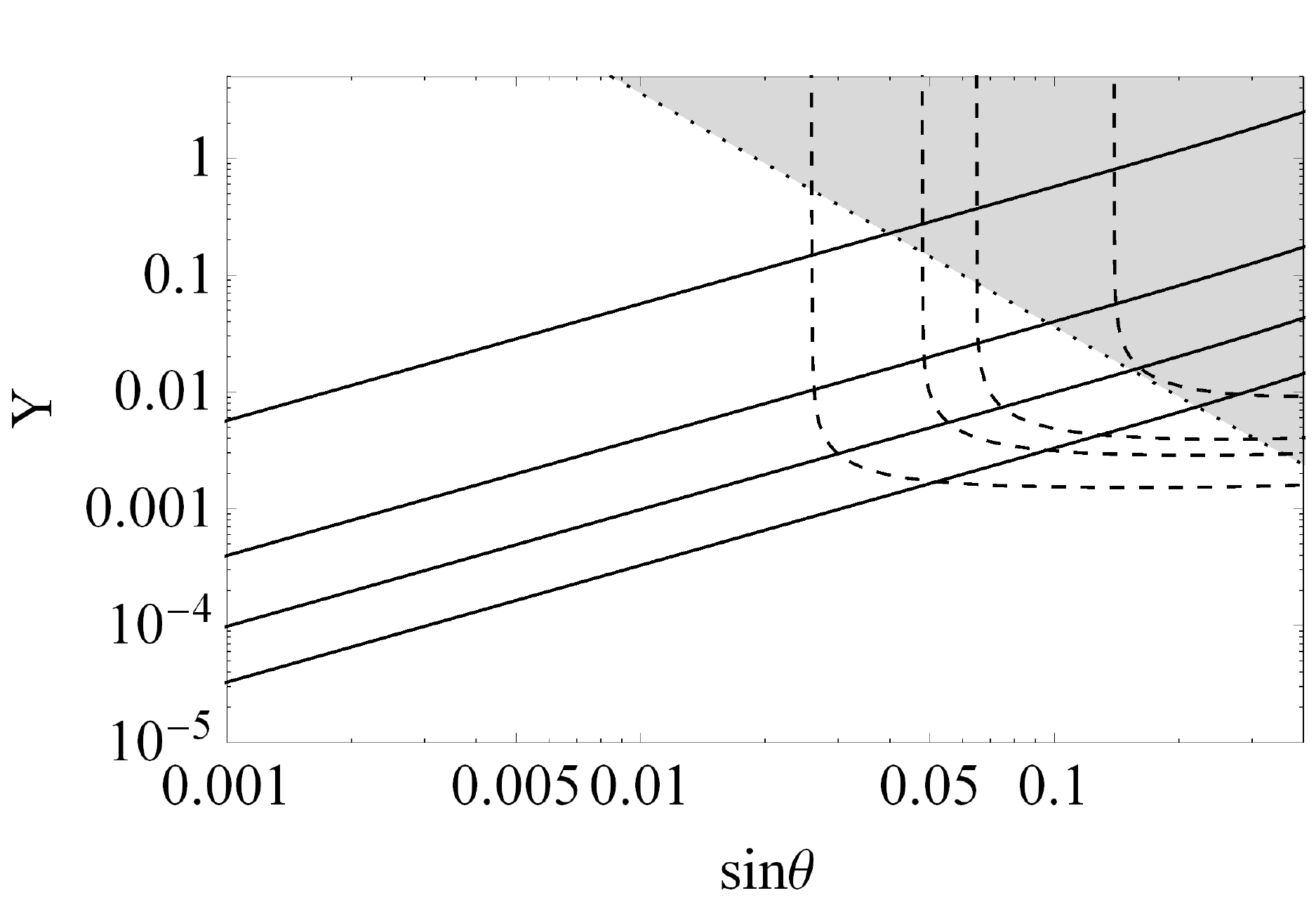}
 \end{center}
\caption{For fixed $m_\phi = 150$ GeV and $m_N = 40$ GeV and for different $m_{\rm lightest}$ values, the plots show the parameter space for displaced vertex search at  HL-LHC when the RHNs are produced from: SM Higgs decay (left) and $B-L$ decay (right). For both the panels, from left to right, the dashed lines correspond to $m_{\rm lightest} \simeq 8.90\times10^{-2}, 10^{-2}, 5.00\times 10^{-3}$, and $10^{-3}$ eV, or equivalently 
$c\tau = 273, 54.0, 27.2$, and $3.10$ m, respectively. 
In both panels, from top to bottom, the solid lines are correspond to fixed $BR(\phi \to NN) = 99.99\%$, $98\%$, $75\%$, and $25\%$, respectively.  
Similarly, the gray shaded region are excluded region by the SM Higgs boson invisible decay searches.}
\label{fig1: yVsin_mphi150}
\end{figure}

In Fig.~\ref{fig1: yVsin_mphi150}, we show our results corresponding to the top-left and bottom-left panels in Fig.~\ref{fig: mphi200}.
The left panel corresponds to the the top-left panel of Fig.~\ref{fig: mphi200}.
Here, we consider the heavy neutrino production from the SM-like Higgs decay and the search reach at the HL-LHC. 
The diagonal dashed lines from left to right are results correspond to 
  $m_{\rm lightest} = 8.90\times10^{-2}$, $10^{-2}$, $5.00\times 10^{-3}$ and $10^{-3}$ eV, 
  or equivalently $c\tau = 3.10$, $27.2$, $54.0$ and $273$ m. 
The solid diagonal lines denote the relations between $Y$ and $\sin \theta$ 
  to yield $BR(\phi \to NN) = 99.99\%$, $98\%$, $75\%$, and $25\%$, respectively, from top to bottom.   
The gray shaded region is excluded by the LHC constraint on the invisible Higgs decay. 
The right panel corresponds to the the bottom-left panel of Fig.~\ref{fig: mphi200}. 
Here, we consider the heavy neutrino production from the $B-L$ Higgs decay. 
The line coding are the same as the left panel. 
The parameter region for $m_{\rm lightest} \lesssim 10^{-3}$ eV is already excluded.

\begin{figure}[h!]
\begin{center}
\includegraphics[width=0.49\textwidth, height=5.7cm]{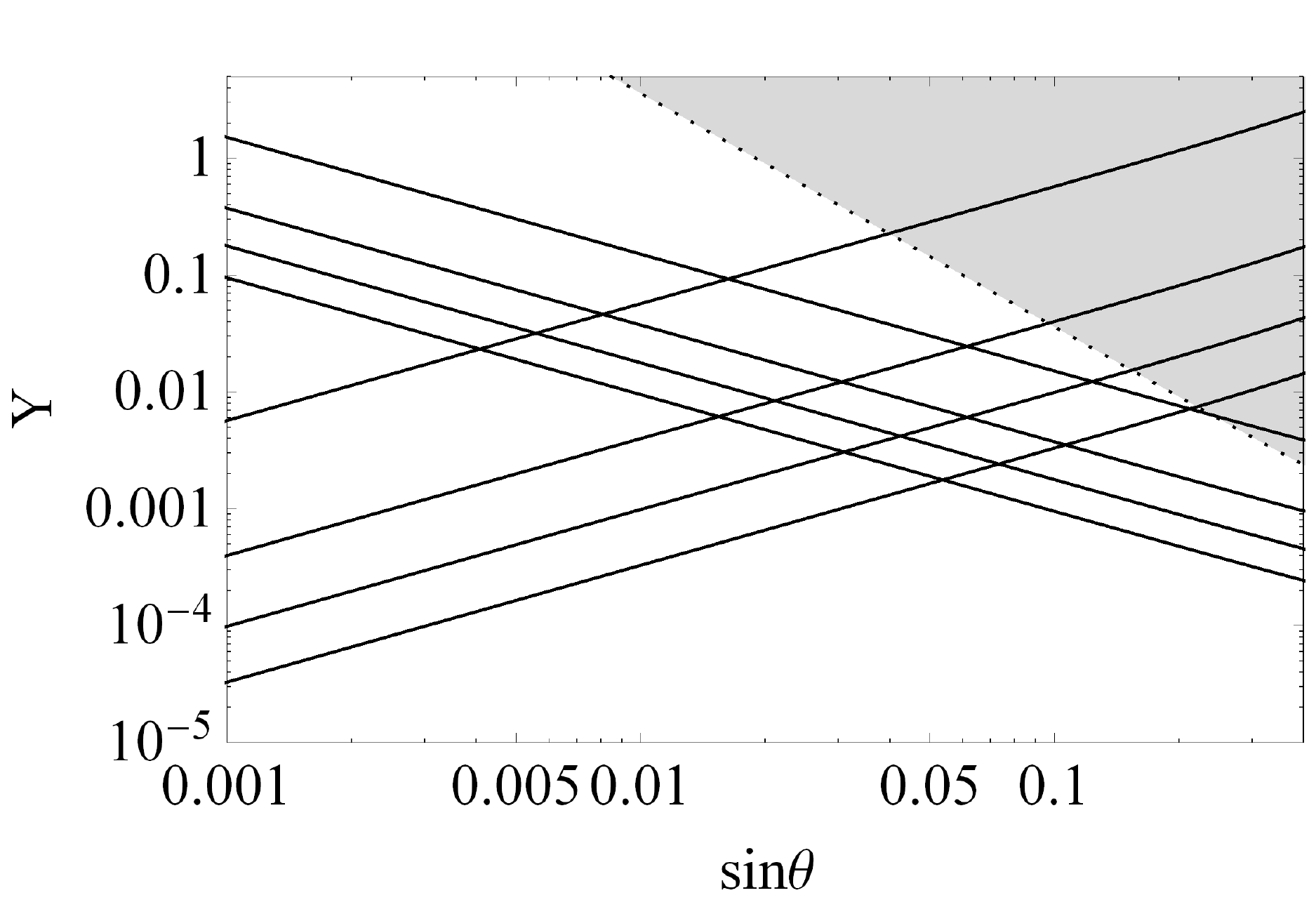}\;
\includegraphics[width=0.49\textwidth, height=5.7cm]{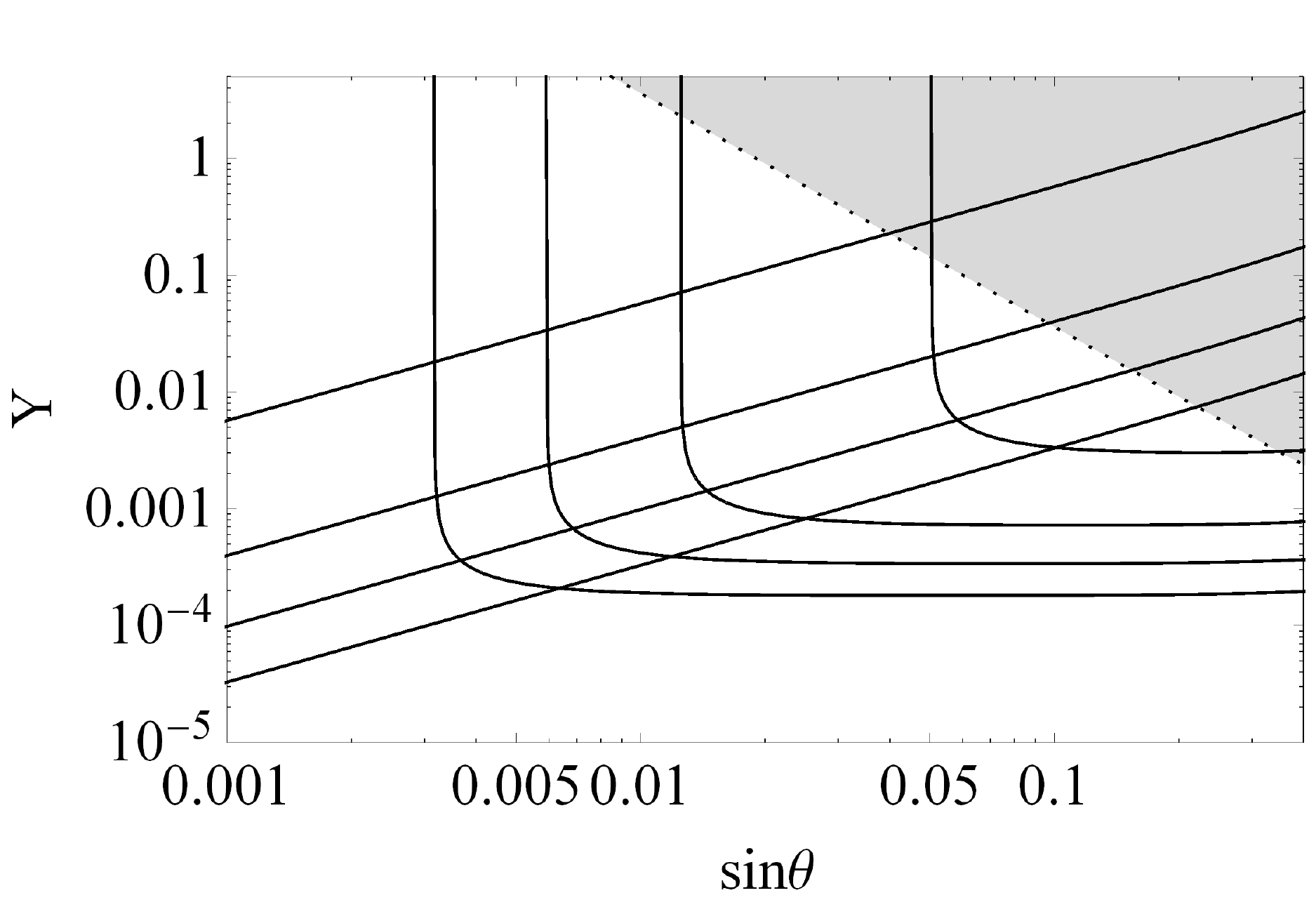}
 \end{center}
\caption{For fixed $m_\phi = 150$ GeV and $m_N = 40$ GeV and for different $m_{\rm lightest}$ values, the plots show the parameter space for displaced vertex search at MATHUSLA when the RHNs are produced from: SM Higgs decay (left) and $B-L$ decay (right). For both the panels, from left to right, the dashed lines correspond to $m_{\rm lightest} \simeq 10^{-1}, 4.61\times10^{-3}, 5.00\times 10^{-4}$, and $10^{-4}$ eV, or equivalently 
$c\tau = 2.73\times 10^{3}, 59.1, 545$, and $2.73$ m, respectively. 
The line coding for the remaining curves and regions are same as Fig.~\ref{fig1: yVsin_mphi150}.}
\label{fig2: yVsin_mphi150}
\end{figure}

Same as Fig.~\ref{fig1: yVsin_mphi150} but for the search reach at the MATHUSLA  
  is shown in Fig.~\ref{fig2: yVsin_mphi150}. 
Solid diagonal lines with negative slope in the left panel and the solid curves in the right panel 
  correspond to $m_{\rm lightest} \simeq 10^{-1}$, $4.61\times10^{-3}$, $5.00\times 10^{-4}$ and $10^{-4}$ eV, 
   or equivalently  $c\tau = 2.73$,  $59.1$, $545$ and $2.73\times 10^{3}$ m, 
   from left to right. 
The line coding for the other lines and the shaded region are the same as Fig.~\ref{fig1: yVsin_mphi150}. 

\section{Complementarity to neutrinoless double beta decay search}
\label{sec:6}

\begin{figure}[t]
\begin{center}
\includegraphics[width=0.45\textwidth, height=5.6cm]{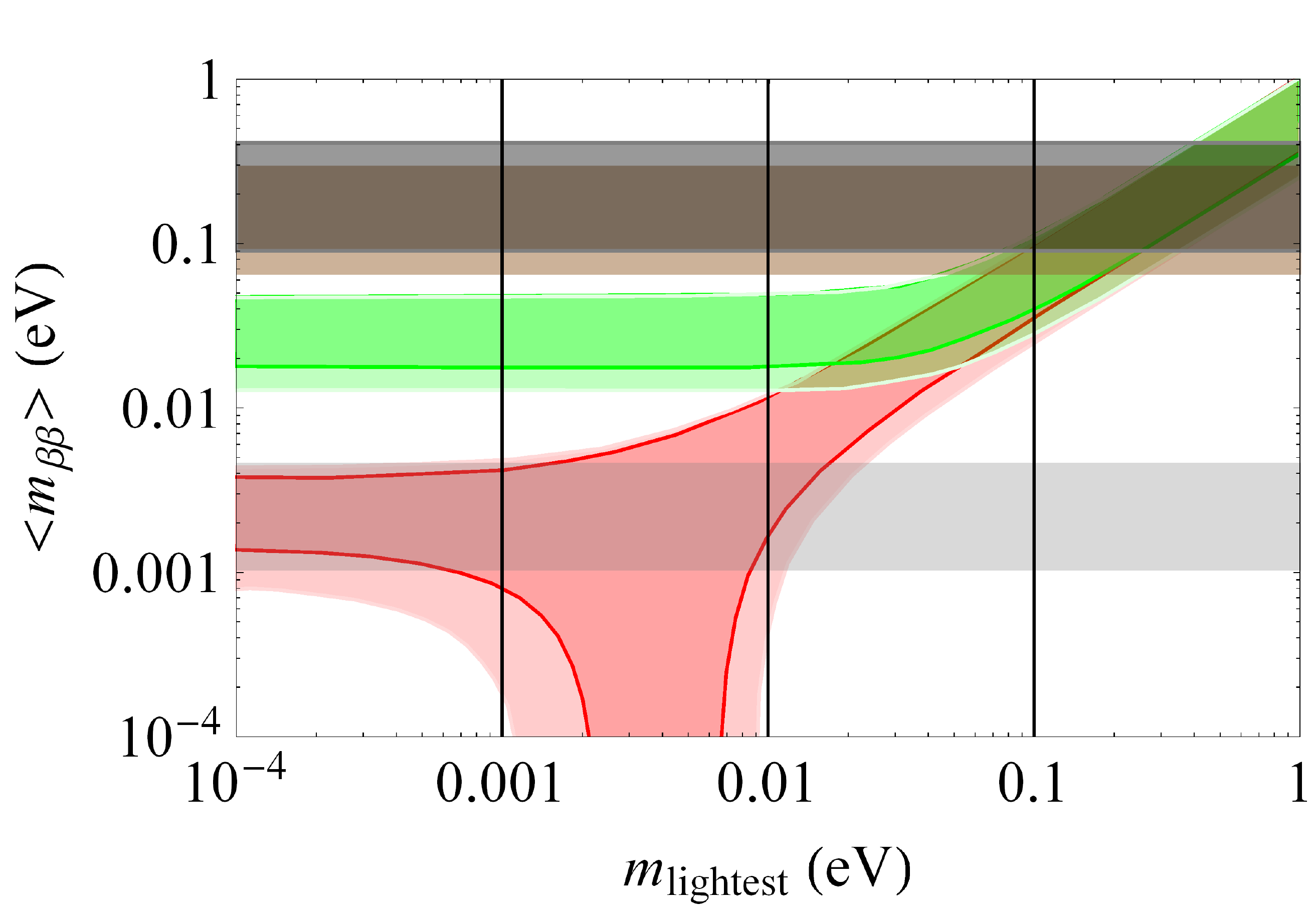}\;\;
\includegraphics[width=0.49\textwidth, height=5.6cm]{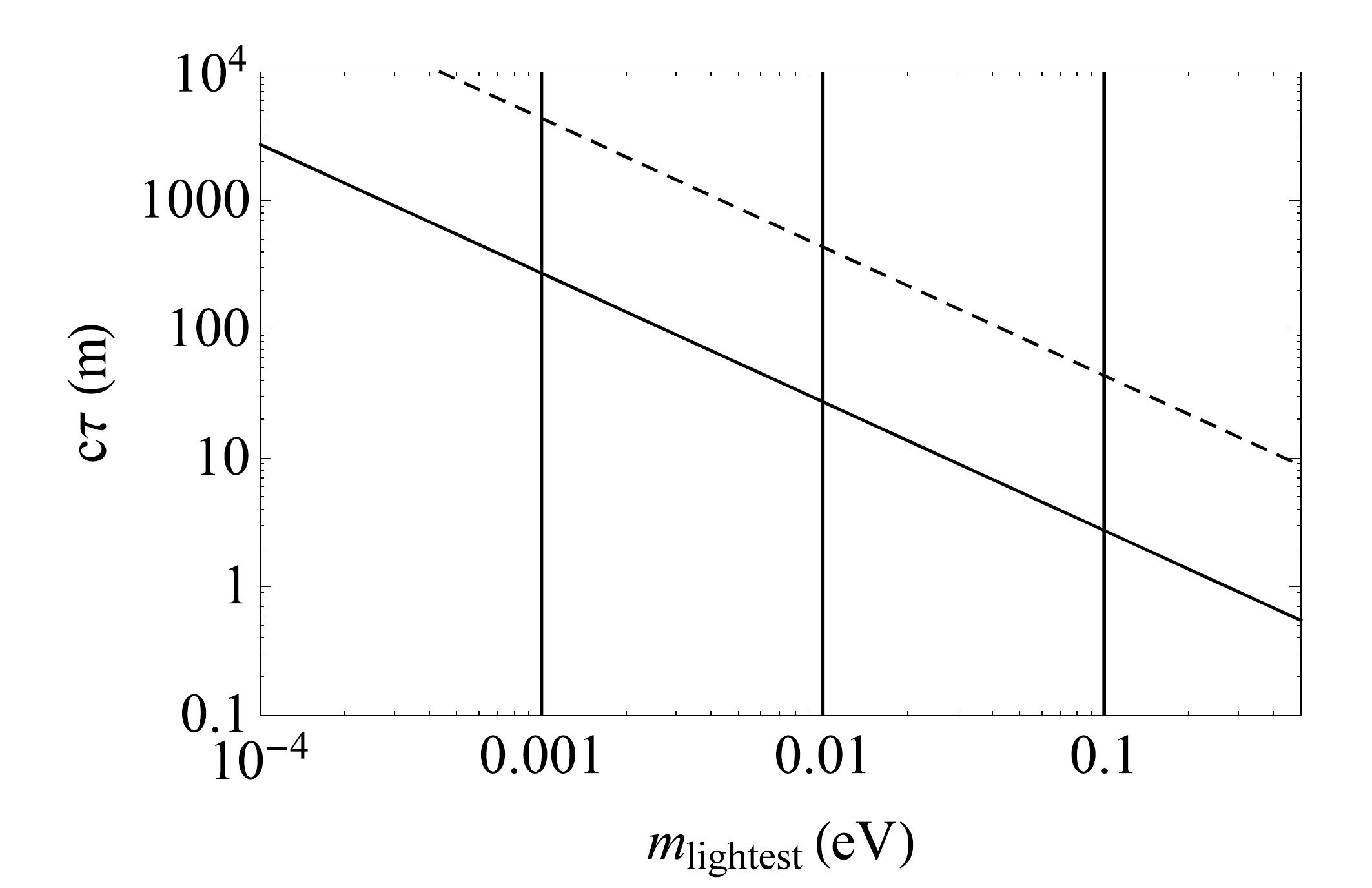}
 \end{center}
\caption{In the left panel, the red and the green shaded region correspond to the constraint on effective netrino mass ($\langle m_{\beta\beta}\rangle$) for the NH and IH, respectively. 
The horizontal shaded regions from the top to bottom, correspond to the current EXO-200 experiment and the future reach of EXO-200 phase-II, and nEXO experiments, respectively \cite{DellOro:2016tmg}.
In the right panel, the solid line depicts the total decay length of RHN plotted against the mass of the corresponding lightest light-neutrino mass. The dashed (solid) line correspond to fixed RHN mass of $20$ ($40$) GeV.  In both the panels, vertical solid lines correspond to the three benchmark points for the lightest neutrino masses for the NH and the IH, namely, $m_{\rm lightest} = 0.1, 0.01,$ and $0.001$ eV.}
\label{betadecay}
\end{figure}

The neutrinoless double beta decay of heavy nuclei is a ``smoking-gun" signature of the Majorana nature of neutrinos. 
In this process, two neutrons in nuclei simultaneously decay into two protons plus 
  two electrons without emitting neutrinos, and hence the lepton number is violated by two units. 
The neutrinoless double beta decay process is characterized by an effective mass 
  of the light neutrino ($\langle m_{\beta\beta}\rangle$) defined as 
\bea
\langle m_{\beta\beta}\rangle = \left| \sum_j m_j U_{ej}^2\right|,
\label{eq: dbeta_decay}
\eea
where $U_{ej}$ is the $(e,j)$-element of the neutrino mixing matrix $U_{\rm MNS}$, see \cite{DellOro:2016tmg} for review of  neutrinoless double beta decay. 
Employing the current neutrino oscillation data, 
   the effective mass is described by the lightest light neutrino mass.  
The left panel in Fig.~\ref{betadecay} depicts the relation between $\langle m_{\beta\beta}\rangle$ and  $m_{\rm lightest}$ 
   for the NH (red shaded region) and IH (green shaded region) cases. 
In this panel, the current upper limit by the EXO-200 experiment (upper horizontal shaded region) 
   and the future reach by the EXO-200 phase-II and nEXO experiments (lower horizontal shaded region) 
   are also shown. 

As we have investigated in the previous section, 
   the decay lengths of the heavy neutrinos are controlled by $m_{\rm lightest}$. 
In the right panel of Fig.~\ref{betadecay}, we show the decay length of the heavy neutrino $N^3$ 
  for the NH case as a function of $m_{\rm lightest}$. 
The dashed and solid lines correspond to the heavy neutrino masses of $m_N=20$ and $40$ GeV, respectively. 
In Fig.~\ref{betadecay}, our benchmarks of $m_{\rm lightest}=0.1$, $0.01$ and $0.001$ are depicted 
  by  the vertical lines. 
It is interesting to compare the two panels. 
If the displaced vertex from a heavy neutrino decay is observed 
   and the heavy neutrino mass is reconstructed from its decay products, 
   $m_{\rm lightest}$ is determined. 
If the neutrinoless double beta decay is observed, $\langle m_{\beta\beta}\rangle$ is measured. 
However, if $\langle m_{\beta\beta}\rangle$ is measured to be around $0.03$ eV or $0.002$ eV, 
   $m_{\rm lightest}$ is left undetermined. 
Hence, the observations of the displaced vertex and the neutrinoless double beta decay 
   are complementary with each other.

\section{Conclusions}
\label{sec:7}

It is quite possible that new particles in new physics beyond the SM are completely singlet under the SM gauge group. 
This is, at least, consistent with the null results on the search for new physics at the LHC. 
If this is the case, we may expect that such particles very weakly couple with the SM particles 
  and thus have a long lifetime.
Such particles, once produced at the high energy colliders, provide us with the displaced vertex signature, 
   which is very clean with negligible SM background. 
In the context of the minimal gauged $B-L$ extended SM, 
  we have considered the prospect of searching for the heavy neutrinos of the type-I seesaw mechanism 
  at the future high energy colliders. 
For the production process of the heavy neutrinos, we have investigated  
  the production of Higgs bosons and their subsequent decays into a pari of heavy neutrinos. 
With the parameters reproducing the neutrino oscillation data, 
  we have shown that the heavy neutrinos are long-lived and 
  their displaced vertex signatures can be observed at the next generation displaced vertex search experiments, 
  such as the HL-LHC, the MATHUSLA, the LHeC, and the FCC-eh.
We have found that the lifetime of the heavy neutrinos is controlled by the lightest light neutrino mass, 
  which leads to a correlation between the displaced vertex search and  
  the search limit of the future neutrinoless double beta-decay experiments.
 \\

{\bf Note added:} 
While completing this manuscript, we noticed a paper by F. Deppisch, W. Liu and M. Mitra \cite{Deppisch:2018eth}
   which also considers the displaced vertex signature of the heavy neutrinos.

\section*{Acknowledgement}
\vspace{-0.2cm}
S.J. thanks the Fermilab Theoretical Physics Department for their warm hospitality during the completion of this work. 
The work of N.O. and D.R. is partially supported in part by the U.S. Department of Energy (DE-SC0012447). 
The work of S.J. is supported in part by the US Department of Energy Grant (dDE-SC0016013) 
    and the Fermilab Distinguished Scholars Program.


\section{Appendix}

The partial decay widths of various decay modes of the SM-like Higgs boson 
  of mass $m_h$ is given by \cite{Gunion} : \\

\noindent(i) SM fermions (f):
\bea
\Gamma_{h\to f{\bar f}} &=& \cos^2{\theta} \times \frac{3 N_f m_h m_f^2}{8\pi v_{SM}^2} \left( 1- \frac{4m_f^2}{m_h^2} \right)^{3/2}, \nonumber
\eea
where $f$ are the SM fermions and $N_f = 1$ and $3$ for SM leptons and quarks, respectively. \\

\noindent(ii) on-shell gauge bosons (V = W or Z):
\bea
\Gamma_{h \to V V} &=&  \cos^2{\theta} \times \frac{C_V}{32\pi} \frac{m_h^3}{ v_{SM}^2} \left(1-4 \frac{m_V^2}{m_h^2}\right)^{1/2}\left(1-4 \frac{m_V^2}{m_h^2}+12 \left(\frac{m_V^2}{m_h^2}\right)^2\right),
\nonumber 
\eea
where $C_V = 1$ and $2$ for $V = Z$ or $W$ gauge boson, respectively. \\

\noindent(iii) gluon (g) via top-quark loop:
\bea
\Gamma_{h \to gg} &=&  \cos^2{\theta} \times \frac{\alpha_s^2 m_h^3 }{128\pi^3 v^2} (F_{1/2}(m_h))^2,
\nonumber 
\eea
where 
\bea
F_{1/2}(m_h) &=& -2 \frac{4m_t^2}{m_h^2} \left[ 1- \left( 1 - \frac{4m_t^2}{m_h^2} 
{\rm arcsin}^2\left( \frac{m_h}{2m_t} \right) \right) \right].
\nonumber 
\eea

\noindent(iii) one off-shell gauge boson :
\bea
\Gamma_{h \to WW^*} &=&  \cos^2{\theta} \times \frac{3m_W^4 m_h }{32\pi^3  v_{SM}^4} G\left( \frac{m_W^2}{m_h^2} \right),\nonumber \\
\Gamma_{h \to ZZ^*} &=&  \cos^2{\theta} \times \frac{3m_Z^4 m_h }{32\pi^3  v_{SM}^4} 
\left( \frac{7}{12} -\frac{10}{9} \sin^2 \theta_W + \frac{40}{9} \sin^4 \theta_W \right)
G\left( \frac{m_W^2}{m_h^2} \right),
\eea
where $\sin^2\theta_W = 0.231$ and the loop functions are given by  
\bea
G(x) &=& 3 \frac{1-8x+20x^2}{\sqrt{4x-1}} {\rm arccos}\left( \frac{3x-1}{2x^{3/2}} \right) 
- \frac{|1-x|}{2x}(2-13x+47x^2) \nonumber\nonumber \\
&& -\frac{3}{2} (1-6x+4x^2)\log (\sqrt x), 
\eea
where $1/4 < x < 1$, for energetically allowed decays. 
For the $B-L$ Higgs boson $\phi$, $\cos{\theta}$ will be replaced by $\sin{\theta}$.


\end{document}